\documentclass[useAMS,usenatbib]{mn2e}
\usepackage{amssymb,amsmath,graphicx}
\usepackage{longtable}[1]
\usepackage{graphics}
\usepackage{lscape}
\usepackage{color}
\title[Young SCs in ESO 185]{The Massive Star Clusters in the Dwarf Merger ESO\,185-IG13:  is the Red Excess  Ubiquitous in Starbursts?}

\author[A. Adamo et al.]{A. Adamo$^{1}$\thanks{E-mail:
adamo@astro.su.se}, G. \"Ostlin$^{1}$, E. Zackrisson$^{1}$, and M. Hayes$^{2}$\\
$^{1}$Department of Astronomy, Stockholm University, Oscar Klein Center, AlbaNova, Stockholm SE-106 91, Sweden\\
$^{2}$Observatoire Astronomique de l'Universit\'{e} de Gen\`{e}ve, 51, ch des Maillettes, CH-1290 Sauverny, Switzerland}

\newcommand{\araa}{ARA\&A}
\newcommand{\apj}{ApJ}
\newcommand{\aj}{AJ}
\newcommand{\mnras}{MNRAS}
\newcommand{\aap}{A\&A}
\newcommand{\pasp}{PASP}

\newcommand{\aaps}{A\&AS}
\newcommand{\apjl}{ApJ}
\newcommand{\msun}{M_{\odot}}

\begin{document}

\date{Accepted December 23, 2010. Received November 4, 2010}

\pagerange{\pageref{firstpage}--\pageref{lastpage}} \pubyear{2011}

\maketitle

\label{firstpage}

\begin{abstract}
We have investigated the starburst properties of the luminous blue compact galaxy ESO 185-IG13. The galaxy has been imaged with the high resolution cameras onboard to the {\it Hubble Space Telescope}. From the UV to the IR, the data reveal a system shaped by hundreds of young star clusters, and fine structures, like a tidal stream and a shell. The presence of numerous clusters and the perturbed morphology indicate that the galaxy has been involved in a recent merger event. Using previous simulations of shell formation in galaxy mergers we constrain potential progenitors of ESO 185-IG13. The analysis of the star cluster population is  used to investigate the properties of the present starburst and to date the final merger event, which has produced hundreds of clusters younger than 100 Myr. We have found a peak of cluster formation only 3.5 Myr old. A large fraction of these clusters will not survive after 10-20 Myr, due to the "infant mortality" caused by gas expulsion. However, this sample of clusters represents an unique chance to investigate the youngest phases of cluster evolution. As already observed in the analog blue compact galaxy Haro 11, a fraction of young clusters are affected by a flux excess at wavelengths longer than 8000 \AA. Ages, masses, and extinctions of clusters with this NIR excess are estimated from UV and optical data. We discuss similarities and differences of the observed NIR excess in ESO 185-IG13 clusters with other cases in the literature. The cluster ages and masses are used to distinguish among the potential causes of the excess. We observe, as in Haro 11, that the use of the IR and the  (commonly used) I band data results in overestimates of age and mass in clusters affected by the NIR excess. This has important implications for a number of related studies of star clusters.  
\end{abstract}


\section{Introduction}

ESO\,185-IG13 (hereafter,  ESO 185) is a luminous ($M_V=-19.4$) blue compact galaxy (BCG) at a distance of  80 Mpc (NED\footnote{http://nedwww.ipac.caltech.edu/}). The galaxy is now experiencing an active starburst episode. The relatively low extinction and metallicity across the galaxy ($E(B-V) \leq 0.16$, \citet{2001A&A...374..800O}) makes this irregular system very bright at blue wavebands. The galaxy shows morphological and dynamical signatures of a recent merger (\citealp{1999A&AS..137..419O}; \citealp{2000ASPC..211...63O}; \citealp{2001A&A...374..800O}): a tail in the north-east direction, numerous bright star clusters. The Fabry-Perot H$\alpha$ velocity field revealed two distinct dynamical components, where the secondary component is counter-rotating and with an estimated mass an order of magnitude lower than the mass of the main galaxy.  The total stellar mass is of $\approx 7 \times 10^9 \msun$ \citep{2001A&A...374..800O}.  However the observed rotation velocity cannot support more than a quarter of this mass, and even after adding the contribution from the velocity dispersion the galaxy seems to have a lower kinematical than stellar mass, suggesting that it is not in dynamical equilibrium.
ESO 185 has also been detected by IRAS at 60 $\mu $m at 0.4247$\pm$21\% Jy, and has an integrated H{\sc i} mass of  $3 \times 10^9 \msun$. The present starburst, which has produced hundreds of star clusters, is probably the result of a minor merger between a more massive and evolved system and a gas rich dwarf galaxy, which has supplied the resulting system with new fuel. 

\begin{figure*}

\resizebox{0.99\hsize}{!}{\rotatebox{0}{\includegraphics{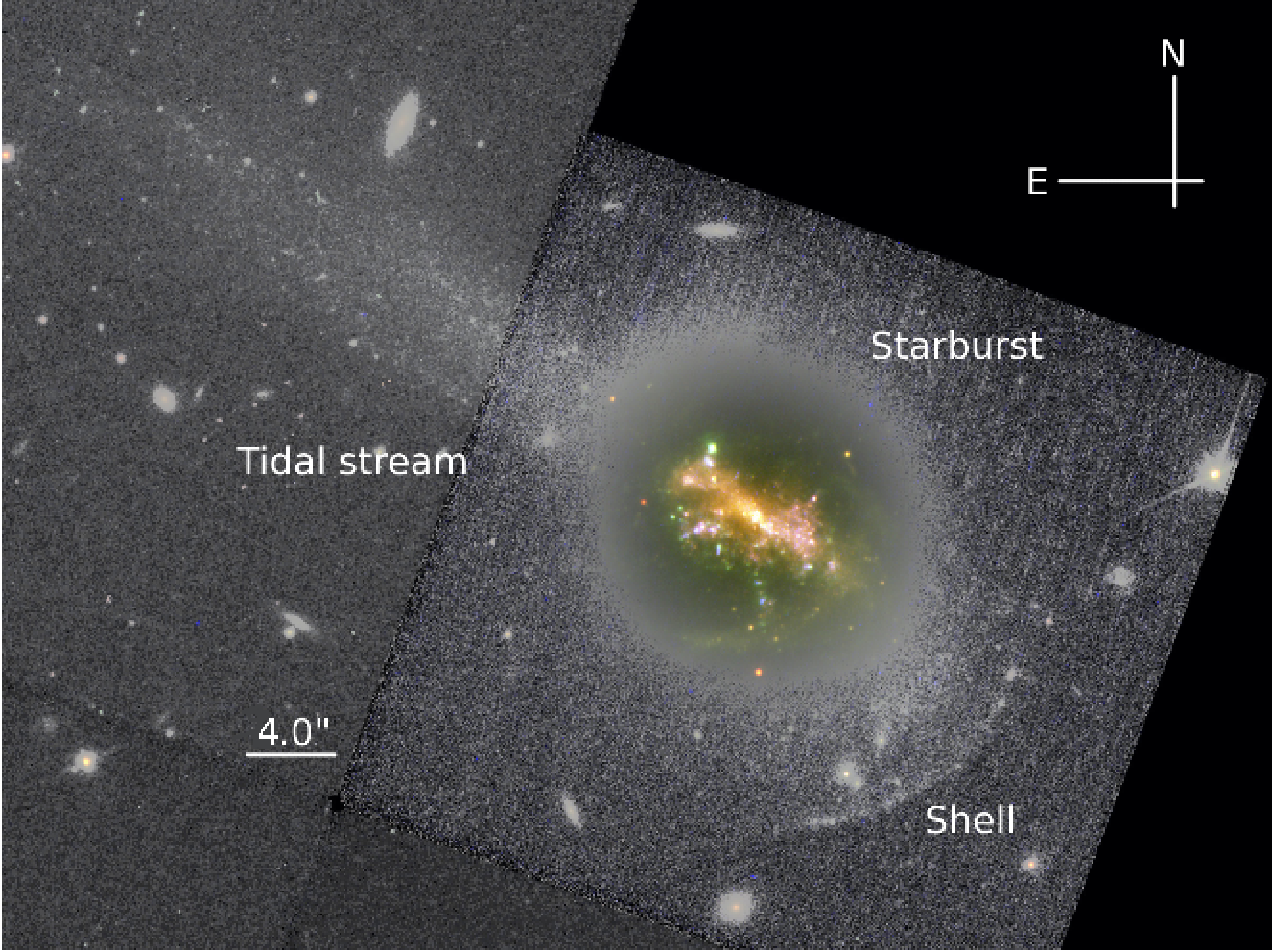}}}\\
%\resizebox{0.41\hsize}{!}{\rotatebox{0}{\includegraphics{eso185_ctr.ps}}}
\resizebox{0.47\hsize}{!}{\rotatebox{0}{\includegraphics{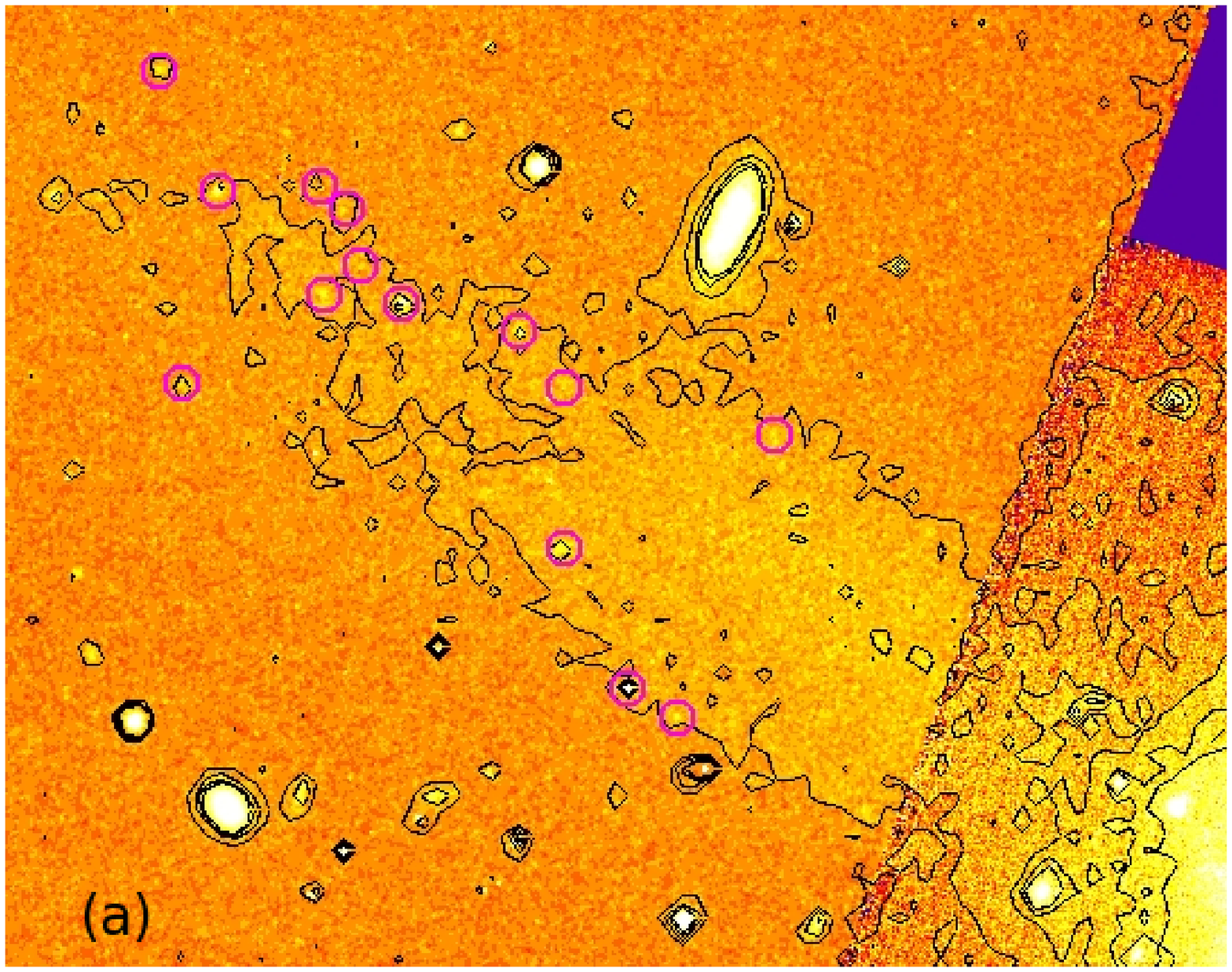}}}
\resizebox{0.52\hsize}{!}{\rotatebox{0}{\includegraphics{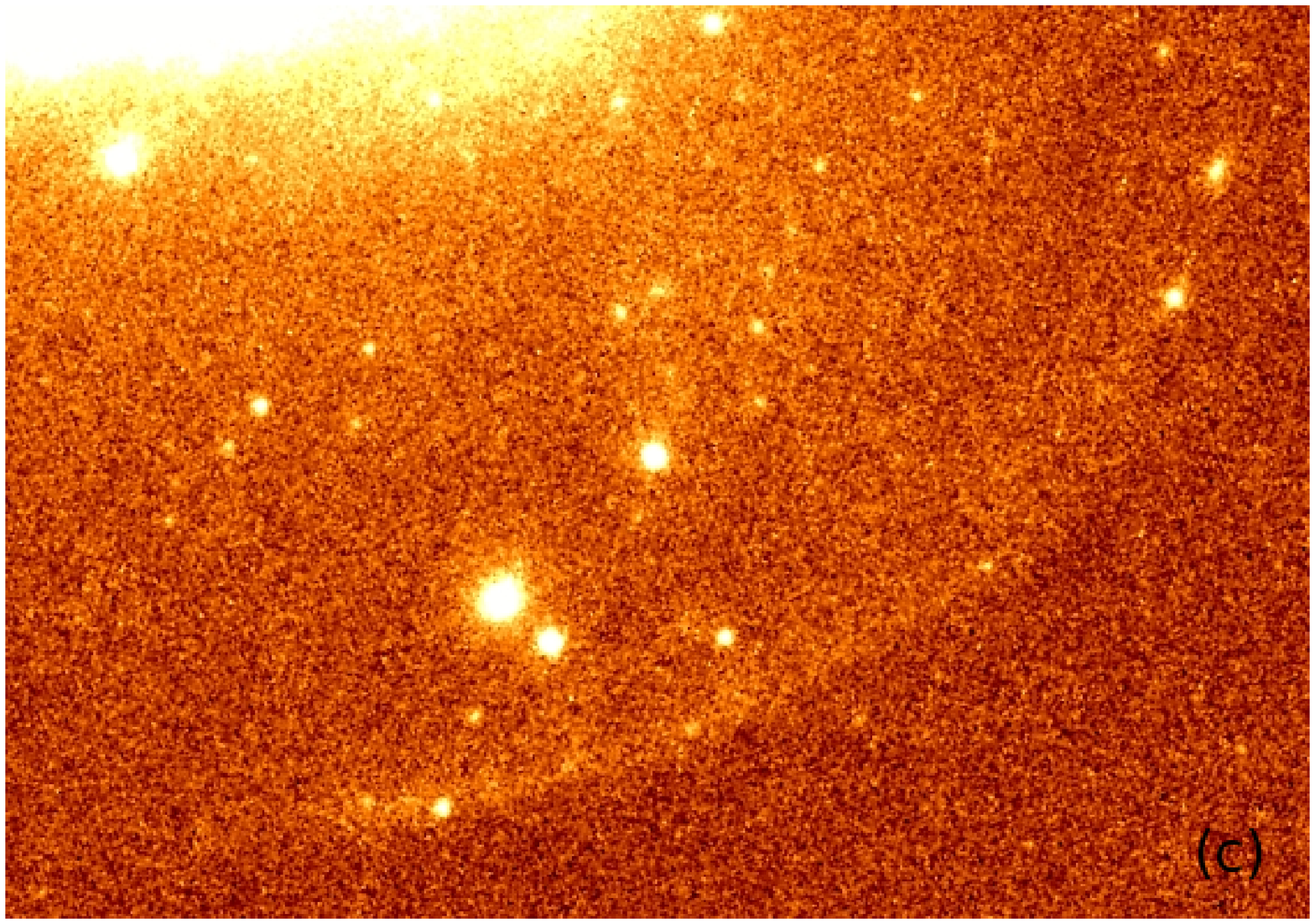}}}
\caption{Composite of the starburst galaxy ESO 185-IG13. On the top left panel, a saturated {\it WFPC2} F606W image, with a 3-color inset of the central starburst, clearly shows the position of the shell and of the tidal stream. The galaxy has been sampled with the high resolution Planetary Camera, while a part of the extended plume, in the Northeast side, falls onto the WF2 chip. Three different regions of the galaxy are outlined and showed in the corresponding panels. Tidal stream: Faint surface brightness contours reveal the shape of the plume as shown in Region (a). The purple circles indicate the position of the star clusters that we detected only in the two F606W and F814W deepest exposures. Starburst, Region (b): it is shown in Figure \ref{h11b}. Shell: In Region (c), we show a particular of the tidal-shell feature visible in the Southwest side of the galaxy. Several clusters are located between the central region and the shell. A few are also visible across the shell.  See the main text for more details.}
\label{h11}
\end{figure*}

\begin{figure*}
%\resizebox{0.8\hsize}{!}{\rotatebox{0}{\includegraphics{eso185_sat_zoomin.ps}}}\\
\resizebox{0.6\hsize}{!}{\rotatebox{0}{\includegraphics{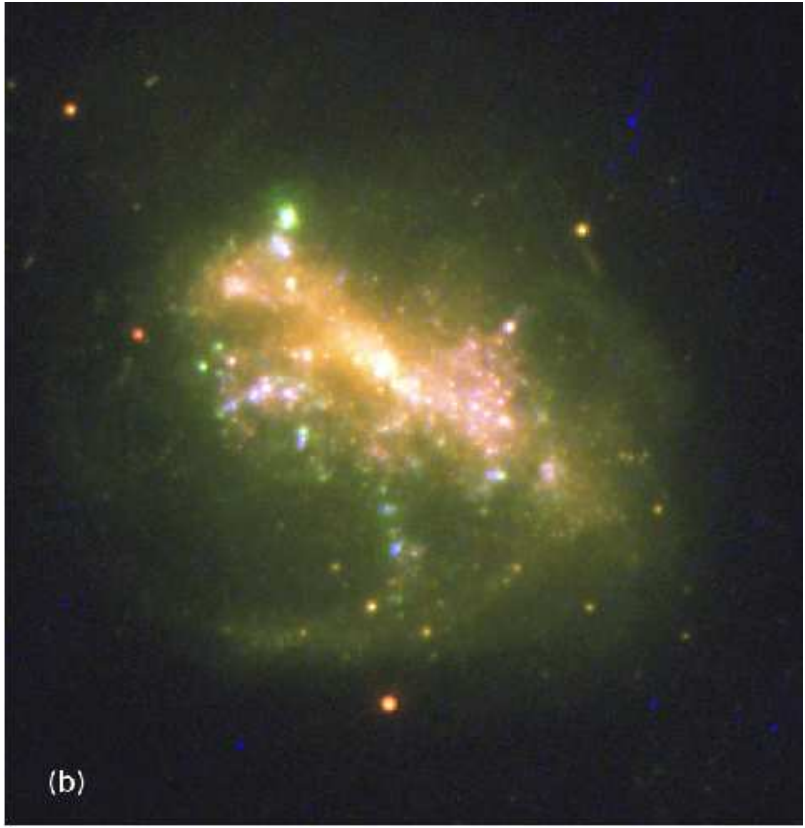}}}
%\resizebox{0.48\hsize}{!}{\rotatebox{0}{\includegraphics{shell.ps}}}
%\resizebox{0.48\hsize}{!}{\rotatebox{0}{\includegraphics{plume.ps}}}
\caption{Three-color image of the central starburst: {\it WFPC2} F814W filter in red, F606W in green and F336W in blue.  A barred-like structure with hundreds of SCs characterizes the nuclear region. We clearly see a hint of spiral arm features. Extended H$\alpha$ emission (transmitted by the F606W filter) surrounds the young starburst as shown by the prominent green halo. See the main text for more details.}
\label{h11b}
\end{figure*}

New high-resolution {\it Hubble Space Telescope} ({\it HST}) imaging data\footnote{Based on observations made with the NASA/ESA
Hubble Space Telescope, obtained at the Space Telescope Science
Institute, which is operated by the Association of Universities for
Research in Astronomy, Inc., under NASA contract NAS 5-26555.} (\# GO10902, PI: {\"O}stlin) of the galaxy allowed us to investigate the formation and evolution of the system, through the analysis of the cluster population. A composite image of the galaxy is showed in Figure \ref{h11}. The top panel shows the galaxy in the deep WFPC2 F606W frame with a 3-colors inset of the central starburst region. Three different regions are outlined and showed in the other two panels and in Figure~\ref{h11b}. The Figure~\ref{h11b} shows a 3-colors composition of the galactic main body. A barred-like structure crosses the center of the galaxy and is formed by hundreds of star clusters and several starburst knots (cluster complexes). Arm paths are traced by star clusters and surrounded by H$\alpha$ emission (green halo in the image). The H$\alpha$ emission is prominent. \citet{1999A&AS..137..419O} estimated a H$\alpha$ luminosity of $L_{H\alpha} \sim 80 \times 10^{33}$ W and an ionised gas mass of few times $10^8 \msun$.  In the bottom left panel (Figure~\ref{h11}) in false colors, we show the tidal stream or plume (both terms will be used in the text) extracted from the F606W image. The plume is approximatively 32.0" long. The contours draw the shape of the plume at $2-3\sigma$ levels from the background. It is a low density structure. We detect clusters (purple circles)  at the end of  filament-like structures. We have, for the first time in this galaxy, detected a shell structure 8.0" away from the starburst region (13.5" radius from the galactic center) in the southwest region of the galaxy. Numerous star clusters are observed in the region between the shell and the main galactic body and on the shell. 

Shells have been observed in many early-type post-starburst galaxies and are typical signatures of recent  mergers (\citealp{1992ApJ...399L.117H}; \citealp{1997ApJ...481..132K}). Using self-gravitating systems of stars and dissipative gas clouds, \citet{1997ApJ...481..132K} simulated a minor merger event between an elliptical system and a gas-rich satellite. They found that the gas of the "sinking satellite" was not only confined to the inner regions of the new formed system, but also distributed in the fine structures formed during the merging, like shells and ripples. Interestingly, they observed short starburst episodes, abruptly reduced due to the dissipative nature of the gas. Millimeter observations of NGC 5128 (CenA, \citealp{2000A&A...356L...1C}), revealed the presence of molecular gas across two of the many shells surrounding the galaxy. Similar features were also observed in the Medusa merging system (NGC4194, \citealp {2001A&A...372L..29A}). These observations confirmed the findings by \citet{1997ApJ...481..132K} and suggested the possibility of cluster formation in shells. Recently, from the analysis of the globular clusters (GCs) in the post-starburst  shell galaxy AM 0139-655, \citet{2007AJ....134.1729M} found some of the youngest GCs ($\sim 0.4$ Gyr) located in the shell. They discussed the possibility that the clusters and the shell may have formed in the same dissipative merger event, and that the star formation in the galaxy was shut  down after this last interaction. 

The formation of star clusters is usually enhanced in interacting/merging systems (the Antennae,\citealp{1999AJ....118.1551W}; M51, \citealp{2005A&A...431..905B}; the Bird, \citealp{2008MNRAS.384..886V} among many others). They have also been found in tidal tails (\citealp{2003AJ....126.1227K}; \citealp{2007ApJ...658..993T}) and collisional rings \citep{2010AJ....139.1369P}. Starburst dwarf galaxies have also proved to be very efficient in producing massive young clusters (\citealp{2002AJ....123.1454B};  NGC 1569, \citealp{2004MNRAS.347...17A};  NGC 1709, \citealp{2009AJ....138..169A}; to cite a few examples). The BCGs belong to this wide group of irregular and low mass galaxies. Deep imaging studies of the cluster populations in BCGs, like ESO 338-IG04 \citep{2003A&A...408..887O} and Haro 11 \citep{A2010}, have revealed very high cluster formation rates (CFRs) during the last 30-40 Myr, and a peak of cluster production less than 10 Myr ago. The analysis of the star cluster population in BCGs can be used to: 1) reconstruct the formation history and the environmental properties of the host galaxy; 2) investigate the complex and short embedded phase, during which more than 90  \% of the clusters will be destroyed (the so-called "infant mortality", \citealp{2003ARA&A..41...57L}). 

The embedded phase is a transient period in the cluster evolution. The clusters form from the collapse and fragmentation of giant molecular clouds (GMCs). The strong UV radiation, produced by the young massive stars, ionize the intracluster gas and create H{\sc ii} regions around the clusters. However a large fraction of stars is still accreting material from their dusty disks (young stellar objects, YSOs) or contracting in the pre-main sequence phase (see \citealp{2010ApJ...713..883B} for an example of different components of a massive star-forming region in the Milky Way). During this partially embedded phase, which could last a few Myr \citep{2009arXiv0911.0779L}, the gas expulsion is the main mechanism causing the early dissolution of the clusters.  NIR studies of very young unresolved clusters in nearby galaxies have revealed fluxes that are impossible to reconcile with evolutionary models of stellar continuum and photoionized gas. In Haro 11, we uncovered a very prominent population of clusters affected by this NIR excess \citep{A2010}. In other starburst galaxies (\citealp{2009MNRAS.392L..16F}; \citealp{2005A&A...433..447C}; \citealp{2005ApJ...631..252C}),  the NIR excess found in very young clusters has been explained as due to a significant flux contribution from young stellar objects (YSOs) or from other sources of hot dust. The use of IR data clearly reveals a much more complex scenario than the one predicted by single stellar population models and more investigations are required to increase our understanding of the star formation process in clusters.

The paper is organized as follow: In Section \ref{data-sample} we describe the data reduction and the photometric analysis, including completeness tests. Section \ref{prop_sc} contains a detailed description of the spectral energy distribution (SED) analysis  of the star clusters, and how we constrain the properties of clusters affected by the red excess. The final age, mass, and extinction distributions are shown in this section, along with Monte Carlo simulation tests. We discuss the possible mechanisms causing the NIR excess in young clusters in Section \ref{rex-cause}. In Section \ref{starb}, we discuss the environmental properties of the host starburst galaxy as revealed by the star clusters population. In the last paragraph of this section, we investigate the formation and possible evolution of the galactic system. The conclusions of this work are summarized in the last section.

\section[]{Data sample and multi-band photometry}
\label{data-sample}
\subsection{Observations \& Reduction}

{\it HST} multiband high-resolution imaging of ESO\ 185 were carried out in 2007 (associated with programs \# GO 10902, PI:
G. \"Ostlin.). The galaxy was sampled from the FUV ({\it ACS}/F140LP) to the NIR ({\it NIC3}/F160W) in order to analyze the stellar and gas components. 

\begin{table*}
  \caption{{\it HST} observations carried out within the framework of program \# GO 10902 (PI G. \"Ostlin). For each filter we list in the table: the total exposure time; the corresponding ZPs  (Vega magnitude system); the aperture corrections, a$_c$; the number of objects detected with a $S/N\geq5$; and the corresponding magnitude limit. $a$) Both the {\it HST} filter names and the abbreviated  nomenclature indicated in bracket are used thereafter in the text.  ${b}$) The frames in the WF2 chip are used for the analysis of the plume (Region (a) in Figure \ref{h11}).}
\centering
  \begin{tabular}{|c|c|c|c|c|c|c|c|}%{@{}llll|@{}}
  \hline
  Instrument &Filter$^{a}$ & Camera &  Exposure time & ZP (mag)&a$_c (mag)$ &N($\sigma\leq 0.2$)&  mag limit\\
   \hline
 ACS&F140LP (FUV)&SBC&2584 s&20.92&-0.54$\pm$0.05&210&24.8\\
  \hline
WFPC2& F336W (U)&PC&1200 s&19.43&-0.42$\pm$0.05&96&23.4\\
WFPC2& F439W (B)&PC&800 s&20.88&-0.36$\pm$0.05&121&24.7\\ 
WFPC2& F606W (R)&PC&4000 s&22.89&-0.61$\pm$0.09&420&27.5\\
& &WF2$^{b}$&4000 s&22.92&-0.31$\pm$0.05&11&27.9\\
WFPC2& F814W (I)&PC&4500 s&21.64&-0.73$\pm$0.04&420&26.7\\ 
& &WF2$^{b}$&4500 s&21.66&-0.35$\pm$0.05&11&28.1\\
\hline
 NICMOS&F160W (H)& NIC3 &4992s&21.88&-2.45$\pm$0.37&126&25.9\\
  \hline
\end{tabular}
\label{table-obs}
\end{table*}

The angular size of the starburst region is quite small (major axis about 26.0") and easily fitted into the field of view of the Planetary Camera (PC), the Solar Blind Channel (SBC), and the NIC3 camera. In a previous short exposure of the galaxy (GO 5479: PI M. Malkan) the plume fell outside the detector. In the images we got in 2007 the position of the galaxy was set so that the plume could be sampled by the WF2 camera. The same position was used for all the data taken with the {\it WFPC2} instrument. We find that the plume is around 32.0" long, too extended to fit into the SBC field of view. The plume was only partially imaged in the NIC3 frames.

The frames were reduced, drizzled and aligned using the {\tt MULTIDRIZZLE} task
(\cite{2002hstc.conf..337K}; \cite{2002PASP..114..144F}) in {\tt
PyRAF/STSDAS}\footnote{STSDAS and PyRAF are products of the Space
Telescope Science Institute, which is operated by AURA for NASA}. The final {\it WFPC2} and {\it ACS} frames were rescaled to $0.025 "/pixel$, before combining. To improve the large plate scale of the NIC3 camera ($0.2"/pixel$)  the galaxy was imaged with a double NIC-SPIRAL-DITH path in 3 dithered pointing. This allowed us to reach a diffraction limited sampling, $0.067"/pixel$, of the final science frame. A more detailed description of the {\it NICMOS} data reduction is given in \citet{A2010}. A list of the filters, total exposure time, zero points (ZPs), and other observational properties related to the reduction of the data are listed in the Table \ref{table-obs}.

\subsection{Cluster catalogue and photometry}
\label{cluster-det}

The extraction of the position of the clusters was done with {\tt SExtractor} \citep{1996A&AS..117..393B}. We cross-checked the detections in the two deepest exposures ({\it WFPC2} F606W and F814W). Crowded regions inside, and isolated ones surrounding the starburst, were treated with different set of parameters (see \citealp{A2010}). {\tt SExtractor} was also run on the WF2 frame to extract potential cluster candidates in the plume. We established by eye the center of the starburst region ($\alpha = 19:45:05.2$; $\delta = -54:16:01.1$) and considered candidate detections inside a circle of radius 15.12". For the plume we delimited the studied region in the WF2 frame to a box of 18.85"$\times$20.8". These chosen projected dimensions were based on the drop of the S/N outside the delimited regions. A fraction of the plume falls on the PC camera frame, however this region is still part of the outskirts of the starburst area. Therefore, we identify plume  clusters as the ones detected in the WF2 frame (see Figure \ref{h11}). This cross-checked catalogue contained 1265 cluster candidates of which 14 are in the plume (shown in panel (a) of Figure \ref{h11}). Using this catalogue we performed aperture photometry ({\tt IRAF/PHOT} task) in all the available science frames. The aperture radius for doing photometry was fixed to 0.1" (similar to the full width half maximum of the point spread function, PSF, in the UV and optical frames, see \citealp{A2010} for details) and a local sky annulus of 0.05" wide at the distance of 0.125" from the center was used. This set of parameters was the same for all the frames from the UV to the IR. We estimated then a mean aperture correction for each frame to take into account the flux lost by the fixed radius used and the different PSFs. More details regarding the aperture correction estimates are given in \citet{A2010}, since Haro 11 and ESO 185 were observed with {\it ACS}/SBC, {\it NICMOS}, and {\it WFPC2} setups. The aperture corrections and their associated uncertainties are listed in the Table \ref{table-obs}.
The cluster candidates in the plume were only detected in the two deepest F606W and F814W exposures. The two F336W and F439W frames are too shallow and heavily affected by charge transfer efficiency (CTE) problems, therefore the plume and clusters were undetected. We performed photometry in the two  F606W and F814W WF2 frames, using an aperture radius of 0.2" (comparable with the full width half maximum of the PSF in these two frames) and a local sky annulus of 0.05" wide at the distance from the center of 0.225". 

For the {\it WFPC2} images, we applied CTE corrections to the observed fluxes using a modified  version of the 
algorithm by \cite{2009PASP..121..655D}. As already described in \citet{A2010}, we chose not to use the median background
value, as prescribed by Dolphin's formula, but a
"local'' background for each source. 

\subsubsection{Final catalogue}
\label{fin-cat}

\begin{figure}
\resizebox{\hsize}{!}{\rotatebox{0}{\includegraphics{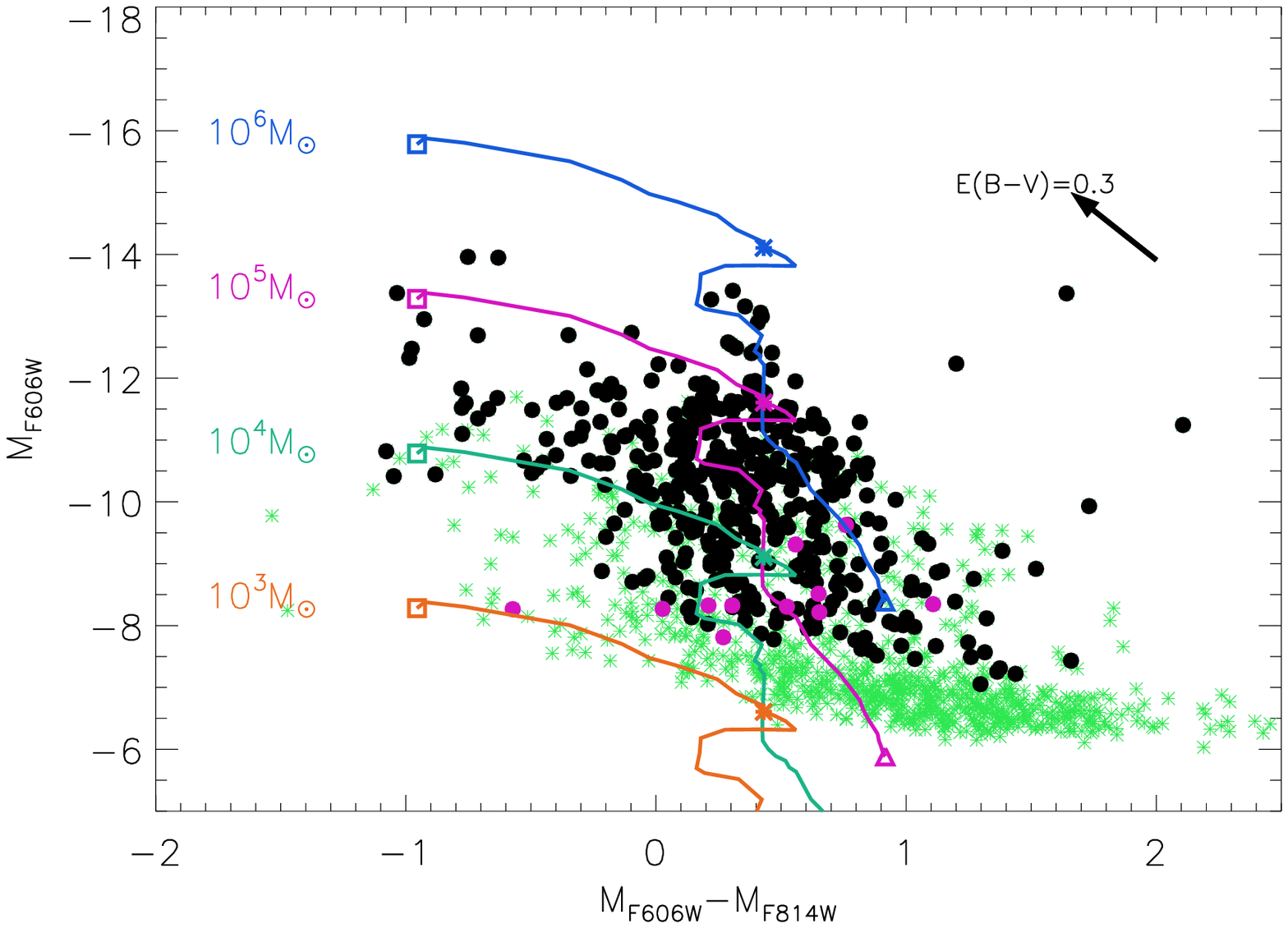}}}
\resizebox{\hsize}{!}{\rotatebox{0}{\includegraphics{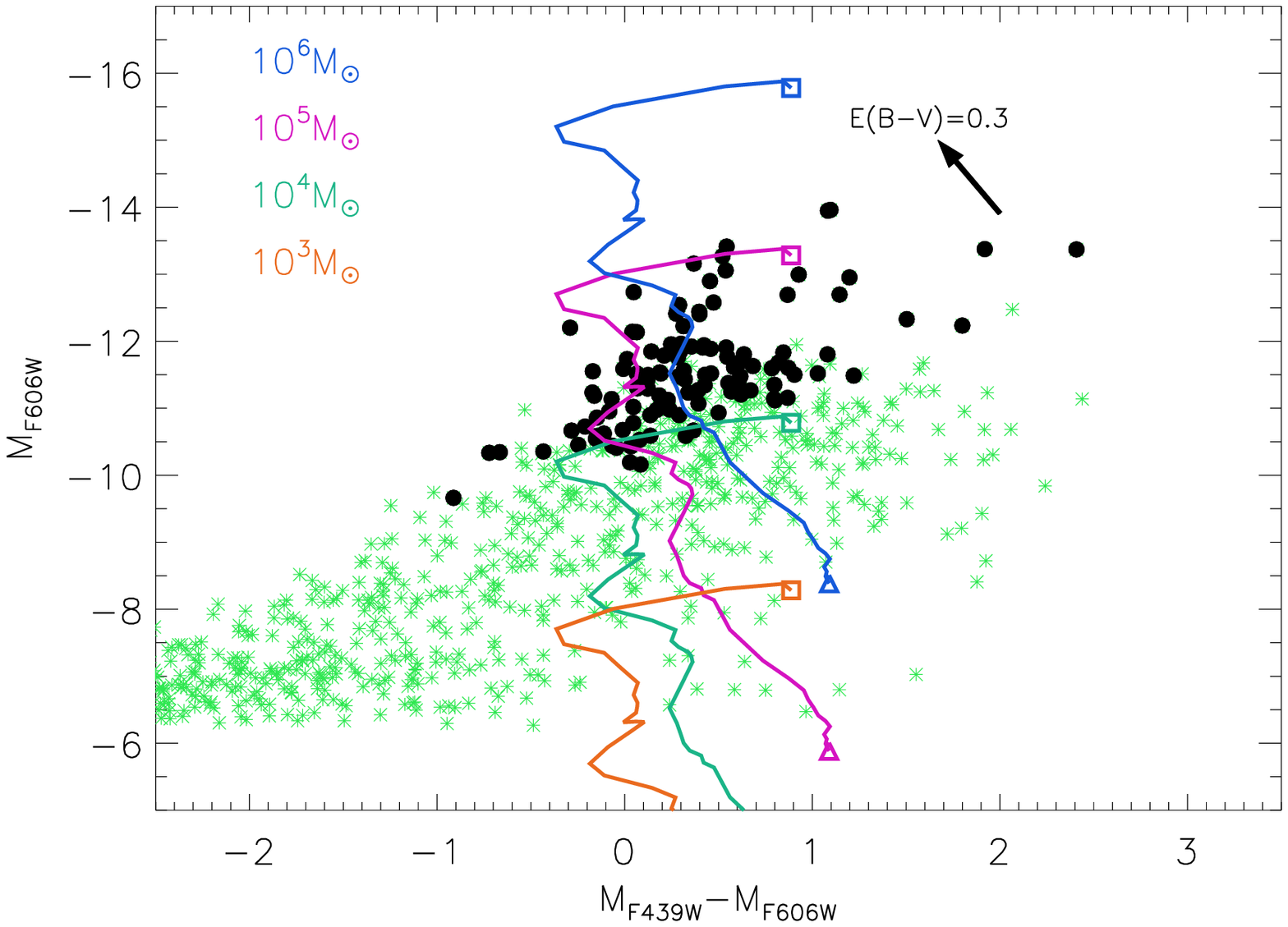}}}
\caption{Color magnitude diagrams of the cluster population. Top panel: the small green (grey, in a B/W printed version) asterisks show the 1265 cluster candidates detected in the starburst region and the plume. The black filled dots represent the 420 clusters with photometric errors in I and R, $\sigma_m \leq 0.2$. The purple (grey) filled dots are the 11 clusters detected in the plume. Bottom: the green asterisks show the candidates which have been detected in R, I, and B band. The filled dots are the clusters with $\sigma_m \leq 0.2$ in R, I, and B band. The extinction vector indicates in which direction the clusters move if corrected for extinction. The evolutionary track from Z01 are for Z=0.008 and several values of the mass. See main text for details.}
\label{CMD}
\end{figure}

The final catalogue was created applying the following criteria:
\begin{description} 
\item i) the output photometric error associated with each measurement was used to reject estimates with magnitude errors $\sigma_m > 0.2$. In Table \ref{table-obs} we show the corresponding magnitude limits and the number of objects retained in each frame after this selection;
\item ii) the catalogue includes only objects detected in at least 3 filters;
\item iii) the final error associated with the integrated flux in each filter was the combined quadrature of the photometric error and the uncertainty on the aperture correction determination (see Table  \ref{table-obs});
\item iv) we corrected the fluxes for the foreground galactic extinction \citep{1998ApJ...500..525S}.
\end{description}
In total the final catalog contains 290 objects\footnote{The complete
photometric catalogue is available on request from the
authors.} with detection in at least 3 filters. All the detections in the plume, except one (which is also detected in the H band) are excluded from this final catalogue. Looking at the Table \ref{table-obs}, it is possible to see that in the R and I bands the number of clusters with  $\sigma_m \leq 0.2$ is significantly higher (420 in the starburst and 11 in the plume). We show this numerous population in the color-magnitude diagram (CMD) on the top panel in Figure~\ref{CMD}. We also include a second CMD, replacing the deep I band exposure with the shallower B filter, which clearly demonstrates  the difference in the quality of the available data. Sources with a S/N larger of 5 in B band correspond to clusters more massive than $10^4 \msun$ or brighter than $-10.0$ mag in absolute magnitude.

\subsubsection{Completeness limits}
\label{completeness}

\begin{figure}
\resizebox{\hsize}{!}{\rotatebox{0}{\includegraphics{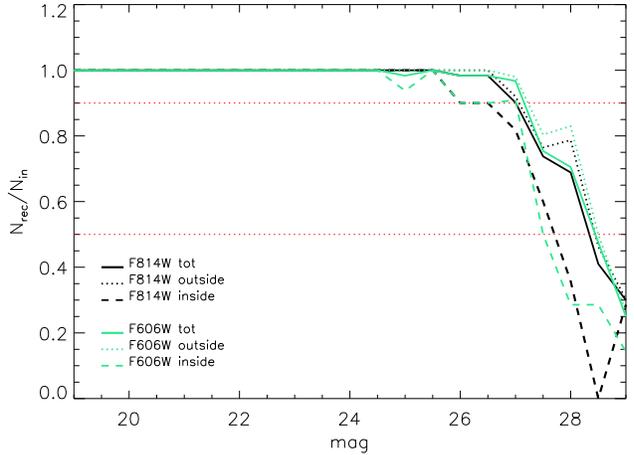}}}
\caption{Completeness fraction as function of the object magnitudes. The outputs of the F606W frame are plotted in green (grey), while in black we show the fraction for the F814W frame. See the main text for the definition of "inside" and "outside" regions. The red dotted straight lines show the 90 \% and 50 \% completeness limits.}
\label{compl_test}
\end{figure}

We performed completeness tests on the two R and I {\it WFPC2} frames to trace the sensitivity limits reached in the analysis. In \citet{A2010}, we give a detailed description of the method used to construct the completeness test. Since the R and I imaging have the same properties as the ones used for the Haro 11 analysis, we used the two PSF samples constructed in \citet{A2010} to make grids of fake sources with {\bf mksynth}  \citep{1999A&AS..139..393L}. We created three different grids of positions and for each of them the magnitudes were assigned randomly, to avoid systematics in the test.  As function of the magnitude bin, we estimated the average recovered fraction of objects (allowing a shift smaller than 3 pixels). The number of recovered objects depended on the crowding of the region. Similarly to what we did for Haro 11, we defined a radius of 6.5" to distinguish between the crowded central region and the surroundings with low background. In Figure \ref{compl_test}, we show, for each filter, the total fraction of recovered detections, and how this fraction changes as function of the environment where the sources have been detected. The 90 \% completeness is reached at 27 mag for both filters (continuos black and green lines). Inside the crowded region 10 \% of objects with $25-26$ mag are already missed and the fraction increases for fainter ones. On the other hand, objects located outside the starburst area have been easier to detect. This result confirms that low mass young objects and more evolved faint clusters have higher chances of being detected around the intense starburst region, while in crowded regions, we are not able to efficiently de-blend low brightness sources.  

\section{Analysis of the cluster properties}
\label{prop_sc}
\subsection{Spectral Evolutionary Synthesis Models and $\chi^2$ SED fit }
\label{chi2}
To estimate the ages, masses, and extinctions of the clusters we used the
\citet[][hereafter Z01]{Zackrisson et al. a} spectral synthesis
model. This model predicts the combined SED of both stars and
photoionized gas. The gas around the star cluster, photoionized by the young massive stars, contributes both to the continuum and with emission lines to the integrated broadband fluxes of the young systems \citep[e.g.][]{Krüger et al., Zackrisson et al. a, Anders & Fritze-Alvensleben, 2008ApJ...676L...9Z, R2009, A2010, A2010b}. In \citet{A2010b} we show that nebular emission non-negligibly affects  the SEDs of the clusters during the first $10-15$ Myr of cluster evolution. The contribution becomes smaller after 6 Myr on the blue side of the cluster spectrum, but lasts longer in the NIR wavebands. This is a very important property to keep in mind when both optical and IR fluxes are used to constrain the cluster properties.

The use of the Z01 models and the $\chi^2$ - algorithm are thoroughly presented in \citet{A2010}. For the analysis of the ESO 185 cluster population we used the same parameter set as for Haro 11: Kroupa's IMF \citep{2001MNRAS.322..231K}, standard H\,{\sc ii} region values for the gas , but a slightly higher metallicity for the gas and the stars ($Z=0.008$). The spectral
synthesis model of \citet[][hereafter M08]{Marigo et al.} has been used to test the robustness of
some of our results. 
\begin{figure*}
\resizebox{0.48\hsize}{!}{\rotatebox{0}{\includegraphics{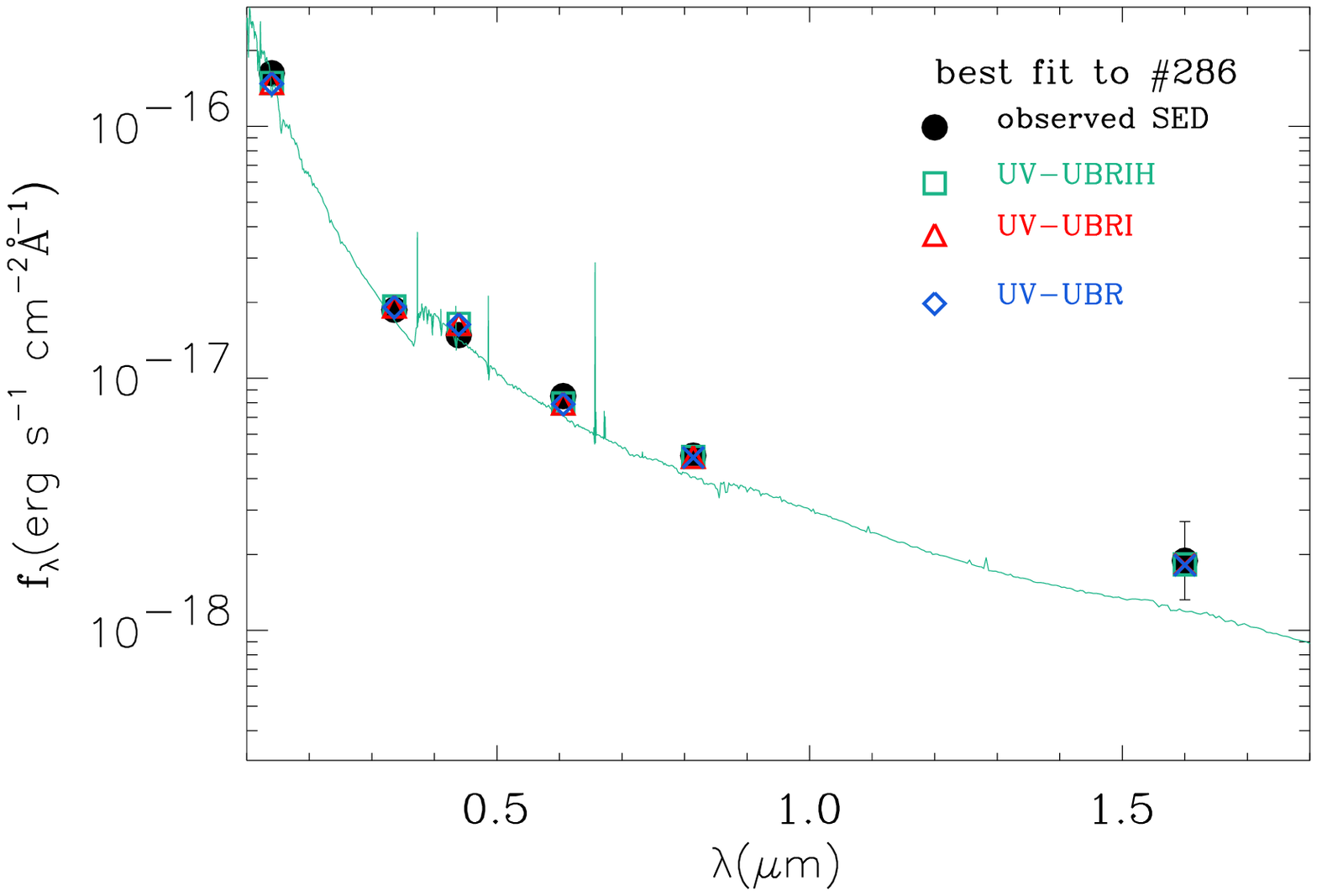}}}
\resizebox{0.48\hsize}{!}{\rotatebox{0}{\includegraphics{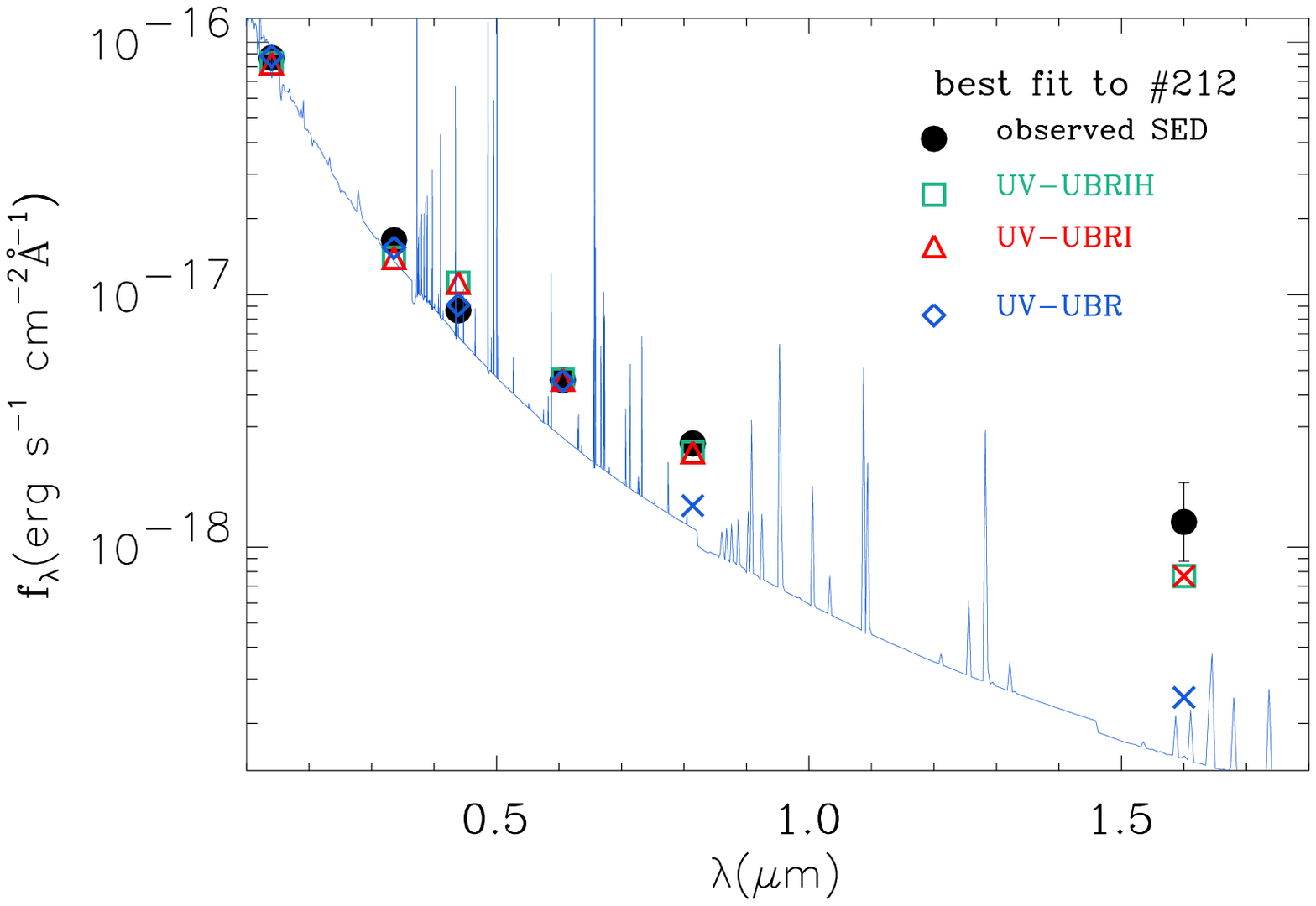}}} \\

\resizebox{0.48\hsize}{!}{\rotatebox{0}{\includegraphics{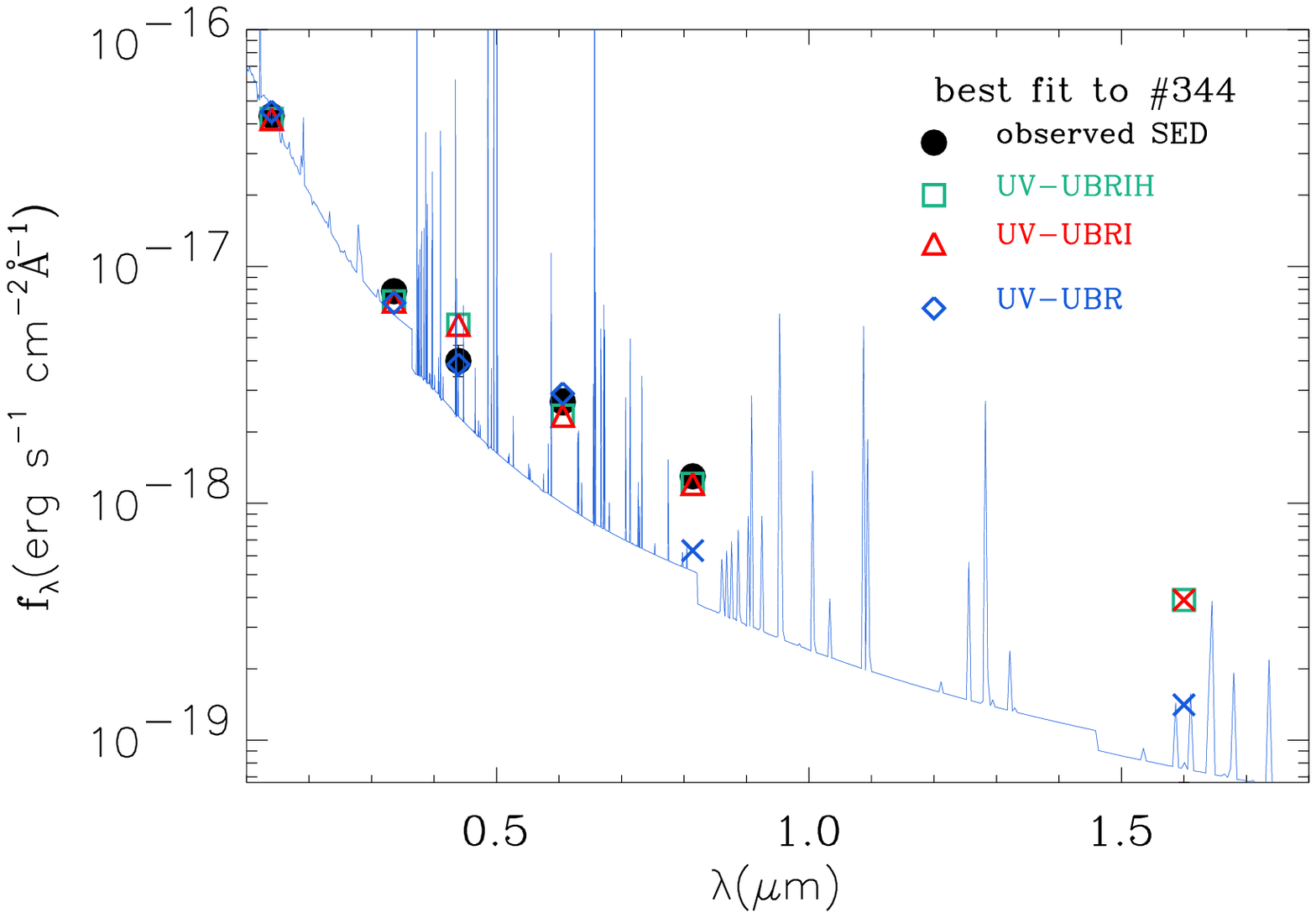}}}
\resizebox{0.48\hsize}{!}{\rotatebox{0}{\includegraphics{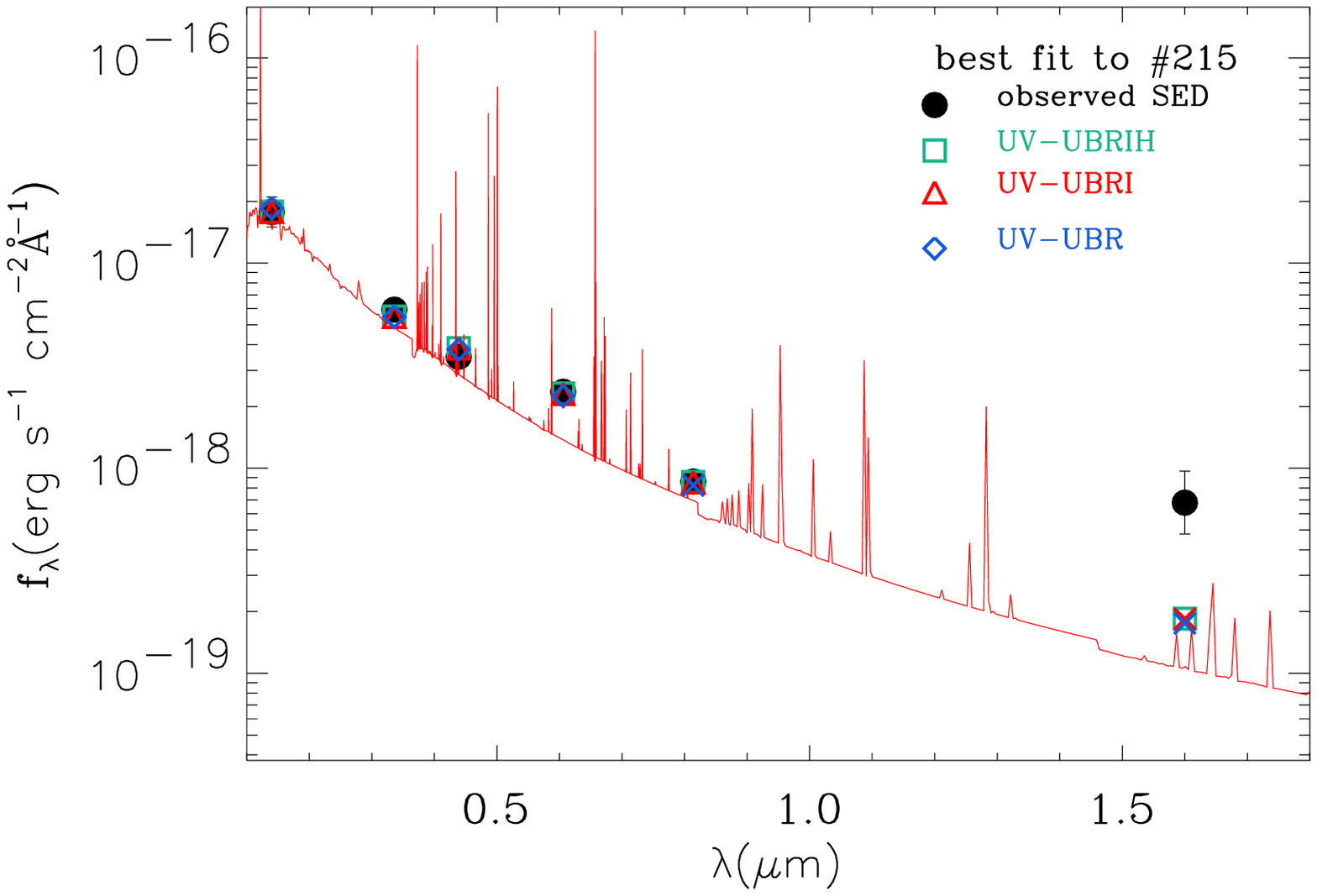}}} \\
\caption{SED analysis of 4 clusters. The filled black points indicate
the observed fluxes of each cluster. The integrated model
fluxes are labelled with different symbols for each set of fit as
indicated in the plot insets. The crosses in the {\it UV-UBR fit} and {\it UV-UBRI fit} show which detections have not been included in the fits. We plotted the spectrum of the final best
fitting model with the colour corresponding to the best fit run. The integrated model fluxes (symbols) sit above the model continuum because of the contribution by emission lines.}
\label{spec}
\end{figure*}

We performed $\chi^2$-fit of SEDs with detections in at least 3 filters. The extinction was treated as a free parameter. We used the Calzetti attenuation law \citep{2000ApJ...533..682C}, which includes a more general treatment of all the complex environments in the galaxy (see \citealp{A2010} for a comparison between the Calzetti and the LMC extinction law). The cluster mass was estimated from the normalization factor between the best fitting model and the observed SED.

\subsection{Constraining the SED of the clusters; with or without red excess}

In \citet{A2010} we discovered that several young clusters have a flux excess in the I, H, and Ks bands. The inclusion of those filters in the $\chi^2$ fit produced high residuals, and a clear overestimation of the cluster age. Due to the similarity between Haro 11 and ESO 185 in terms of optical luminosity, metallicity content and patchy extinction, we expected to find young clusters possibly showing the same excess.  

In the first run of $\chi^2$-fits we included detections in all the available filters ($\sigma \leq 0.2$). Thereafter, we refer to this run as the {\it UV-UBRIH fit}. However, not all the objects in the catalogue were detected in the 6 available frames. Of the 290 candidates, 50 \% are detected in 3 filters, 21 \% in 4 filters, and the remaining 29 \% in 5 or 6 filters. 
In the second run of $\chi^2$-fits, the {\it UV-UBRI fit}, we excluded the H band detections. In this case, the number of fitted SEDs diminished from 290 to 242. Finally the third set of  $\chi^2$-fits, the {\it UV-UBR fit}, only included the blue optical and UV filters. In this final run, only 117 clusters could be fitted. The small number of clusters is due to the shallow B and U frames.

Using plots as the ones shown in Figure \ref{spec} and comparing the reduced $\chi^2_r $ and corresponding $Q$ values in the 3 sets of fits we could distinguish whether a cluster is affected by some kind of excess. In general, the exclusion of a data point generates a smaller $\chi^2_r$,   however the best fitting model should be the same if the observed fluxes are self-consistent. In our exercise we found that several objects didn't behave as expected. As already observed in many clusters in Haro 11, there are systems which show a better fit to the UV and blue optical data points when the IR and I bands fluxes are excluded. Systematically, the excluded observed fluxes sit above the best fitting models. It is clear that self-consistent stellar and nebular evolutionary models are not enough to interpret the NIR luminosity properties of these clusters and other mechanisms need to be addressed (see Section \ref{rex-cause} of this work and Section 5 in \citealp{A2010}).  

In Figure \ref{spec}, we show a representative sample of cluster SEDs in ESO 185. The $\chi^2$ outputs of the 3 runs are collected in Table \ref{fit-output}. More objects are included in the Appendix (Figure \ref{spec_app1} and  \ref{spec_app2} and Table \ref{tab-chi2}). In Table \ref{tab-phot} of the Appendix we list the magnitudes, in the Vega system, of all the shown cluster SEDs.

\begin{table}
  \caption{Final outputs given by the three sets of SED fits as shown in Figure~\ref{spec}. In bold we show the final age, mass and extinction assigned to the clusters.}
\centering
  \begin{tabular}{|c|c|c|c|c|c|}
  \hline
  \hline
 IDs&$\chi_r^2$& $Q$&Myr&$10^5 \msun$&E(B-V)\\
 \hline
 \hline
 &\multicolumn{4}{|c|}{{\it UV-UBRIH} fit} \\  
 \hline  
     286     &  {\bf2.5}   &   {\bf0.109}&  {\bf9.5}     &   {\bf2.6}  &    {\bf 0.0}  \\
     212   &    8.4  & 5.e-05   &19.5    &   4.5 &  0.03\\
     344     &  5.0   & 0.018  & 19.5    &   2.3   &  0.03\\
     215    &   2.0    &  0.205  & 3.5     &  0.4   &  0.19\\
\hline
 &\multicolumn{4}{|c|}{{\it UV-UBRI} fit} \\  
 \hline  
     286  &    3.8  &     0.056  &   9.5    &    2.7     &    0.0   \\
     212   &        12.0  &   2.e-05  &   19.5    &     4.5  &    0.03\\
     344   &5.0    &   0.018  &   19.5     &    2.3   &    0.03\\
     215    &  {\bf0.9 } &      {\bf0.632}  &    {\bf3.5}   &      {\bf 0.4 }   &    {\bf 0.19}\\
\hline
 &\multicolumn{4}{|c|}{{\it UV-UBR} fit} \\  
 \hline  
     286  &     7.4   &    0.024   &  9.5    &     2.6   &      0.0\\
     212 &      {\bf1.6}    &     {\bf0.444}  &    {\bf3.5 }    &     {\bf0.5 }  &     {\bf0.05}\\
     344     &     {\bf1.8}   &     {\bf0.401 }  &  {\bf 2.5 }   &      {\bf0.3  }  &    {\bf0.05}\\
     215    &       1.7    &    0.430  &   3.5     &    0.4  &     0.18\\
\hline
 \hline
\end{tabular}
\label{fit-output}
\end{table}

Cluster \#286 is an example of an "expected" regular cluster SED. The top left panel of Figure \ref{spec} shows that the 3 sets of fits were all reasonable successful. Moreover, from Table \ref{fit-output} we see that the best age, mass, and extinction do not change from one fit to another. On the other hand, the {\it UV-UBRIH fit} fails to reproduce the SED of the clusters \# 212, \# 344, and \# 215. Cluster \# 212 has a clear excess in H band, and the best model produced by the {\it UV-UBRIH fit} cannot properly fit the U and B data points. The exclusion of the H band data, in the {\it UV-UBRI fit}, does not provide an acceptable fit (outputs in Table \ref{fit-output}). Only the exclusion of the I band allows us to find a best fitting models able to fit the U and B data. The outputs of the {\it UV-UBR fit}, in Table \ref{fit-output}, show that the cluster is younger and less massive than predicted by the other two sets of fits. Cluster \# 344 is similar in properties to \# 212, except that this cluster is not detected in the H band. Finally, cluster \# 215 has a clear excess in the H band but not in the I band. 

As already done in the analysis of the Haro 11 clusters, we estimate the properties of the clusters affected by NIR excess using only the UV and blue optical SEDs, or in other words with the {\it UV-UBRI} and {\it UV-UBR fits}. From the initial 290 analyzed objects, 197 clusters do not  have any NIR flux excess and have been fitted by a {\it UV-UBRIH fit}. A total of 37 clusters have an I and an H band (if detected) excess, similar to clusters \# 212 and \# 344; and 30 objects show properties similar to \# 215. The remaining 26 objects, mainly detected in only 3 filters have been excluded because of inconsistent fits. Among them, 4 objects are not detected in the UV-U bands but show an increasing brightness from the R to the H filters. We will discuss the fit to these four sources in the following section.
\begin{figure}
\resizebox{\hsize}{!}{\rotatebox{0}{\includegraphics{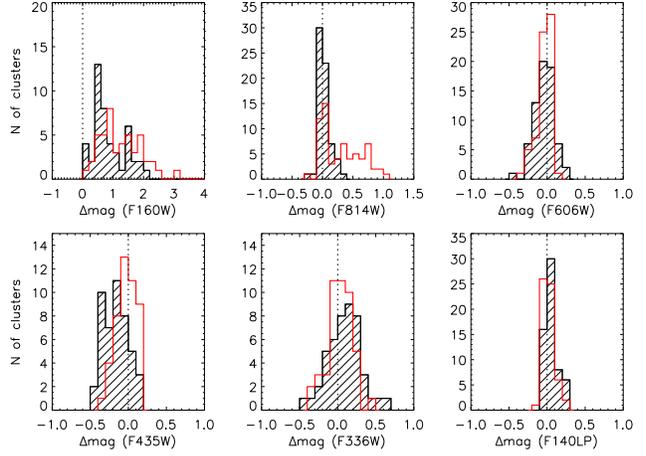}}}
\caption{Residuals in all the filters of clusters with H band excess. The showed sample includes also sources with excess in I and H. The hatched histograms show the residuals, $\Delta m = m_{\textnormal{mod}}-m_{\textnormal{obs}}$, produced by the best fitting models determined with the {\it UV-UBRIH fit}. The red (grey) line histograms show the residuals when the {\it UV-UBRI} and {\it UV-UBR fits} (depending if the excess is only in H or also in I band) are used. The vertical thick dotted lines
indicates the perfect match between model and observation.
The number of objects differs between plots because of the different
detection thresholds in each band. See text for more details. }
\label{res}
\end{figure}

In Figure \ref{res} we show an analysis of the residuals produced by clusters with a flux excess. We define the residual as $\Delta m = m_{\textnormal{mod}}-m_{\textnormal{obs}}$, i.e. $\Delta m $ gives a quantitative measurement of the offset between the flux provided by the best fitting model and the observed data point. When clusters affected by a NIR excess (I and/or H band) are fitted by a  {\it UV-UBRIH fit} the output residuals (see the distributions shown in hatched histograms) have large scatters around zero. In particular, the residuals in the H band are systematically offset versus positive values. The red histograms shows the re-estimated residuals produced by the {\it UV-UBRI} and {\it UV-UBR fits} (i.e. excluding the data points affected by red excess). We clearly notice that the distributions in the H and I bands become even wider. On the other hand, in all the remaining filters, we obtain smaller residuals centred around zero.

\subsection{Testing uncertainties on the determination of the cluster properties}
\label{mc_sim}

We described in \citet{A2010} the method used to create cluster populations using Monte Carlo simulations. Here, we performed the same exercise  using the set of filters available for ESO\,185. Combining three values of masses [$10^4$,$10^5$,$10^6$] $\msun$ and $E(B-V)=[0.05,0.2,0.3]$, we created for each of the 51 time steps of the model  nine different clusters, giving a total  input mock cluster population of $9\times 51= 459$ objects. For each of these clusters we simulated 1001 realisations by stochastically adding offsets in accordance with 
the observed error distribution, for which we fitted SEDs to determine ages, masses, and extinctions. We compared the output age, mass
and extinction distributions with the input parameters to test the uncertainties we introduced in our analysis.

\begin{figure*}
\resizebox{0.48\hsize}{!}{\rotatebox{0}{\includegraphics{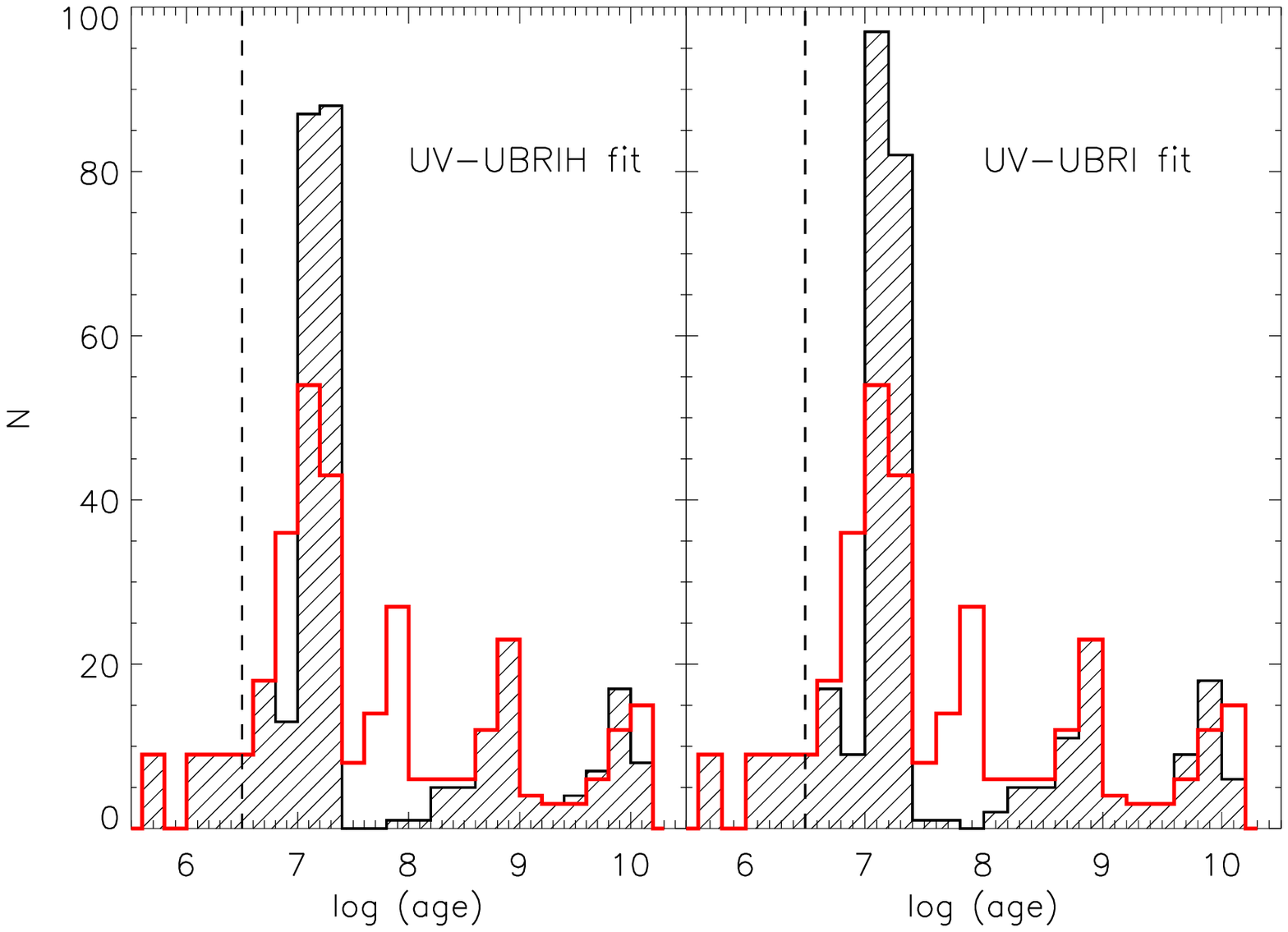}}}
\resizebox{0.48\hsize}{!}{\rotatebox{0}{\includegraphics{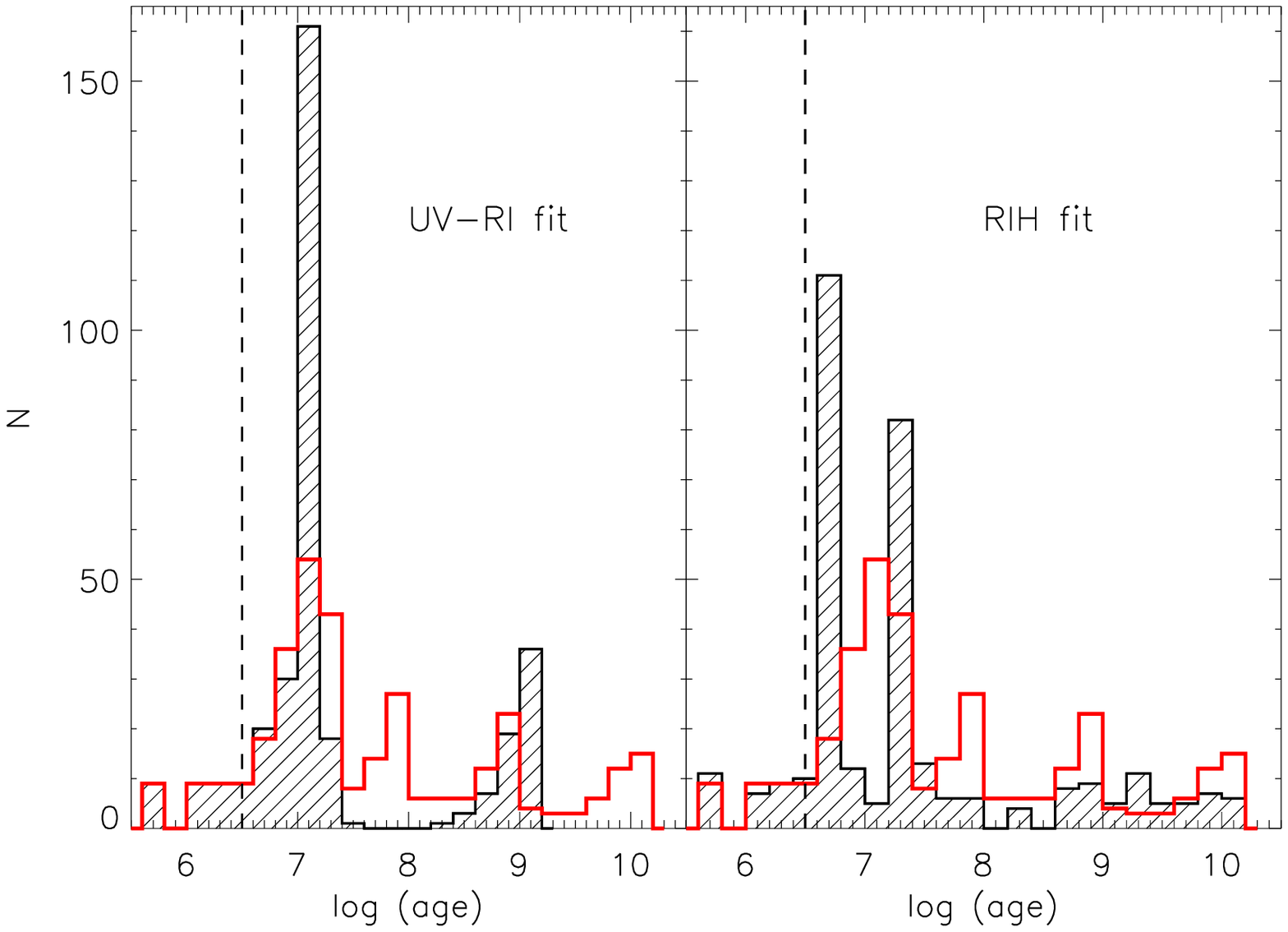}}}\\
\resizebox{0.48\hsize}{!}{\rotatebox{0}{\includegraphics{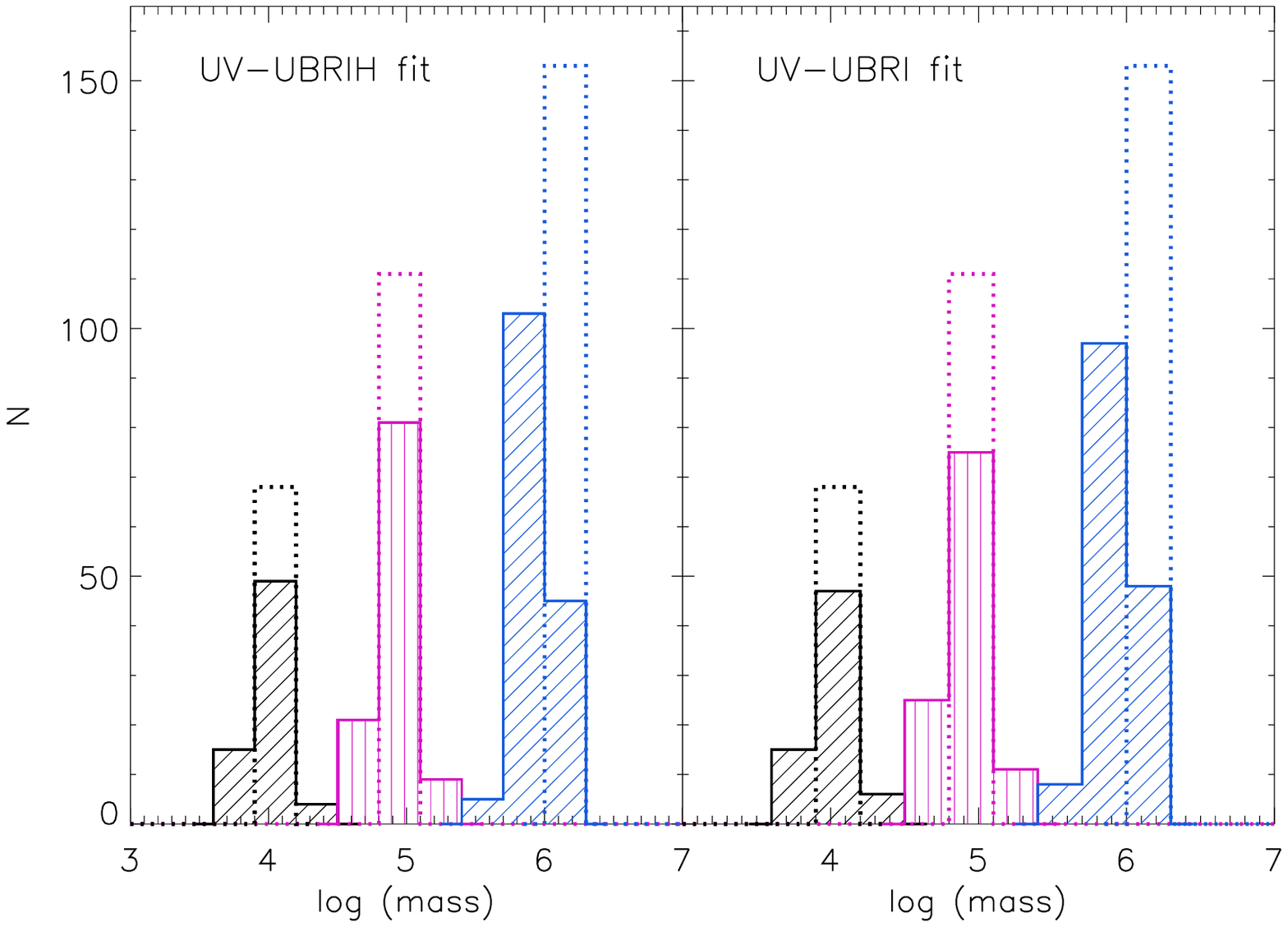}}}
\resizebox{0.48\hsize}{!}{\rotatebox{0}{\includegraphics{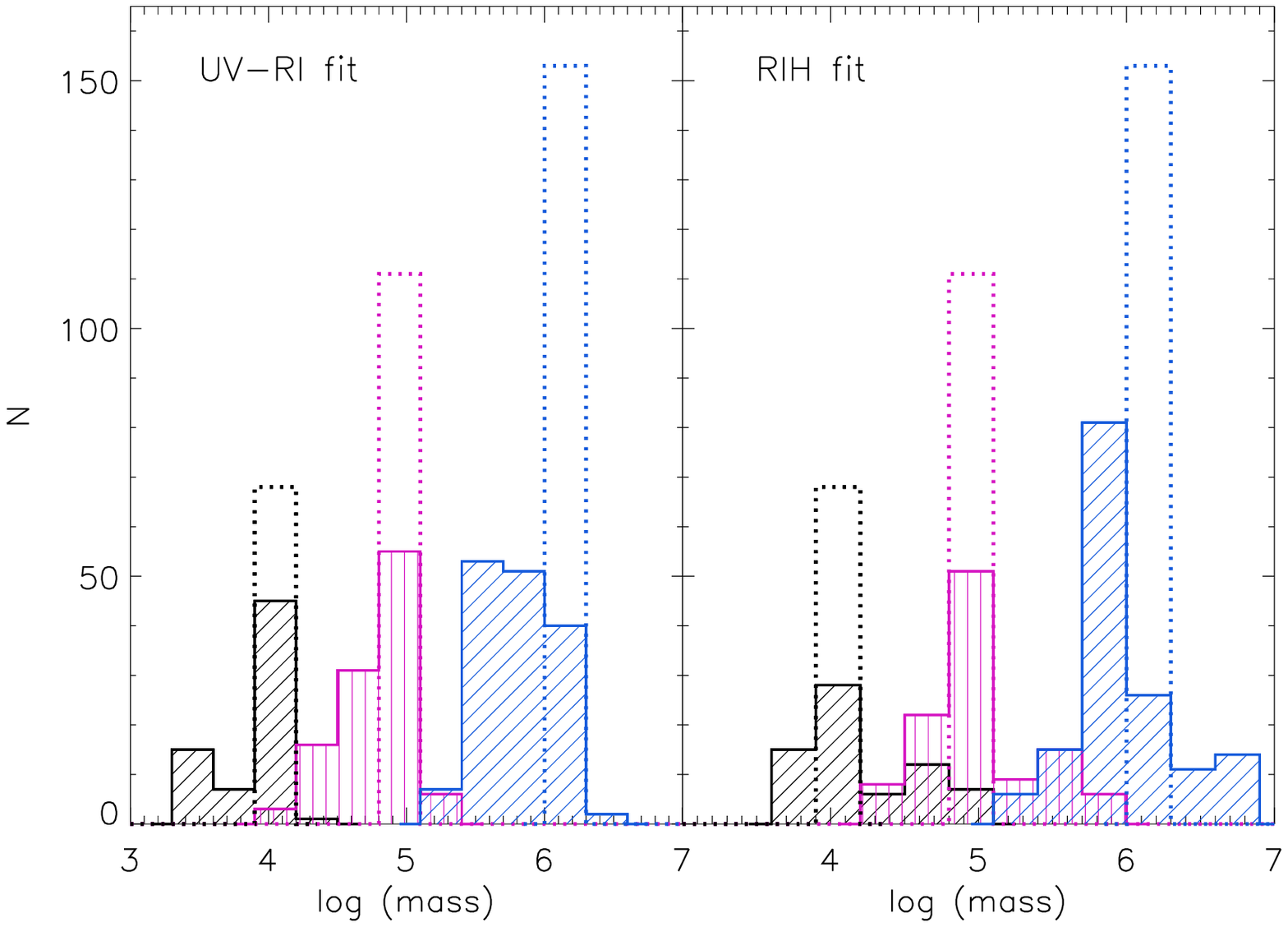}}}\\
\resizebox{0.48\hsize}{!}{\rotatebox{0}{\includegraphics{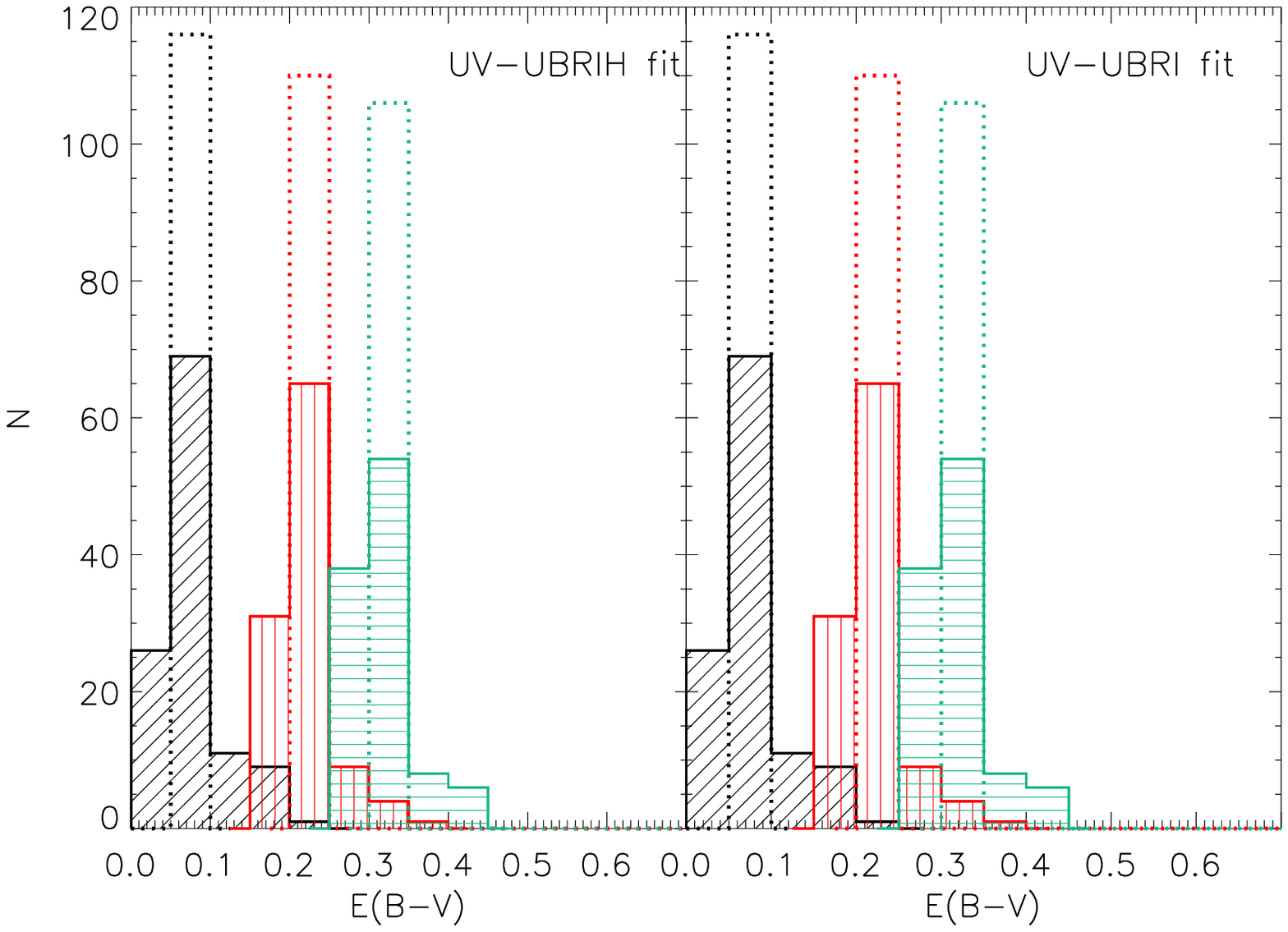}}}
\resizebox{0.48\hsize}{!}{\rotatebox{0}{\includegraphics{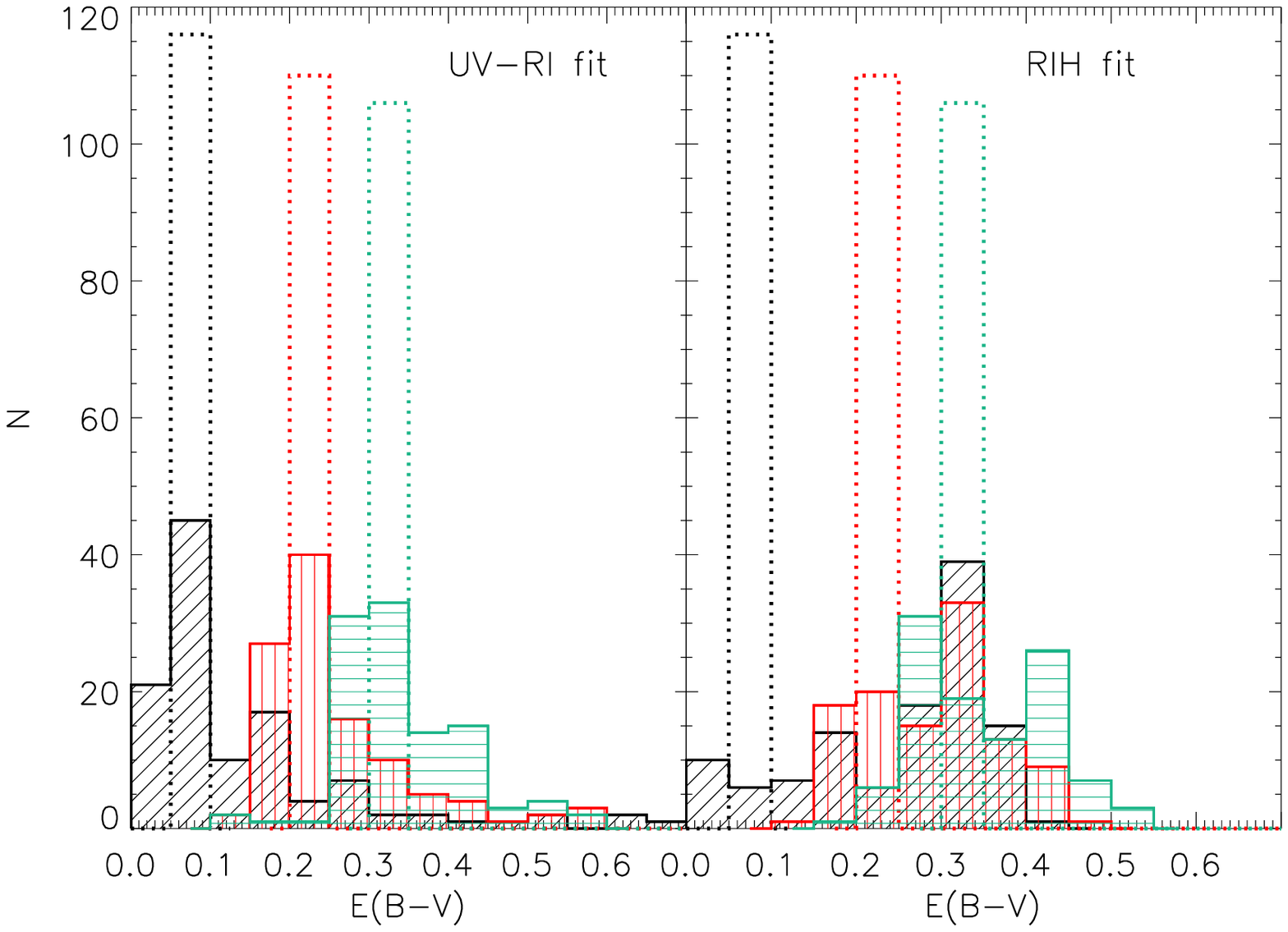}}}\\
\caption{Results from our Monte Carlo simulations of the uncertainties in the SED modelling. We compare the input distribution to the output (represented by the median of 1001 realisations). Top panels: Recovered ages (black hatched histograms). The original input age distribution is shown with a red (grey) line. Central panels: Recovered masses are represented with the hatched histograms. Different colors (and line inclinations, for a B/W printed version) are given to each input mass (dotted histograms): black shows $10^4 $, purple $10^5$, and blue $10^6 \msun$ clusters. The recovered masses are shown with the corresponding colors. Bottom panels: Recovered extinctions are again plotted with hatched histograms, while the input values were drawn in dotted lines. Black (inclined lines) correspond to the initial value of $E(B-V)=0.05$ mag, red (vertical lines) is for 0.2 mag; and green (horizontal) for 0.3 mag. The combination of filters used in the SED fit are indicated in the plots.}
\label{mc_sim_out}
\end{figure*}

\begin{figure*}
\resizebox{0.75\hsize}{!}{\rotatebox{0}{\includegraphics{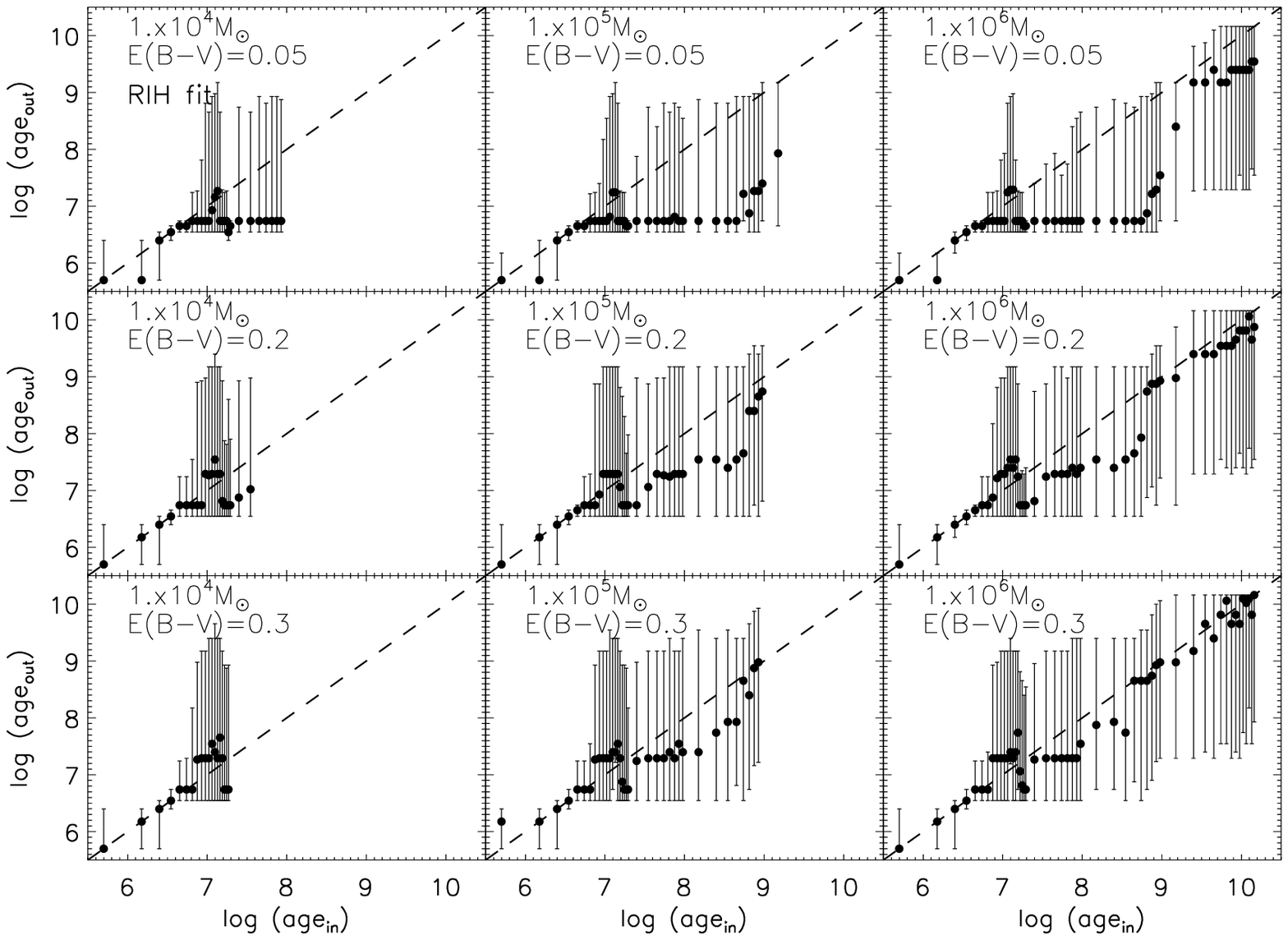}}}
\caption{Recovered ages as function of the input ones, for a RIH fit.  This filter combination produces very poor fits for clusters with age between 10 Myr and 1 Gyr. Clusters fainter than 27.5 mag have been removed from the analysis, therefore clusters with different physical parameters cover varying age ranges. Each of the panels is for a different value of mass and extinction. The dots are the median values of the 1001 realizations, and the error bars show the quartiles of the distributions. A dashed line in each panel draws where, in the plane, the input and output ages are equal. }
\label{sim_age}
\end{figure*}

In Figure \ref{mc_sim_out} we briefly summarize the result of our test. Of the 459 clusters, we included in the analysis only clusters with apparent magnitude, m$_R \leq 27.5$ mag. This cut is similar to the one performed on the real catalogue so we could recreate a sample with properties comparable to our data. We performed four different sets of fits: a fit of the whole SEDs, using the {\it UV-UBRIH fit}; in the second run the {\it UV-UBRI fit} where we excluded H band; two more runs with a UV-RI and RIH fit, since many of the final clusters were fitted with these 2 combinations of 3 filters.  

In the top panels we compare the input age distribution (red line) with the recovered ages (black hatched histogram) of the simulated clusters using the 4 sets of fits. 

Fitting all the available filters or excluding the H band (see top left panels) give quite trustable results except in the region of 10-100 Myr. Clusters with input ages of 8.5 Myr ($\log$(age)$=6.9$) are redistributed partially at slightly older ages, 10-12 Myr. More important, in our analysis we tend to underestimate the ages of clusters between 35 and 100 Myr. We find, in fact, that the age bins at $\log$(age)$=7.5$ to 8.0 are emptied while two overpopulated regions are created between $\log$(age)$=7.1$ and 7.3. This effect has previously been noted   in other similar multi-band analyses \citep{2009ApJ...699.1938M}. The combination of  available filters for ESO\,185 was not able to fully break the  resulting age degeneracies. In our analysis of Haro 11, the addition of more filters, like the F220W (NUV filter), F550M (narrow V-continuum), and Ks (IR), proved to be less susceptible to this effect (see Figure 17 in \citealp{A2010}). 

When only UV, R, and I bands are used, we observe that the gap at 30-100 Myr becomes more extended creating a single peak at 12 Myr. This happens because the model track from  $\sim$10 Myr to 14 Gyr moves parallel to the extinction vector, creating a degeneracy between age and extinction. This effect can be also seen on the right panel at the bottom, where the recovered extinctions for the UV-RI fit tend to be overestimated.  The same mechanism moves input clusters with ages around 14 Gyr down to 1 Gyr. The use of this combination of 3 filters could introduce important uncertainties in the final age distribution. We observe in Figure \ref{age-mass-ext} that the estimated cluster ages using these 3 filters are between 3 and 25 Myr. There is, therefore, a high chance that those 12 Myr old clusters are in reality much older and less extinguished.

Finally, for the RIH fits, we observe significant displacements of the output age distribution (the most right top panel), as well as in the recovered masses and extinctions. In Figure \ref{sim_age} we show the recovered ages for the RIH fit as function of the input ones, for different combinations of masses and extinctions. For all combinations of masses and extinctions, we observe that input ages $\leq 6$ Myr are well reproduced. Ages older than 1 Gyr are well recovered in more than the 50 \% of the cases. Why does the RIH fit fail in recovering ages between 6 Myr and 1 Gyr? The evolutionary tracks between 6 and 1 Gyr move parallel to the extinction vector and make some complex loops (Figure~\ref{fuzzy}). This behaviour lies behind the displacements we see for the lowest extinctions, at all simulated masses (see top panels in Figure \ref{sim_age}). At the same time, in Figure \ref{mc_sim_out} bottom right panel, ones sees that the values of the recovered extinctions, for input $E(B-V)=0.05$ are widely spread up to output values of 0.4 mag. These clusters contribute to the spurious peak at  roughly 6 Myr ($\log$(age)$=6.8$). For higher input extinction values the situation is less dramatic. The secondary peak at 20 Myr ($\log$(age)$=7.3$) is created both by objects of input age 10 Myr for which the age is overestimated and objects in the usual 30-100 Myr input age range.

Due to the large uncertainties associated with the RIH fit, we could have excluded objects detected in only these three filters from the final analysis. However, as discussed in the next section, the output age distribution for observed clusters produced by the RIH fit (see Figure \ref{age-mass-ext}) has two peaks, one at very young ages, and a second one at very old ones. For the young age bin the recovered properties are consistent with the input ones. Clusters as old or older than 1 Gyr have more than a 50 \% chance to be consistent with the real ages. Therefore, we consider that the age for most of the clusters observed in the RIH bands can be trusted.

In Figure \ref{mc_sim_out}, the central panels  shows that there is a tendency in our analysis to underestimate the masses up to a factor of 2. This behaviour is related to the recovered extinctions which are also underestimated (bottom panels). That the mass offset increases with  input mass is due to the fact that we only include clusters with a simulated magnitude brighter than $m_R=27.5$, and to enter the sample at high ages they need to be massive.    Clearly, the mass  and extinction uncertainties increase when only 3 filters are used.

\subsection{Results: From the embedded phases to the old cluster population}

\begin{figure}
\resizebox{\hsize}{!}{\rotatebox{0}{\includegraphics{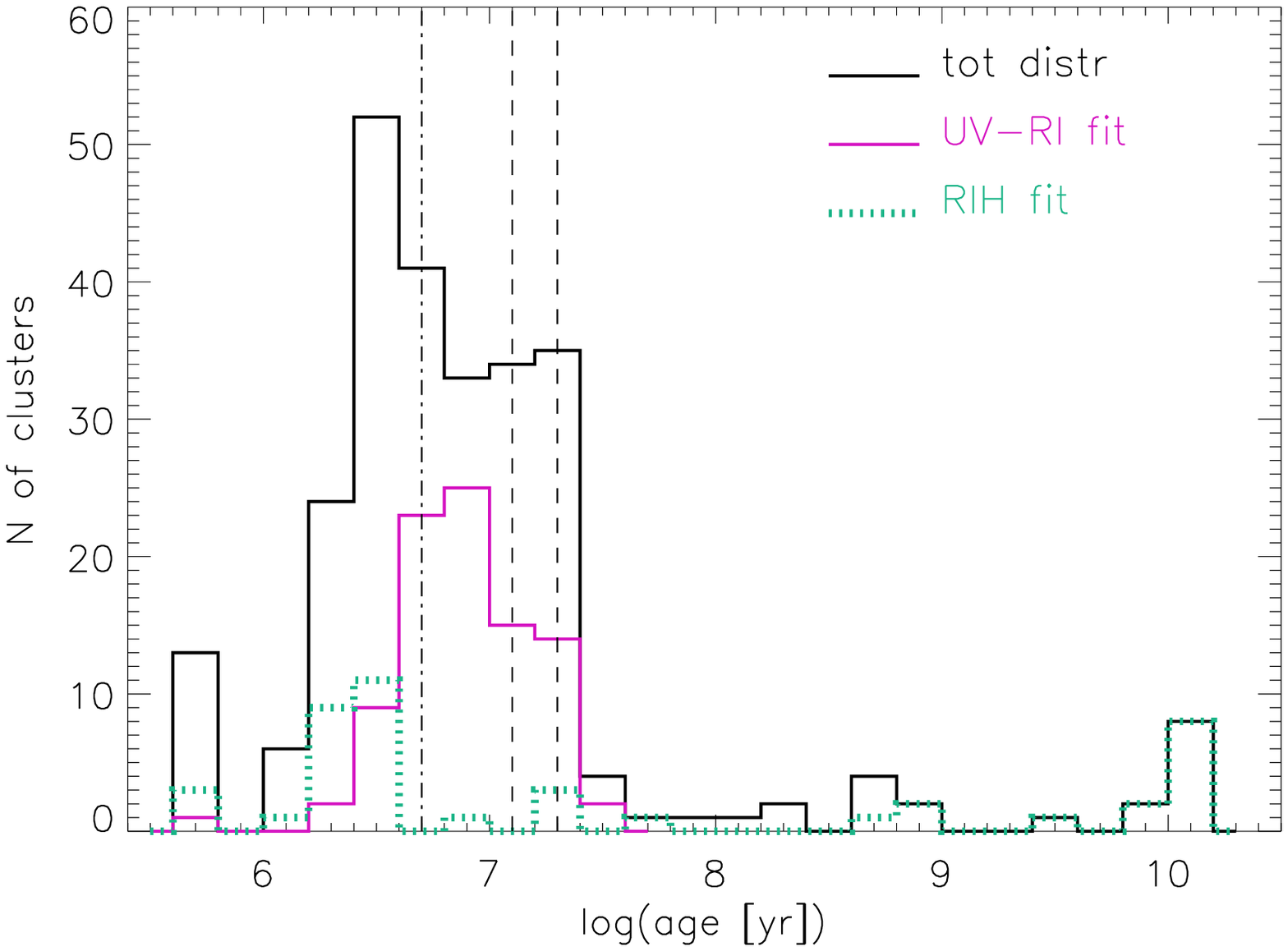}}}\\
\resizebox{\hsize}{!}{\rotatebox{0}{\includegraphics{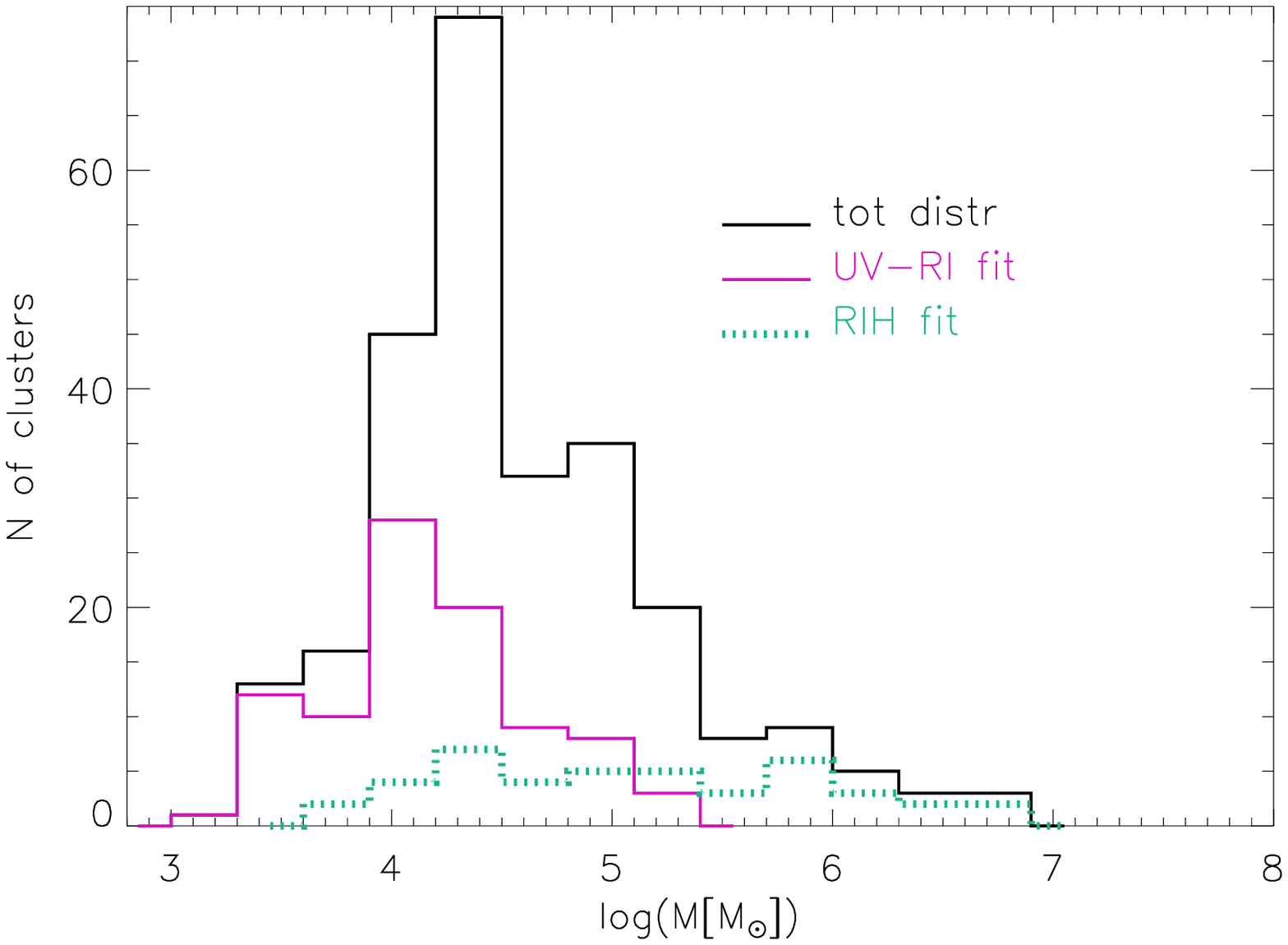}}}\\
\resizebox{\hsize}{!}{\rotatebox{0}{\includegraphics{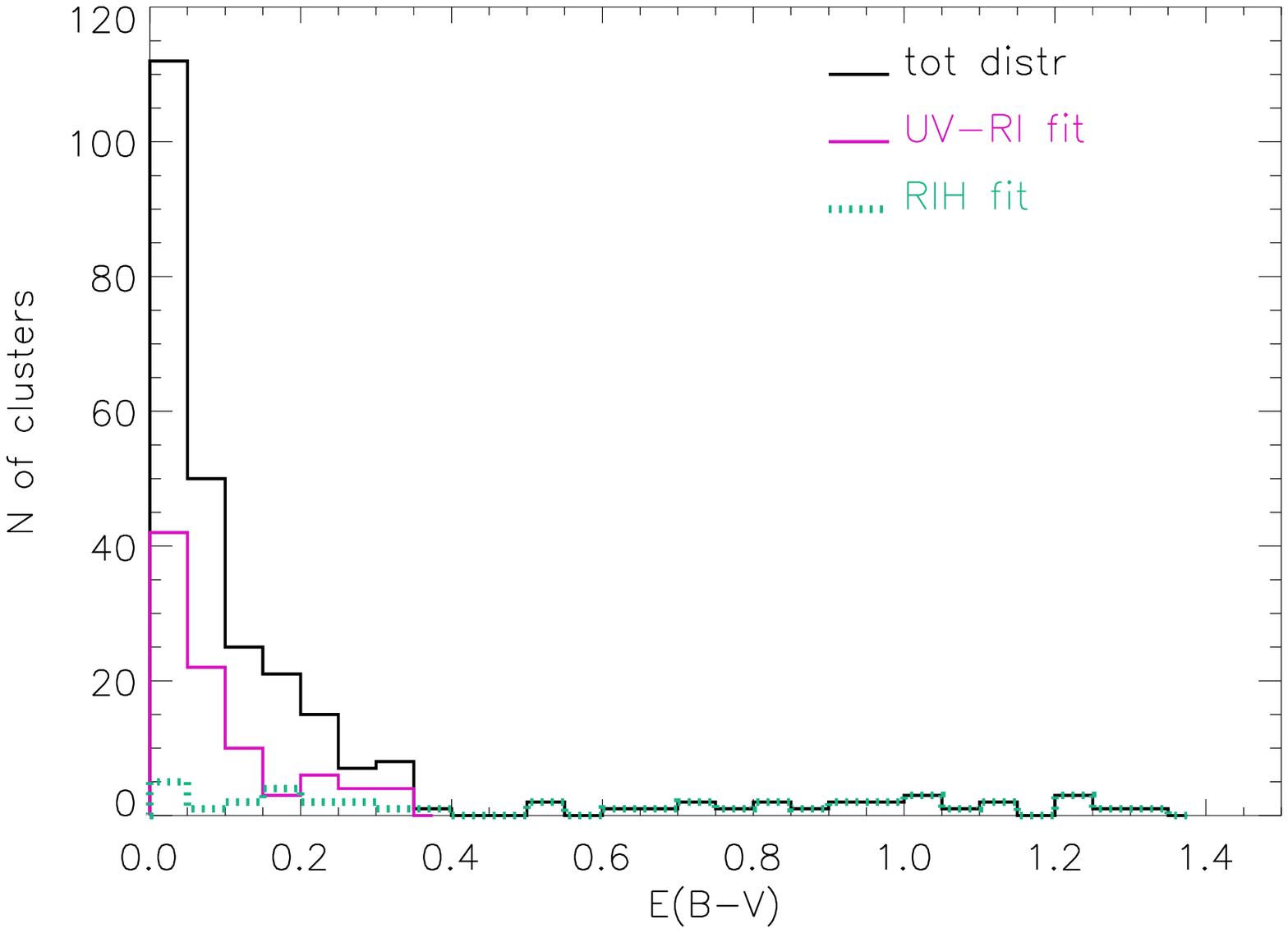}}}\\
\caption{Final age, mass, and extinction distributions for 264 clusters in solid black line histograms. The plots include also the individual distributions of the 43 clusters detected in R, I, and H bands (dotted thick green line) and the 91 clusters for which properties were find using FUV, R, and I detections (solid purple line). See the main text for details.}
\label{age-mass-ext}
\end{figure}

The final distributions of age, mass, and extinction for the 264 clusters are shown in Figure \ref{age-mass-ext}. In total, the properties of $\sim$50 \% of the sources have been estimated from detections in only 3 filters: 43 clusters have detections in the R, I, and H bands only; 91 ones are observed  in  FUV, R, and I only. Uncertainties due to the use of only 3 filters have been assessed by Monte Carlo simulations (addressed in Section \ref{mc_sim}).  

\subsubsection{Cluster age and their reliability}
The cluster population of ESO 185 shows a peak at an age of  3.5 Myr ($\log$(age)$=6.5$; top panel of Figure \ref{age-mass-ext}). The strength of this peak is quite robust, since at this age bin the light of the cluster is dominated by massive blue stars %like the luminous blue variables
 which have short life times and create unique color features (see \citet{2009ApJ...699.1938M} for more details). We tested our models and the uncertainties produced by the  $\chi^2$ fit (see the previous section), and none of the combinations of filters we used in this analysis would produce a spurious peak at 3.5 Myr. In the plot, we point out with vertical lines which age bins are affected by uncertainties. A fit performed on a combination of the R, I, and H filters produces a spurious narrow peak at  $\log$(age)$=6.7$ (5 Myr; dotted-dashed line), and leads to the underestimation of the real ages. However the green histogram in the top panel of Figure \ref{age-mass-ext} shows that not a single cluster in our sample of the R, I, and H detections has an estimated age in that age bin.  We noticed from our simulations that all the used combinations of filters produce fake peaks at the age bins $\log$(age)$=7.1$ and 7.3, while emptying the age bins between 30 and 100 Myr (from $\log$(age)$=7.5$ to 8.0). This effect is due to a loop in the color-color space model of the tracks and is commonly known to create features of this type. In this age range we found 70 objects, so there is a possibility that  a fraction of them has older ages (up to 100 Myr). This uncertainty is taken into account in the further discussions in this work. The final age distribution (black continuos line distribution in the top panel of Figure \ref{age-mass-ext}) shows that the starburst phase started less then 100 Myr ago and became very active in the last 30 Myr. We also found $\sim 10$ older (age $\le$ 1 Gyr) clusters (globular clusters). Thus, the galaxy is not producing the first generation of stars, but has  been rejuvenated  by the recent merger with a  gas-rich dwarf.
 
\subsubsection{The cluster mass distribution}
The final mass distribution (central plot in Figure \ref{age-mass-ext}) shows a peak at a few times $10^4 \msun$. As the simulations show (Figure \ref{mc_sim_out}, central panels), there is a tendency in our analysis to underestimate the mass of some of the clusters. In SED fits involving more than 3 filters, the difference is not more than 30 \%. On the other hand, if only the FUV, R, and I filters are used, the masses are possibly underestimated up to a factor of 2. The plot in Figure \ref{age-mass-ext} shows that the masses estimated from a UV-RI fit have a peak at $10^4 \msun$. It is likely that a fraction of those clusters are possibly more massive. A counter-tendency is observed in the fit involving the R, I, and H bands (Figure \ref{mc_sim_out}, central right panel), where a third of the low mass cluster population has recovered masses overestimated up to a factor of 2, but the effect is smaller at higher masses. The most massive clusters in the observed mass distribution are determined with a RIH fit (green histogram in the central plot of Figure \ref{age-mass-ext}), implying the possibility that some of them in reality may be a factor of 2 less massive, even if their average mass is more likely to be underestimated. 

\subsubsection{The cluster extinction distribution }
\begin{figure}
\resizebox{\hsize}{!}{\rotatebox{0}{\includegraphics{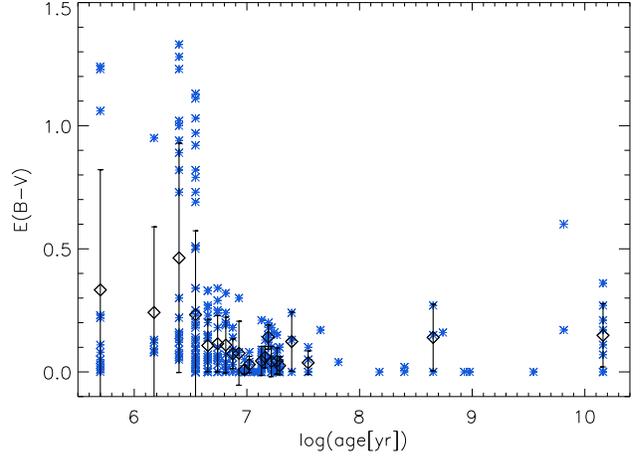}}}\\
\caption{Extinction versus age diagram. The blue (grey) asterisks show each single cluster. Superimposed the mean extinction estimated at each model age step with the corresponding standard deviations.}
\label{age-ext}
\end{figure}

The distribution of the extinction, $E(B-V)$,  peaks close to zero and has a slow decline to values higher then 1.0 (bottom panel in Figure \ref{age-mass-ext}). We observe that values of extinction higher than 0.4 are all constrained with the RIH fit (green dotted line distribution). Using simulations (bottom right panel in Figure \ref{mc_sim_out}) we found that for this type of fit there could be an overestimation of up to a factor of 4 for clusters with an input extinction of 0.05 (black hatched histogram). However, for higher input values ($E(B-V)=0.2$ and 0.3) this effect tended to diminish, with overestimations smaller than a factor of 2.  In Figure \ref{age-ext} the estimated extinctions of the clusters versus the ages are shown. We found that many of the clusters with extinctions higher than 0.6 mag have ages smaller than 3.5 Myr. At this age range clusters are still involved in the partially embedded phase so we expect to observe a high scatter in the extinctions. Even if some  values of $E(B-V)$ are overestimated in our analysis (mainly for young clusters detected in R, I and H) we expect the uncertainty to be smaller than a factor of 2.

\subsubsection{Revealing the properties of the clusters population}
\begin{figure}
\resizebox{\hsize}{!}{\rotatebox{0}{\includegraphics{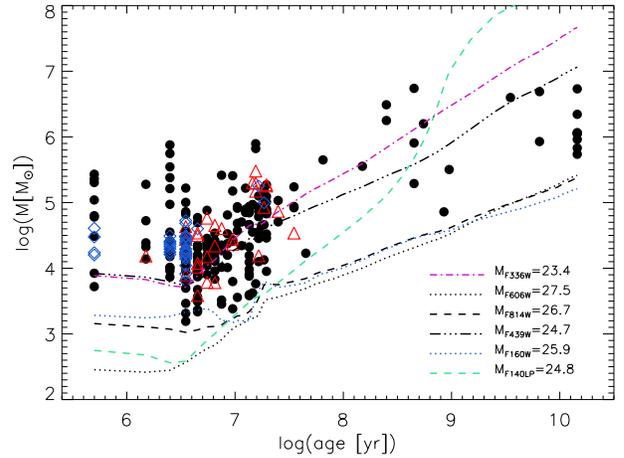}}}\\
\caption{Mass versus age diagram. Clusters which do not show any red excess are represented with filled black dots. Clusters affected by IR excess are showed with red (grey) triangles. The blue (grey) diamonds are for clusters affected by I and IR (where detected) bands excess. The lines (see the inset) show the detection boundaries in ages and masses corresponding to the magnitude limits reached at $\sigma\leq 0.2$ in each filter. As already expected there are not clusters sitting below these limits.}
\label{age-mass}
\end{figure}

In Figure \ref{age-mass} we show the masses of the clusters as function of the ages. Due to our inclusion of only clusters with detections in at least three filter in the SED fit,  all clusters lie above at least 3  detection limit lines.  Most of the clusters, $\sim 60$ \%, are located at ages below 10 Myr and have masses between $10^3$ and $10^6 \msun$. As already observed in Haro 11, we did not find low mass ($10^3-10^4 \msun$) clusters at very young ages (below 3 Myr). Two mechanisms, blending and extinction (partially embedded phase) can remove objects in this region. Clusters in active star-forming environments do not form in isolation but in complexes. At the distance of ESO 185 (similarly to Haro 11), it is possible that less massive clusters are blended. On the other hand, as we observe in Figure \ref{age-ext}, extinction at this age range is quite spread, indicating that the embedded phase lasts several Myr (\citealp{2009arXiv0911.0796L}; \citealp{2009arXiv0911.0779L}). The newborn clusters need time to clean their surroundings from the natal GMCs where they formed. For clusters with masses below $10^4 \msun$, even a moderate extinction of $E(B-V)\approx0.5$ ($A_V \approx 1.6$) would move the object below the detection limits.  In addition, a significant fraction of young clusters are affected by red excess, in which case the H and/or I filters are removed from the SED fit which then has to rely on the shallower (in terms of the mass detection limit) U and B bands. Hence, the apparent dearth of low mass young clusters is probably not real. 

Contrary to what found in Haro 11 \citep{A2010}, the most massive clusters are not in the youngest bins, but at very old ages (Figure \ref{age-mass}). We found significant secondary peaks of clusters with ages around 1 and 14 Gyr, and masses comparable to the most massive globular clusters observed in the Milky Way.  Even if the number of objects is smaller, the total mass in observed intermediate age (0.1--1 Gyr) and old (1--14 Gyr)  clusters is comparable to that of young clusters, in each case a few times $10^7 \msun$. However, the mass detection limit is
higher for the older clusters and the total inferred cluster population mass is therefore higher. 
This suggests that the galaxy has experienced extreme star formation episodes also in the past. 
We will return to this point and discuss implications for the past evolution of ESO\,185 in Section \ref{gal-prop}.

\subsubsection{Cluster misfits or interlopers?}

\begin{figure}
\resizebox{\hsize}{!}{\rotatebox{0}{\includegraphics{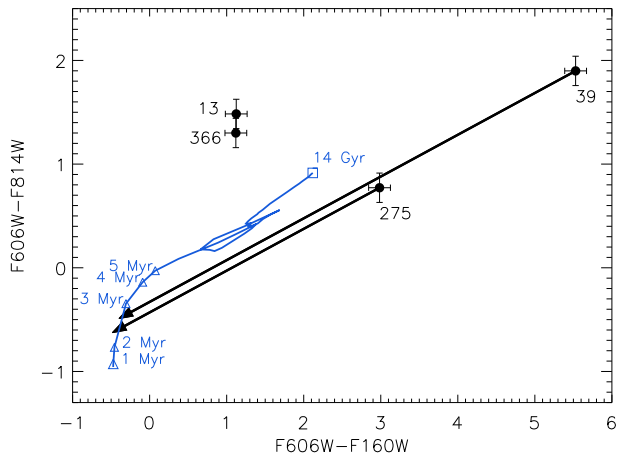}}}\\
\caption{optical ($R-I$) versus NIR ($R-H$) color-color diagram. The position and the associated photometric error of the 4 extended and NIR bright cluster candidates are plotted. The apparent F184W magnitudes of the clusters \# 13, 39, 275, and 366 are $m_I = 19.5, 22.5, 20.3$ and $21.1$, respectively.  The blue solid line is the evolutionary track predicted by \citet{Zackrisson et al. a} models at the redshift and metallicity of the galaxy. Several age steps are indicated. The arrows show where the clusters \# 39 and \# 275 will move if a correction for extinction of $E(B-V)= 2.12$ and 1.24 mag is applied. The other 2 objects have $R-I$ colors impossible to explain with the current models at the redshift of ESO\,185.}
\label{fuzzy}
\end{figure}

We found 4 objects which were undetected in UV-U, only 2 of them partially detected in B, and with increasing brightness from R to H band. Since these objects are resolved and isolated (in the outskirts of the starburst region) we performed new photometry (using an aperture of $1.0"$) and a new fit to the SEDs. Two of the sources, \# 39 and \# 275 (Figure~\ref{fuzzy}), are consistent with being deeply embedded star clusters. They have fitted reddenings of $E(B-V)= 2.12$ and 1.24 (the arrows show how the clusters move in the color-color diagram, due to this extinction), ages between 2 and 3 Myr, and masses of several times $10^6 \msun$, respectively. The other two sources, \# 13 and \# 366, have a $R-I$ color, impossible to reconcile with the evolutionary track of the galaxy. There is hence a high probability that they are background objects (either exceptionally bright $z\sim 3$ Lyman break galaxies or $z\sim 0.5$ ellipticals) and we excluded them from the present analysis. 
They are not included in any of the figures showing our overall results for the cluster population in ESO\,185.

\section{The cluster red excess in ESO 185: differences and analogies with other cases in the literature}
\label{rex-cause}

In this section we will discuss the cause of red excess seen in many ESO\,185 clusters and compare to other galaxies/environments found in
the literature.

At low metallicities ($Z \leq 0.008$), the contribution from photoionized gas around clusters younger than 10 Myr is not negligible (\citealp{Krüger et al.}; \citealp{Zackrisson et al. a}; \citealp{2002A&A...390..891B}; \citealp{2008ApJ...676L...9Z}; \citealp{R2009}; \citealp{A2010b}). If the spectral energy distributions (SEDs) of the clusters would be fitted by models including only stellar continuum, it produces  a considerable observed flux excess at NIR bands. The fraction of nebular emission contributing to the integrated fluxes increases and last longer in the NIR (see \citealp{A2010b}). The red excess found in the young star clusters of NGC 4449 \citep{2008AJ....135.2222R} and SBS 0335-052E \citep{2008AJ....136.1415R} is produced by the nebular emission surrounding the clusters (see \citealp{R2009}; \citealp{A2010b} ). On the other hand, the NIR excess found in Haro\,11 star clusters \citep{A2010} and ESO\,185 sit above models which already include nebular contribution so other mechanisms need to be invoked.

In Figure \ref{age-mass}, we differentiate clusters affected by red excess using different symbols. With red triangles, we show clusters affected by excess in the IR (H band). In blue diamonds we display clusters which possess NIR flux excess, i.e. clusters affected by excess at $\lambda > 8000$ \AA. Clearly, clusters that show a NIR excess are around 5 Myr old or younger, and have masses below $10^5 \msun$. Only two clusters affected by NIR excess are older ($\sim 16-18$ Myr) and more massive ($> 10^5 \msun$). Clusters with only IR excess (i.e. with H-band but without I-band excess) have a wider range of ages (from a few to 35 Myr) and masses (between $10^4$ and few times $10^5 \msun$). 

\begin{figure}
\resizebox{\hsize}{!}{\rotatebox{0}{\includegraphics{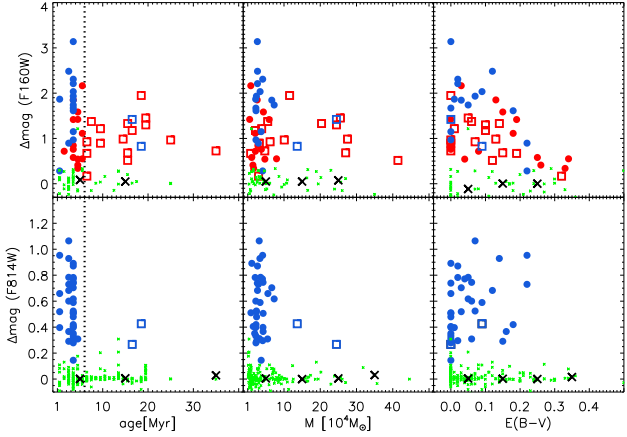}}}
\caption{Strength of the flux excess in H and I bands as function of the cluster properties  derived from fitting the blue side of the SEDs. The vertical dotted lines  in the left side panels separate clusters younger than 6 Myr (filled dots) from older clusters (open squares).  In red we show clusters affected by IR excess (H band), and in blue the ones with NIR excess (I and H bands). The crosses show, for each bin, the median value of the residuals for normal clusters in the sample. %See the text for explanation of the chosen age values ad the analysis.
}
\label{delta-exc}
\end{figure}

A correlation between the strength of the observed flux excess ($\Delta m = m_{\textnormal{mod}}-m_{\textnormal{obs}}$) and the derived ages, masses, and/or extinctions can give valuable information on the underlying mechanisms. In Figure \ref{delta-exc} we show in blue, clusters with a NIR excess and in red,  those with IR excess. Clusters younger than 6 Myr are drawn as filled dots, while the older are showed as open squares. The median values of the residuals in the H and I bands of star clusters not affected by red excess are also included for each age, mass, and extinction bin (black crosses).

Studies of very young clusters in the Milky Way show that for ages below 6 Myr, the cluster is experiencing a rapid and complex evolution from being deeply embedded in the natal giant molecular cloud (GMC) to shining in an environment that has been cleared from dust and molecular gas  (see \citealp{2003ARA&A..41...57L} for a review). During this period the short-lived massive stars emit strong UV-radiation which ionise the ambient gas and heat the dust. The presence of dust in the cluster medium and strong UV-fields can create a rise of the continuum at two different wavelengths. Hot dust usually re-emits at wavelengths much longer than 1 $\mu$m, of course sensitively depending on the dust temperature. At shorter wavelengths, between $0.6-0.9$ $\mu$m a photoluminescence process involving dust grains and hard UV-fields can create the so-called extended red emission (ERE, \citealp{2004ASPC..309..115W}). ERE as been observed as a soft "rise" in the continuum at $0.6-0.9$ $\mu$m of H{\sc ii} regions and gaseous nebula (among many others \citealp{2000ApJ...544..859G}; \citealp{1995A&A...304L..21P}; \citealp{1992A&A...255..271P}). 

Moreover, in a newborn cluster a large fraction of stars may still be in the contracting phase as pre-main sequence (PMS) objects.
 \citet{2005ApJ...630L.177M} estimated that PMS stars contributed between 1 and 17 \% of the integrated H band flux of an unresolved cluster, depending of the stellar population age (from 1 to 10 Myr) and the used IMF. Recently,  \citet{2010ApJ...710.1746G} found direct evidence of PMS stars in NIR spectroscopic data of a young knot (89/90) in the Antennae system. Among the pre-main sequence stars, the young stellar objects (YSOs) show a strong NIR excess caused by their luminous accretion  disks. Interestingly, a large number of YSOs are detected at the edges of young star forming regions, due to triggered star formation \citep{1998ASPC..148..150E}. These transition phases are typically observed in massive star-forming regions of the Milky Way (\citealp{2010ApJ...713..883B}; \citealp{2010A&A...510A..32M})  and very young star clusters in the Small Magellanic Cloud (NGC 602, \citealp{2007ApJ...665L.109C}; NGC 346/N66, \citealp{2010A&A...515A..56G}).   

Resolved nearby clusters give further insights into the main sources of the NIR excess:  ERE, at shorter wavelengths ($\sim 0.8 \mu$m), hot dust  and a large fraction of YSOs and PMS stars in the IR ($> 1.5 \mu$m). The young knots of the nearby starburst spiral NGC 253 \citep{2009MNRAS.392L..16F} show a soft bump at $1-2$ $\mu$m which could be reproduced by an substantial fraction of YSOs. Studies of embedded clusters in nearby dwarf starburst galaxies have also revealed IR excess which has been interpreted as due to hot dust (NGC 5253, \citealp{2005A&A...433..447C}; Henize 2-10, \citealp{2005ApJ...631..252C}). Clusters in Haro\,11 \citep{A2010}, and possibly also in SBS 0335-052 \citep{A2010b}, show an  I-band excess which may be explained by the ERE feature. 

In Figure \ref{delta-exc}, clusters represented by dots (ages below 6 Myr) are the ones which potentially could be affected by these processes. The strength of the excess in the H band appears to be higher for younger,  lower mass and less extinct clusters. However the scatter is large (top panels) and there is a general decrease of the number of clusters with increasing age, mass and extinction (see Figure~\ref{age-mass-ext}). The conclusion is that H-band excess seems to affect an approximately random (with respect to mass and extinction) 10 \% fraction of young (age $\le30$Myr) clusters .  If the excess is caused by hot dust in the cluster interstellar medium we would expect to find higher foreground extinction, resulting from a clumpy distribution in front of the cluster, shielding partially the light produced by the young stars. If there is no/little dust along the line of sight to the clusters (very low extinction), it is likely that other sources are dominating and the best candidates are YSOs. The aperture used to do photometry (radius $35$ pc) is much larger that the expected radius of the clusters ($5-10$ pc). Therefore, inside the region where we do photometry, we are including not only the cluster, but potentially also star formation at its rim, or possibly other newborn clusters, since they form in complexes \citep{2010IAUS..266....3E}. The edges of very young clusters are places for triggered star formation (large fraction of YSOs) and these regions may be
cleared from their dust at a time when the central massive cluster has already been naked for some time. So, there is a fair chance that the excess we are observing in the H band is possibly a combination of YSOs and hot dust. 
The bottom panels show the I band excess as function of ages, masses and extinctions. No particular trend is observed, except that the excess at this waveband is only seen for very young star clusters, showing that the ERE effect is probably the best explanation for this behavior. 

At older ages ($> 7$ Myr) red super giants (RSGs) affect considerably the integrated fluxes at wavelengths redder than 10000 \AA \citep{2008A&A...486..165L}. They may be relevant for the red excess for two reasons: Firstly, at low metallicities, the models have a tendency to underpredict the number of RSGs observed and there is evidence that their number increases at lower metallicity (see \citealp{2002A&A...386..576E} and \citealp{2001A&A...373..555M} for details). Since the RSGs dominate the light in the NIR, if the models underestimate their number the observed integrated fluxes would shown an excess at such wavelengths. 
Secondly, in less massive clusters ($<10^5 \msun$) stochastic effects in the IMF sampling \citep{2004A&A...413..145C} will severely affect 
the colours of clusters, in particular in the IR.  Only clusters with mass  on the order of $10^5 \msun$ can fully sample the IMF\footnote{e.g. for a Salpeter IMF from 0.1 to 100 $\msun$ there will be one star more massive than 80 $\msun$ per $10^4 \msun$ of cluster mass}. 
Such statistical fluctuations are likely to occur and affect the NIR SED when the expected number of RSGs is on the order of unity or less.
In Figure \ref{delta-exc}, clusters plotted as squares (7--35 Myr) are the ones where RSGs could have an impact on the flux excess. The 
masses of some of these clusters are low enough that IMF sampling effects may be important, and the varying number of RSGs  from 
cluster to cluster could explain the spread in $\Delta m_{F160W}$.

\section{The starburst properties of ESO 185 as revealed by the star clusters}
\label{starb}
\subsection{The cluster luminosity function}
\label{CLF} 

The cluster luminosity function (CLF) is a useful tool to study the formation and evolution of clusters in galaxies. It has been shown for many nearby galaxies that the cluster mass function (CMF) can be approximated by a power-law with index approximately $\alpha \approx -2.0$ (M51, \citealp{2003A&A...397..473B}; the Antennae, see  \citealp{2010AJ....140...75W} for a new revisited CLF and CMF of the system; LMC and SMC, \citealp{2003AJ....126.1836H}; M83, \citealp{2010ApJ...719..966C}). Assuming that the CLF is built up of fully sampled CMFs formed at different ages, then the resulting CLF is supposed to have the same power-law shape. In reality the observed CLFs have a large spread of power law indexes ranging from $-2.4 \leq \alpha \leq -1.7$ (among many others \citealp{2002AJ....124.1393L}; \citealp{2003AJ....126.1836H}; \citealp{2003MNRAS.343.1285D}; \citealp{2009A&A...501..949M}; \citealp{2009A&A...503...87C}; \citealp{2010AJ....139.1369P}; \citealp{2010AJ....140...75W}). Interestingly, some galaxies have a double power law CLF, steeper than $-2.0$ at the bright end, and approximately $-2.0$ at the faint end. The bend occurs  between roughly $-8.0$ and $-10.0$ mag (Gieles et al. 2006a,b). The  steepening of the CLF (both for single power laws and for the bright-end of the double ones) could be caused by a truncation at high mass in the CMF \citep{2002AJ....124.1393L}, or in other words, a limit on the maximum mass with which a cluster can be formed. \citet{2006A&A...450..129G} showed that galaxies whith a double power law CLF impose physical limits on the formed maximum cluster mass. A steepening of the CLF as function of the wavelengths is expected if there is a truncation in the CMF, since evolved stellar populations are brighter at red wavelengths. Recently, \citet{2009A&A...494..539L} suggested that instead of a sharp truncation in the CMF, a Schechter function with a characteristic mass, $M_c\approx2\times10^5 \msun$, and an index of $-2.0$ at the low-mass end is a better representation of the CMF in quiescent spiral galaxies. 
Shear and/or regular patterns (like spiral arms) effectively limit the mass of GMCs from which clusters form. In presence of merging events which enhance the compression and the star formation efficiency, the $M_c$ has likely higher values. A Schechter-like CMF could explain the steepening of the CLF at the bright ends, the possibility for galaxies to form very massive clusters (\citealp{2006A&A...448..881B}; \citealp{A2010}), and why the most massive young clusters scale with the global star formation efficiency of the systems \citep{2008MNRAS.390..759B}.
\begin{figure}
\resizebox{\hsize}{!}{\rotatebox{0}{\includegraphics{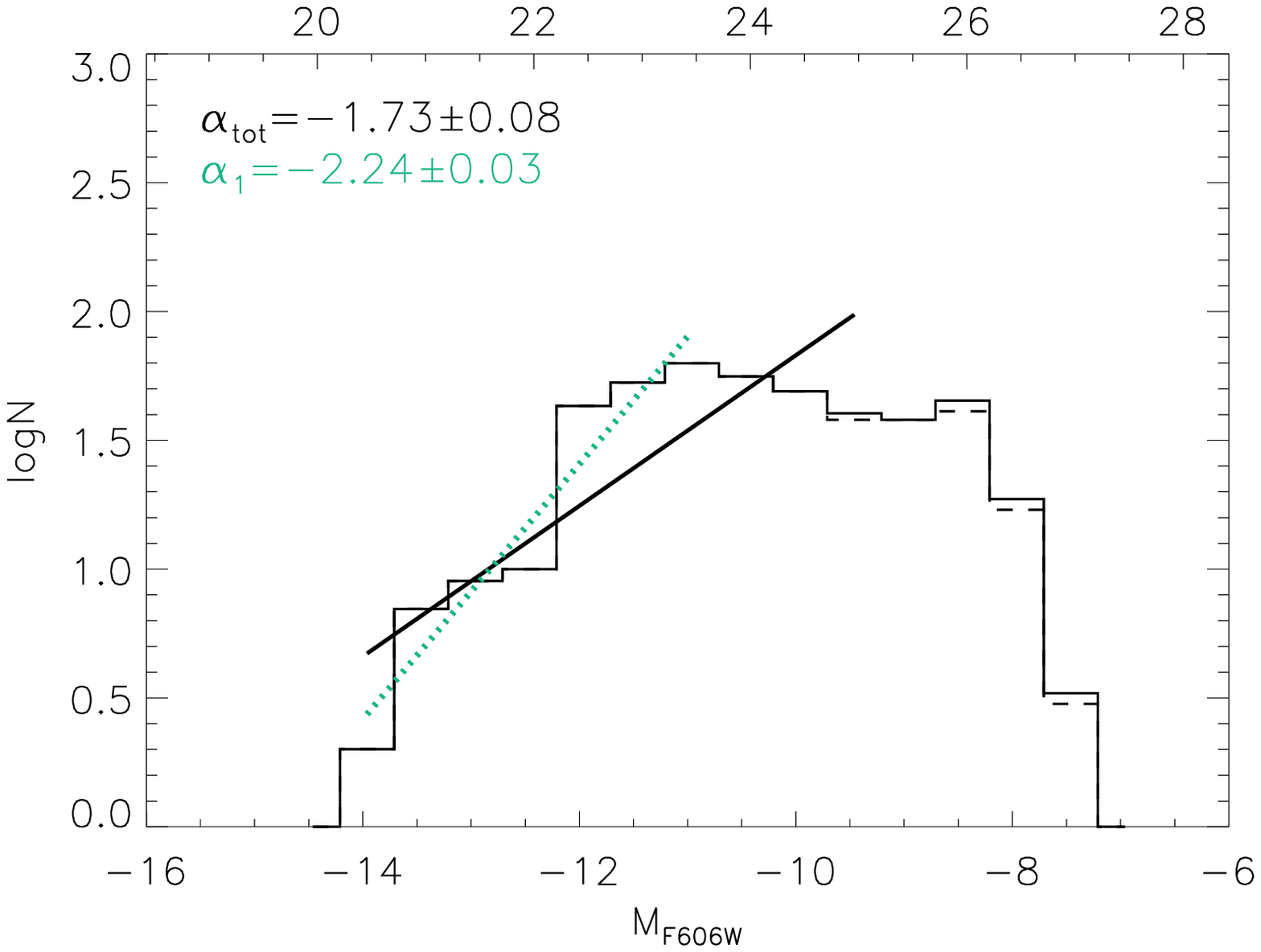}}}
\resizebox{\hsize}{!}{\rotatebox{0}{\includegraphics{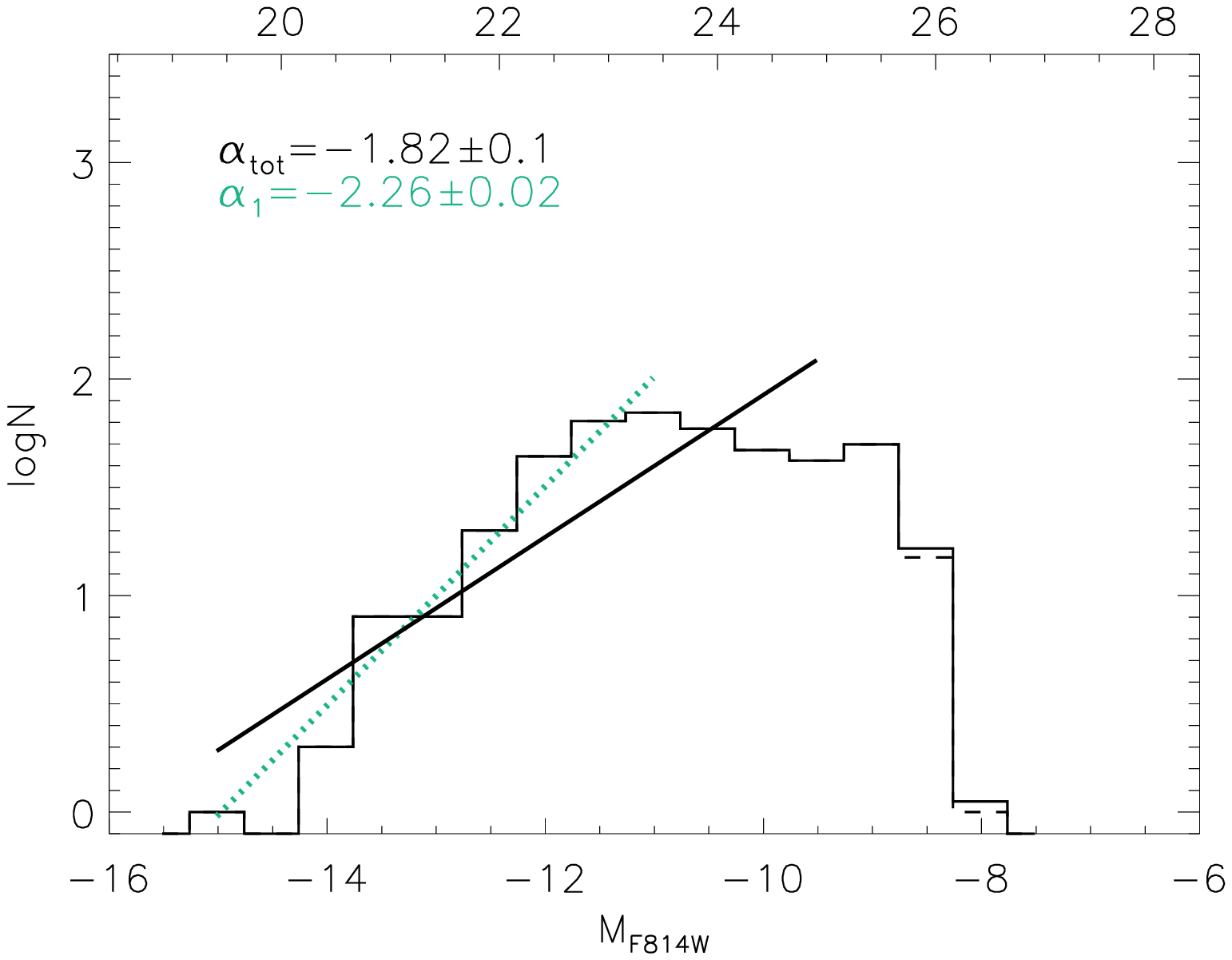}}}
\caption{Cluster luminosity functions in the two deepest frames available for ESO\,185 (R and I filters). The dashed histograms are the observed distributions, the solid ones have been corrected for completeness. The power law fit performed on bins with 100 \% completeness is shown in black. In green (grey) we show the fit to the luminosity bins brighter than -11.0 mag.}
\label{CLF-plot}
\end{figure}

ESO 185 has produced many clusters in the most recent, and still active, burst episode. Using the two deep {\it WFPC2} exposures we detected more than 400 clusters (see Section \ref{fin-cat}). Of these, we were able to derive ages, masses, and extinctions of roughly 50 \%. 
In Figure \ref{CLF-plot} we show the CLFs recovered for the F814W and the F606W filters for the entire population of clusters detected in both  bands. The completeness tests (Section \ref{completeness}) show that in crowded regions, already at 26.0 mag a considerable fraction of objects is not recovered. Using a conservative limit we fitted the CLF using magnitude bins brighter than 25.0 mag in both filters (the bins have been corrected for completeness, black continuum line histogram). This first fit is showed with a thick black line in the two panels of  Figure \ref{CLF-plot}. The power-law indexes found are indicated and consistent with $\alpha \approx -1.8$ in both filters. It is clear that the number of objects decrease rapidly in bins fainter than 25.0 mag. We checked whether including the total number of clusters detected in both filters (1265 objects) would change the slope at the faint ends (magnitude bins $> 25.0$ mag). We recovered a quite flat slope of $\sim -1.2$ at such faint magnitude bins, while the slope at brighter magnitudes was unchanged.  Finally, we tried a fit to the very bright magnitude bins (brighter than $-11.0$ in absolute magnitudes). Bins brighter than $-11.0$ mag are usually scarcely populated. The brightest clusters in the Antennae system are $\sim -11.5$ mag \citep{2010AJ....140...75W}; nearby spiral galaxies, like M83 and M51, barely fill the $-11.0$ magnitude bin (\citealp{2010ApJ...719..966C}; \citealp{2008A&A...487..937H}). \citet{2000ASPC..211...63O} showed that blue compact galaxies are, instead, very efficient in producing clusters brighter than $-11.0$ mag. {\it HST} data have revealed that small starburst galaxies with total stellar masses $10^9 - 10^{10} \msun$ have a large fraction of very massive young clusters (\citealp{2003A&A...408..887O}; \citealp{A2010}; this work). The fit to these bright bins, in ESO 185, produced a slope a factor of 0.5 steeper (green dotted line). Due to the distance of the galaxy ($\sim$ 80 Mpc), even the high resolution power of the PC camera, it's not enough to spatially resolve all the single clusters. We expect, in fact,  that not all the clusters form in isolation. Blending and crowding are affecting our data analysis. Usually, these mechanisms tend to flatten the slope at the faint-ends with a few percent, while the slope at the bright bins is almost unchanged. Fully understanding the observed luminosity distribution of the clusters in ESO 185 (flat at low luminosities and approximately $-2.0$ at bright ends) is challenging. Monte Carlo simulations are needed to investigate in more detail the possible causes (Adamo et al. in prep.). We discuss, here briefly, possible scenarios.
 
Blending, a change of slope in the CMF at low masses \citep{2006A&A...450..129G}, and a mass dependent cluster disruption timescale (low mass clusters dissolve faster than high mass ones; see \citealp{2009Ap&SS.324..183L} for a short review) could all produce a flattening of the CLF. The mass--age diagram in Figure \ref{age-mass}, shows that there are empty regions at low mass ($\sim10^4 \msun$) at very young ages ($1-3$ Myr) and between $20$ Myr and 100 Myr. We already discussed that very young low mass objects could be blended or fall below the detection threshold if affected by moderate extinction ($E(B-V)\geq 0.5$). At older ages, between 30-100 Myr, we expect that we might be seeing to few clusters since many of them have been assigned younger ages ($\sim 20$ Myr) by the SED fit (see Section \ref{mc-sim}). However, the CLF is not affected by systematics in the performed cluster age analysis and represents an independent test of the cluster formation. If we introduced important biases in the age estimates this would be revealed by a normal recovered CLF with index $-2.0$. However, we observe an important flattening of the CLF which suggests that the missing fraction of low mass clusters are either heavily blended or preferentially destroyed. None of the two scenarios can be excluded. Finally, we already noticed in Haro 11 \citep{A2010}  and in ESO 338-IG04 \citep{2003A&A...408..887O} that the CLF in blue compact galaxies tend to be flatter than the typical values (found in spirals and interacting systems) and produce  high fractions of massive clusters. It's not possible to draw any conclusion from the available data, but it appears  that the galactic environment in these dwarf starburst galaxies has favored the formation of massive clusters and/or the disruption of low mass clusters, playing a crucial role in shaping their CLF.

\begin{figure*}
\resizebox{0.48\hsize}{!}{\rotatebox{0}{\includegraphics{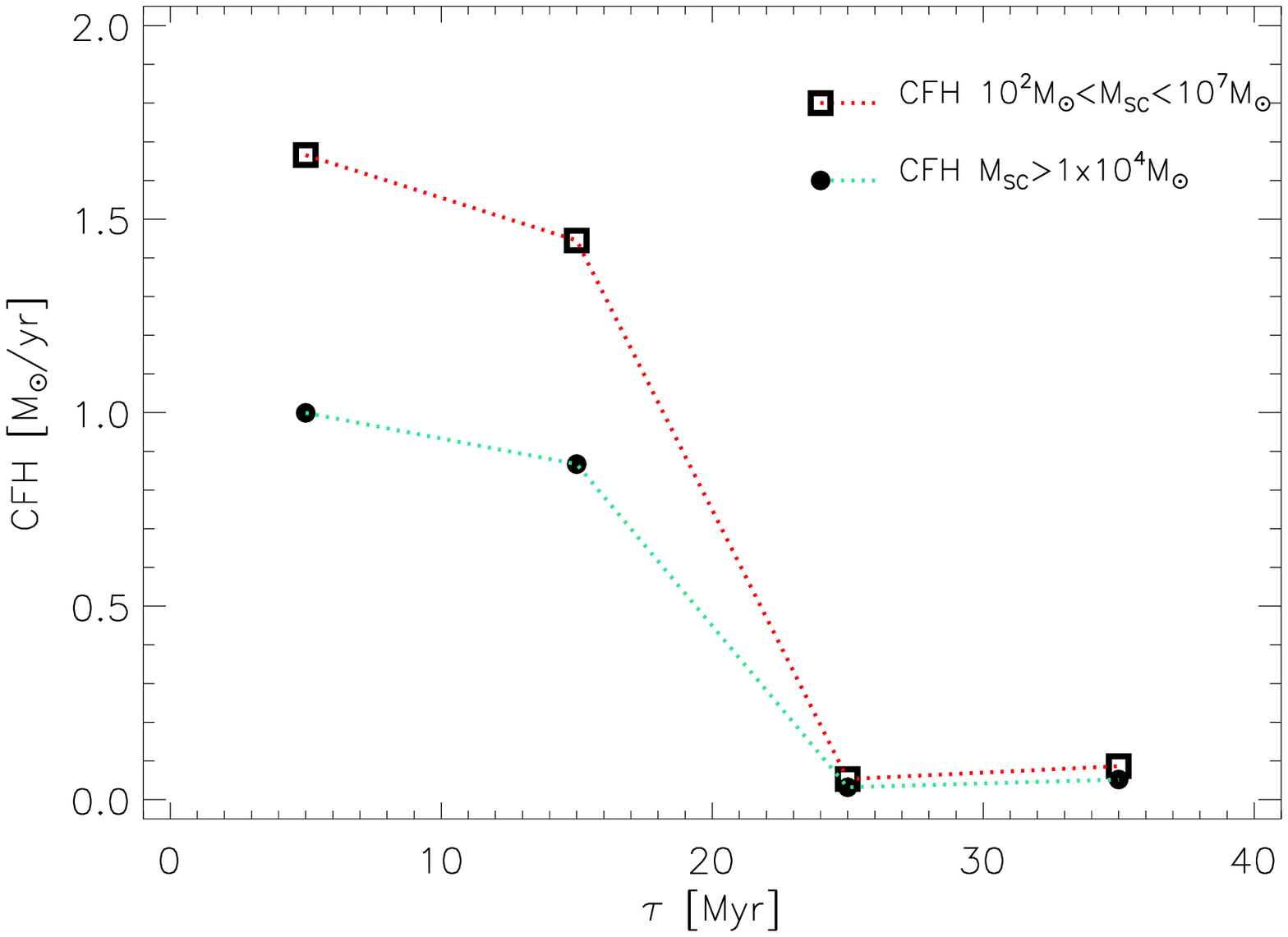}}}
\resizebox{0.48\hsize}{!}{\rotatebox{0}{\includegraphics{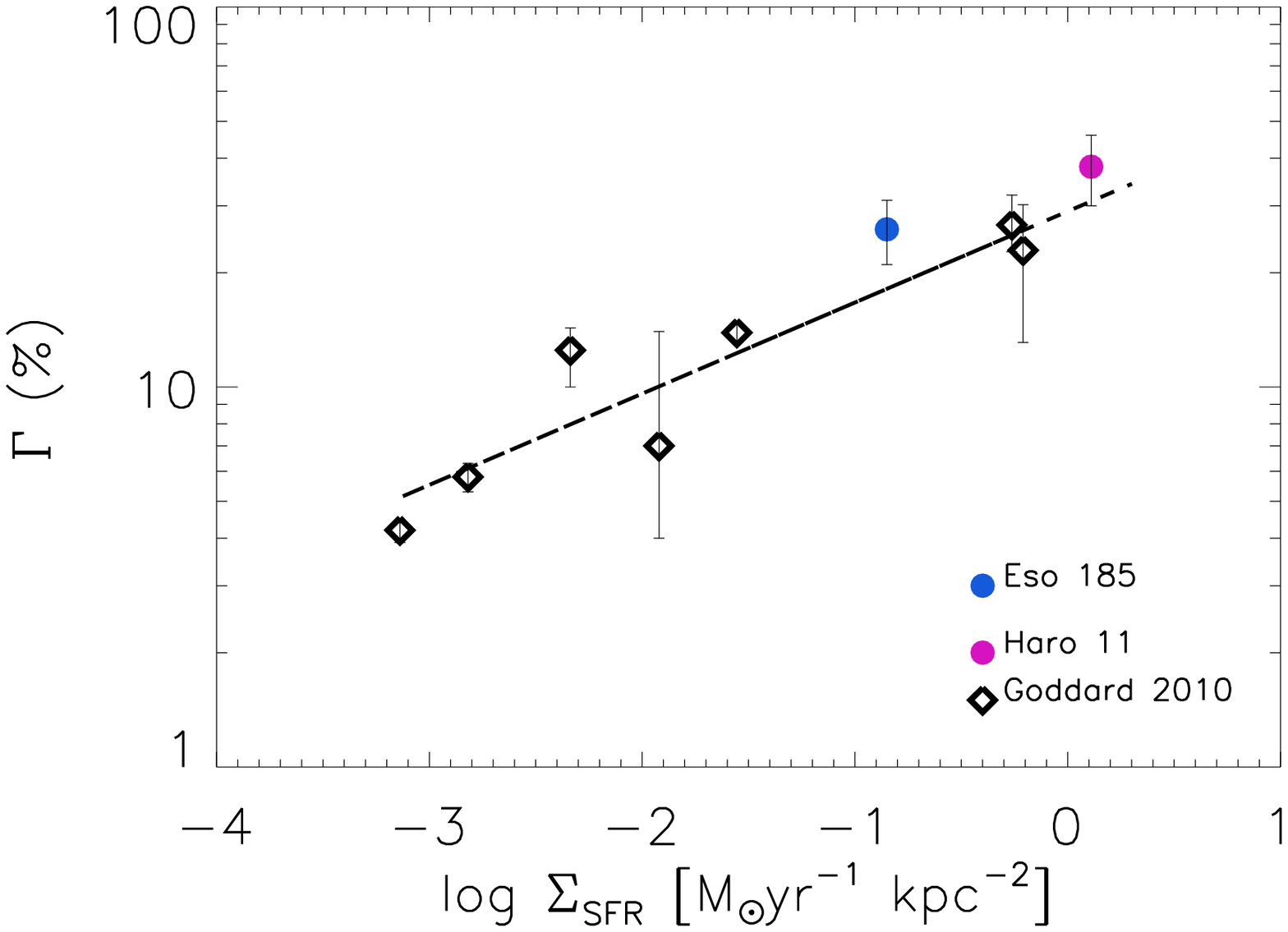}}}
\caption{Left panel: Cluster formation rates during the last 40 Myr of starburst activity. The filled dots connected by the green dotted line show the observed CFR derived from clusters more massive than $10^4 \msun$. The thick squares show the derived CFR, if the total mass in clusters less massive than $10^4 \msun$ is extrapolated using a CMF with index $-2.0$ down to $10^2 \msun$. Right panel: The cluster formation efficiency ($\Gamma $) vs the star formation rate surface density, using data from \citet{2010MNRAS.405..857G} with the  positions of Haro 11 \citep{A2010} and  ESO 185 added. See text for details.}
\label{CFR-plot}
\end{figure*}

\subsection{CFR versus SFR in ESO 185}
\label{cfr_sfr}

Using the total H$\alpha$ luminosity \citep{1999A&AS..137..419O} and the radius of ESO 185 (3.8 Kpc, \citealp{2001A&A...374..800O}), we applied the Kennicutt relation \citep{1998ARA&A..36..189K} to obtain an estimate of the present global star formation rate in the galaxy: SFR$\approx 6.4 \pm 1.0 \msun$yr$^{-1}$; and of the SFR surface density, $\Sigma_\mathrm{SFR}\approx0.14 \msun$yr$^{-1}$pc$^{-2}$. Assuming that a whole cluster population has formed every 10 Myr and that we are complete in detecting all clusters more massive than $10^4 \msun$, we derived the observed cluster formation rate, CFR. We assumed that the clusters have been formed with a power law CMF of index $-2.0$ and 
estimated the total CFR by extrapolating to clusters less massive than $10^4 \msun$ (a similar exercise was done in \citealp{A2010} following \citealp{2010MNRAS.405..857G}). 

Since we know the present SFR and CFR in the galaxy, we can estimate the cluster formation efficiency, $\Gamma$ \citep{2008MNRAS.390..759B}. $\Gamma$ is defined as the ratio between the CFR and SFR and is an estimate of the fraction of star formation happening in compact clusters. For ESO 185 we derived a $\Gamma = 0.26 \pm 0.05$, or in other words, $26 \pm 5$ \% of the star formation is happening in clusters. Although this ratio is smaller than the one found for Haro 11 (see Table 4 in \citealp{A2010}), is still high compared to many other starburst systems. Interestingly, \citet{2010MNRAS.405..857G} showed that the star formation density of the host systems and the corresponding values of $\Gamma$ are positively correlated, i.e.  that higher star formation rates in the host galaxy enable the production of more numerous and more massive clusters. As already done for Haro 11, we checked whether ESO 185 followed the same relation. In the right panel of Figure \ref{CFR-plot}, we see that the position of ESO 185 is only 2$\sigma$ above the expected value predicted by the relation (18 \%).

\citet{2001A&A...374..800O} produced an approximative estimate of the burst mass (i.e. the mass in young stars) in ESO 185: $2.1^{+0.85}_{-0.18} \times 10^8 \msun$. If we consider the extrapolated total mass in clusters younger than 40 Myr ($\sim 3.4 \times 10^7 \msun$) we 
estimate that roughly 16 \% of the burst mass has been produced in bound clusters.

In Figure \ref{CFR-plot}, we also show the cluster formation history in the last 40 Myr of ESO 185. The uncertainties in the estimation of the cluster ages between 20 and 10 Myr make the rate of cluster formation at those ranges not fully reliable. It is clear, however, that the galaxy is experiencing now a peak of cluster formation ($1.7 \msun$yr$^{-1}$).

\subsection{ESO 185 progenitors and the starburst evolution}
\label{gal-prop}

\begin{figure}
\resizebox{\hsize}{!}{\rotatebox{0}{\includegraphics{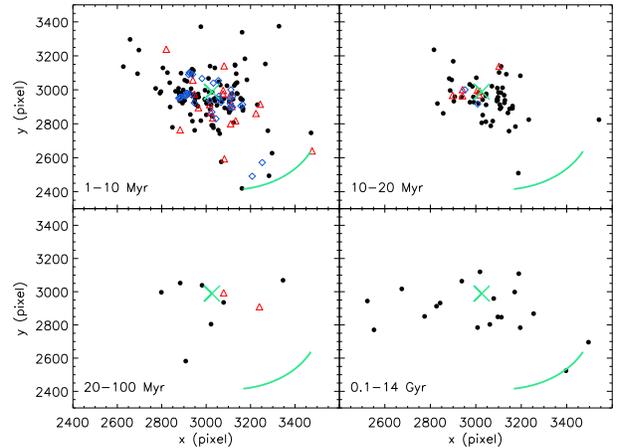}}}
\caption{Position of the clusters in the galaxy as function of their ages. Clusters affected by red excess are showed in red triangles (excess in H band), and in blue diamonds (excess in I and H bands). The center of the galaxy and the position of the shell are shown by a green cross and arc.}
\label{pos-gal}
\end{figure}

The analysis of ESO 185 morphology and gas dynamics reveals a galaxy that has experienced a merger event. 
The recent cluster production in the galaxy was probably caused by the infall of gas in the merger event.
The discovery of a shell structure in the deep {\it WFPC2} images and the numerous clusters in the galaxy could be used to trace the possible progenitors and evolution of the system. From \citet{2001A&A...374..800O} and \citet{1999A&AS..137..419O} we know that a second component in the galaxy is counter-rotating and has a mass which is $\sim1/10$ of the main component. Since the total stellar mass of the system is  $7\times10^9 \msun$ we expect the the merger process has involved dwarf systems. Even though the simulations performed by \citet{1997ApJ...481..132K} did not consider merger between dwarfs, we make a qualitative comparison with their results. 
From their 4 different sets of simulations, the ones produced by radial encounters are most similar to the morphology of ESO 185;  the merging disk produces a shell formed of both stars (from the early type galaxy) and gas (from the sinking satellite), while a tidal stream is left behind and a bar-like structure is formed in the center of the potential. The simulation with a smoothed potential well shows short-lived shell structures. They noticed that when the shells form the star formation in the galaxy is ceased due to the dissipative nature of the gas particles. In the real case of ESO 185, we derived the star formation history in the galaxy, following the formation of clusters.  In Figure \ref{pos-gal}, we show how clusters of different age ranges are located in the galaxy. The most evolved clusters, the globular clusters, older than 0.1 Gyr, appear to be biased towards the south part  of the galaxy. However this could be a projection effect, since we don't know the inclination of the galaxy. Two of these clusters are in the shell and have ages of 6 and 14 Gyr. These globular clusters were most likely part of the early type progenitor and in the oscillations around the potential that formed the shell dragged together with the stars. The panel in the bottom left side show clusters in the age range 20-100 Myr. A fraction of objects from the top right panels should in reality be in this older range. We observed that the clusters at 20-100 Myr and at 10-20 Myr age ranges are more central distributed. Two of the clusters in the shell with detection in more than 2 filters, have ages of 15-20 Myr. Clusters between the shell and the starburst region are younger than 10 Myr and visible in the top left panel of Figure \ref{pos-gal}. This panel shows also that in the last 10 Myr the starburst has rapidly expanded outward from the central region. 
 The final encounter between the galaxies happened between 100 and 20 Myr ago. The high rate at which the clusters have been formed in the last 20 Myr proves this scenario. In fact, the merging event has probably enhanced the pressure in the intergalactic medium and the compression of the gas, making the star formation more efficient. The shell is at a projected distance of only 5 Kpc from the center of the galaxy. Such proximity suggests that it has recently formed. Concentric with the shell and at approximately one third of the distance from the centre another bent narrow structure is visible that could possibly be an inner shell, or possibly a small spiral arm. If we assume that the outer shell (marked in Figure 1) was the first one and note that young clusters are found only out to the shell, but not beyond it, the shell  must be no older than 10 Myr. \citet{1997ApJ...481..132K} noticed that the formation of the shell happens when the system is entering in a post-starburst phase.  Similar conclusions have been reached by \citet{2007AJ....134.1729M}, who observed that the ages of the globular clusters in the shell correspond to the youngest stellar population formed in the galaxy before it went into a more quenching phase. It is plausible that after this burst episode, ESO 185 will experience a more quiet post starburst phase. However, the galaxy has a mass of neutral hydrogen, M$_{HI} \sim 3 \times 10^9 \msun$ (Bergvall et al., in prep.). At the current SFR ($\approx 6.4 \pm 1.0 \msun$yr$^{-1}$), the present burst could be active for another 500 Myr, although the fraction of gas close to the centre and currently available for star formation is unknown.

While the cluster formation in ESO\,185 is now very active  the most massive clusters that we find in this galaxy are  found among the $> 100$ Myr old clusters. The total mass in  clusters younger than 100 Myr and more massive than a mass completeness limit of $10^4\msun$ is $2\times10^7 \msun$. Assuming a power-law mass function with index $\alpha=-2$ for masses below 
we estimate a total mass in young clusters of $3.4\times10^7 \msun$.  We note that no mass has been ascribed to the red excess luminosity.

In the age range 100 Myr to 1 Gyr we find a total cluster  mass of $1.9\times10^7 \msun$ when counting clusters more massive than $10^5 \msun $. If we assume that clusters in this age range has a Gaussian CMF similar to that of Milky Way globular clusters ($M_{\rm peak}=2\times 10^5 \msun$, $\sigma = 0.6 \log(M) [\msun]$) the expected mass in clusters below $10^5$ is negligible. In this age range, and as previously noted, we see what appears to be a dearth of clusters with mass in the range a few times $10^4$ to $10^5$, but this has little impact on the total estimated mass of this sub-population (unless there are in reality many more low mass clusters than seen and accounted for).

For  old (age $>1$Gyr) clusters with properties similar to GCs we see a total mass of $3.1\times 10^7 \msun$ when including clusters with masses down to  $5\times10^5\msun$. Assuming the same Gaussian CMF as in the previous example, the total estimated mass for the
old GC population is $1.1\times10^8\msun$. The stellar mass of the galaxy underlying the starburst component was estimated by 
\citet{2001A&A...374..800O} to $M_{gal}\approx 6\times10^9\msun$ which would imply that the GCs in ESO\,185 make up more than a
percent of the total stellar mass. This value is high in comparison with most galaxies, and a similar result was found for ESO\,338--04  \citep{2003A&A...408..887O} and may indicate a greater survival rate for GCs in low mass systems. If we instead of the cluster population {\em mass} view the {\em number} of old clusters, we get a similar picture. 
The total number of old clusters  is 11. Correcting for completeness we expect 30 (again assuming a gaussian LF identical to that for the MW). 
From this we find a mass normalised specific frequency \citep{1998gcs..book.....A} of $T\approx 5$, which is typical for early type systems,
and consistent with the picture that ESO\,185 as we see it today is the product of a merger between an early type low mass
galaxy and a gas-rich dwarf.

\section{Conclusions}

This work, together with the analysis of ESO 338-IG04 \citep{2003A&A...408..887O}, Haro 11 \citep{A2010}, and MRK 930 (Adamo et al. in prep.) provides a statistical study of the properties of star clusters in blue compact galaxies and of their formation and evolution  in general.

In this paper we have used images obtained with the HST in several passbands ranging from the UV $\lambda_{cen}= 0.15\mu$m to the near IR ($1.6\mu$m)  of the luminous blue compact galaxy ESO\,185. The new HST images confirms the presence of a tidal tail/plume and in addition finds a shell, corroborating that this starburst was triggered by a (dwarf) galaxy merger. We find several hundreds of luminous star clusters which we analyse by means of SED fitting using state of the art models including a nebular gas emission.  
The clusters span a wide range of ages, from newborn $<1$Myr to old ($10$Gyr) globular clusters, and the masses range between a few times $10^3$ to $10^6 \msun$. A recent peak in the cluster formation is visible at an age of 3.5 Myr, and at least 20 \% of the present star formation is happening in compact (unresolved) star clusters. 

The dust extinction for young clusters present a large scatter, but rapidly decrease for ages $>5$Myr.
  This indicates that the embedded phase does not last for a fixed lapse of time, but is a transient phase in the cluster evolution (\citealp{2009arXiv0911.0779L}; \citealp{2009arXiv0911.0796L}). 

We found that a fraction of clusters younger than 6 Myr  displays a flux excess at $\lambda > 8000 \AA$, so-called NIR excess. 
We discuss possible scenarios which could explain why the observed fluxes sit above the prediction of evolutionary models. Hot dust and a considerable contributions from young stellar objects (YSOs) are among the advocated explanations for similar excess cases found in the literature. Resolved stellar populations of young star clusters in the SMC and in the Milky Way, show that even if the central part of the system is gas and dust free, star formation, triggered by positive feedback, is still active at the edges  \citep{1998ASPC..148..150E}. Considering that the very young star clusters in ESO\,185 are quite massive and that the photometric aperture used is spatially bigger than the real size of the compact clusters, we are likely including contributions from the edges of the star forming regions where hot dust and YSOs reside  in our photometry. The clusters in ESO\,185, for which we observe a NIR excess, may therefore be objects that are still in a partially embedded phase. This scenario could also explain why we detect those clusters at optical wavelengths.

Another fraction of clusters with ages older than 6 Myr showed a flux excess only for wavelengths  $\sim 1.6$ $\mu$m, referred to as IR excess. In this case, we conjecture that Red Super Giants (RSGs) may contribute to the  excess.

The  cluster luminosity function (CLF) becomes steeper at brighter luminosity and is very flat at the faint ends. A single power law fit gives an index of $-1.8$, which however is a rather poor representation of the CLF. We discuss possible mechanisms which could cause a flattening of the CLF. Blending and projection effects are addressed, together with a possible missing fraction of low mass clusters at very young ages (below 3 Myr) and around 30 Myr and 100 Myr. 

The tidal stream and  shell characterize the morphology of ESO\,185, and help us to trace back the progenitors of the present system. Using hydrodynamical simulations, \citet{1997ApJ...481..132K} showed that  a minor merger between an elliptical and a gas rich satellite, create shells of stars and gas, and tidal tails. Based on the age of the star clusters, the final encounter likely happened between 100-20 Myr ago. The clusters in and close to the shell  are not older than 20 Myr. The shell is approximatively 5 Kpc away from the center of the galaxy, proving that it has been recently formed. 
\citet{1997ApJ...481..132K}  observed that the formation of the shell and the ceasing of the star formation were coeval. Similarly, \citet{2007AJ....134.1729M} found the younger globular clusters in the post-starburst galaxy AM 0139-655 in the shell. 
 ESO\,185 will likely soon transit into a post burst phase. There is, however, a considerable fraction of HI that, at the current SFR, could support the starburst for the coming 500 Myr. 

\section*{Acknowledgments}
We thank {\it HST-helpdesk} for helpful suggestion. The referee, Peter Anders, is thanked for useful comments and suggestions made on this manuscript. A.A. thanks A. R. Lopez-Sanchez for making the saturated 3-colors image of the target. E.Z. acknowledges a research grant from the Royal Swedish Academy of Sciences. G.\"O. is a Royal Swedish Academy of Sciences research fellow, supported from a grant from the Knut and Alice Wallenberg foundation. A.A., G.\"O. and E.Z. also acknowledge support from the Swedish Research council and the Swedish National Space Board. M.H. acknowledges the support of the Swiss National Science Foundation. This research has made use of the NASA/IPAC Extragalactic Database (NED) which is operated by the Jet Propulsion Laboratory, California Institute of Technology, under contract with the National Aeronautics and Space Administration.

\appendix

\begin{figure*}
\resizebox{0.48\hsize}{!}{\rotatebox{0}{\includegraphics{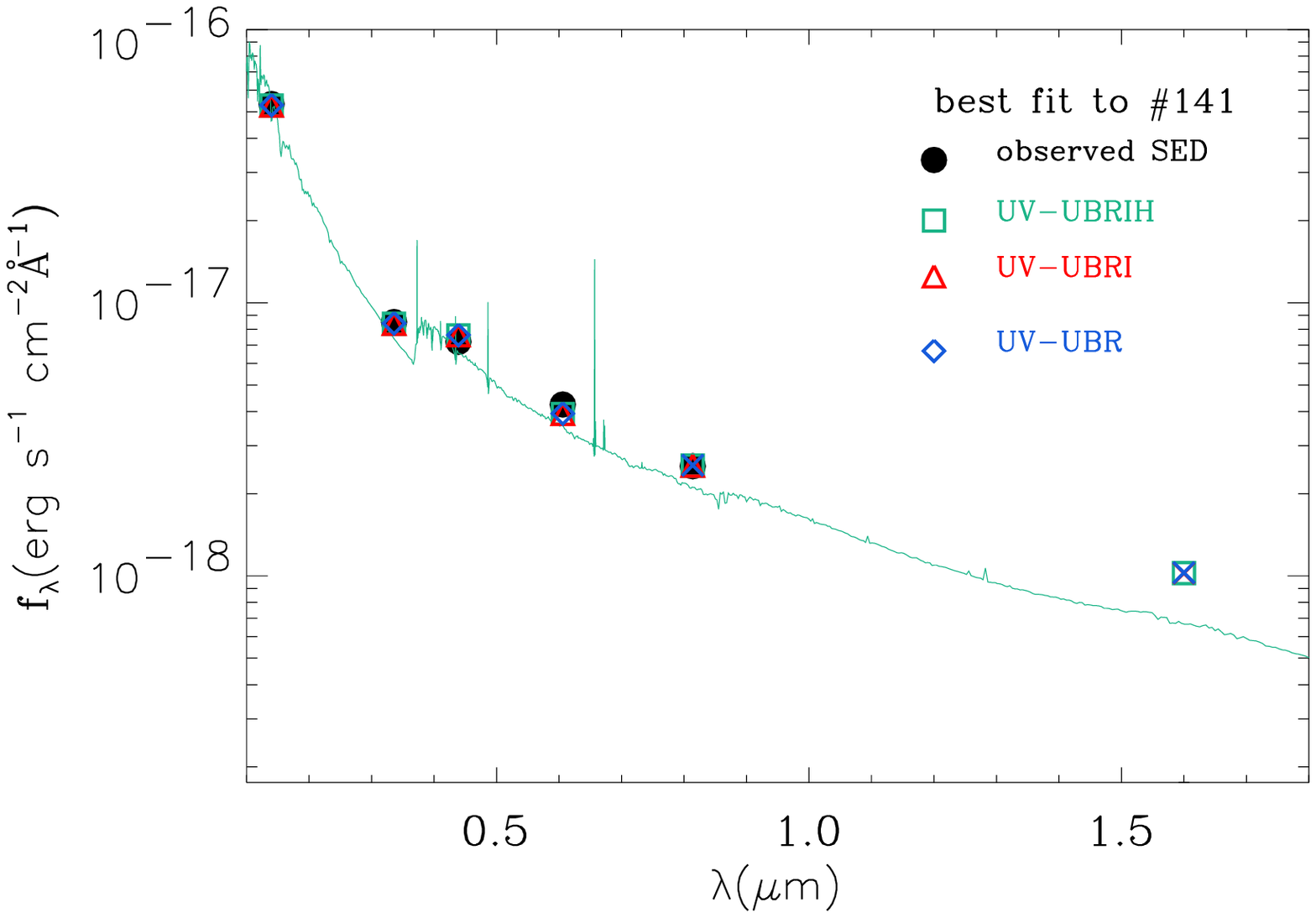}}}
\resizebox{0.48\hsize}{!}{\rotatebox{0}{\includegraphics{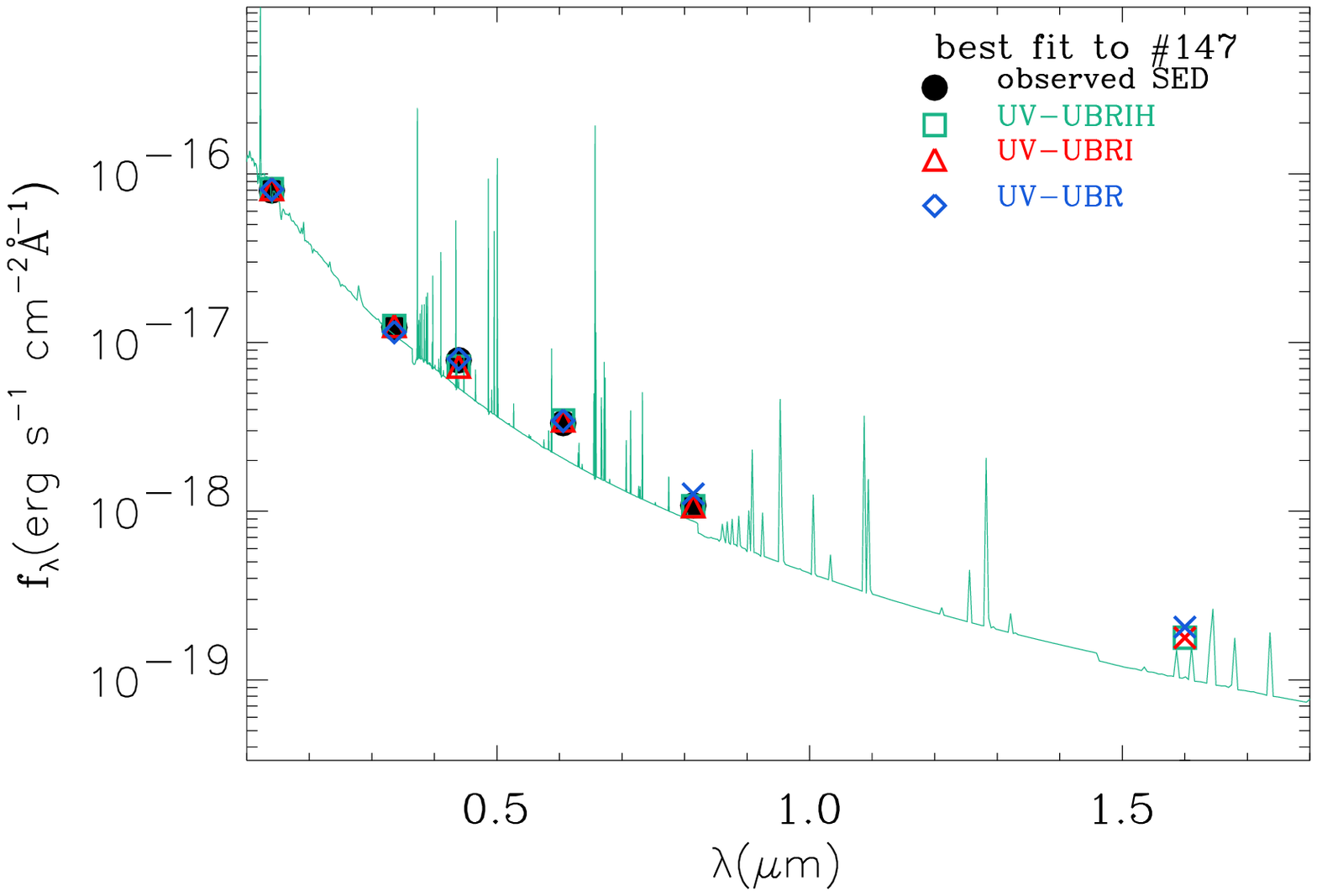}}} \\

\resizebox{0.48\hsize}{!}{\rotatebox{0}{\includegraphics{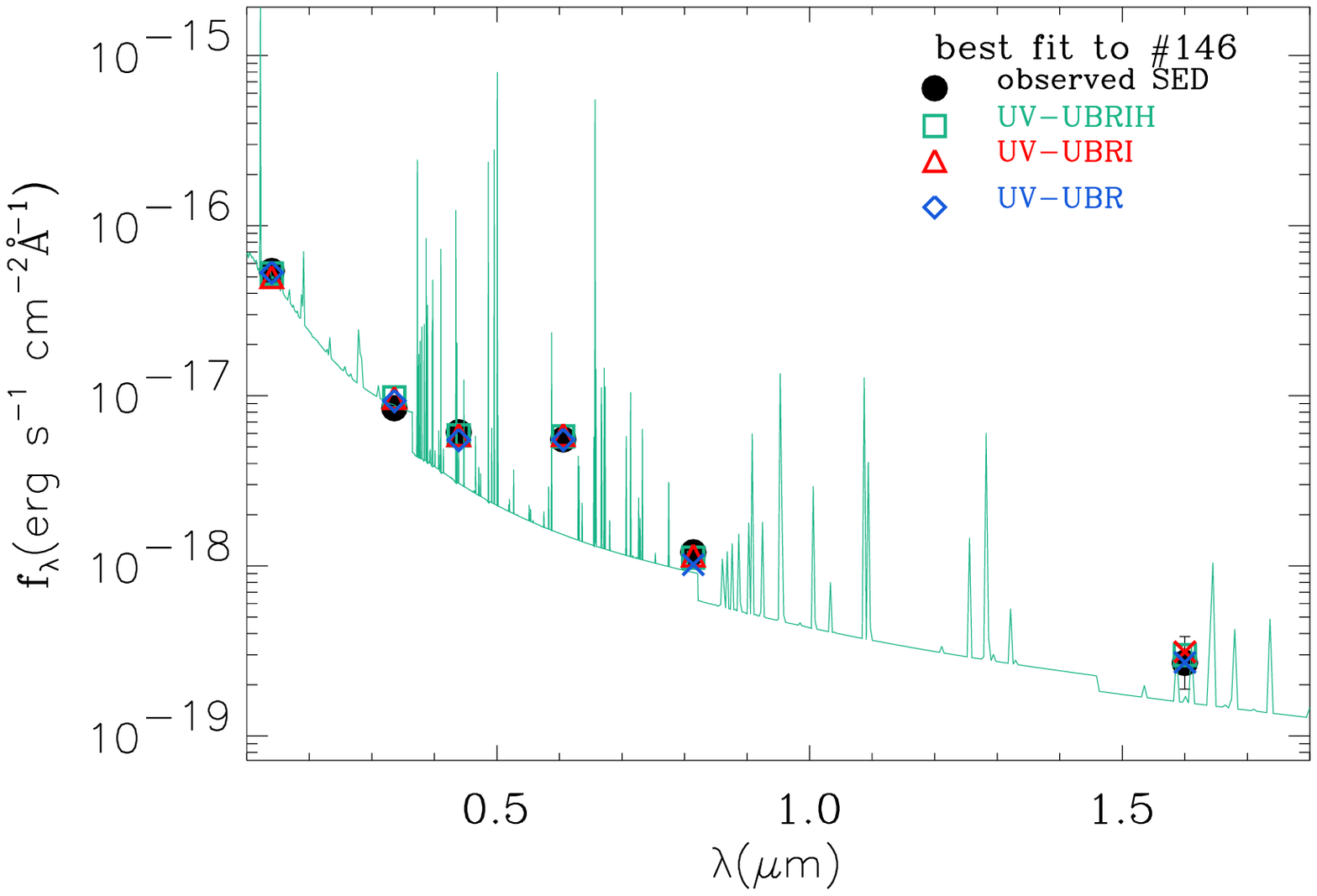}}}
\resizebox{0.48\hsize}{!}{\rotatebox{0}{\includegraphics{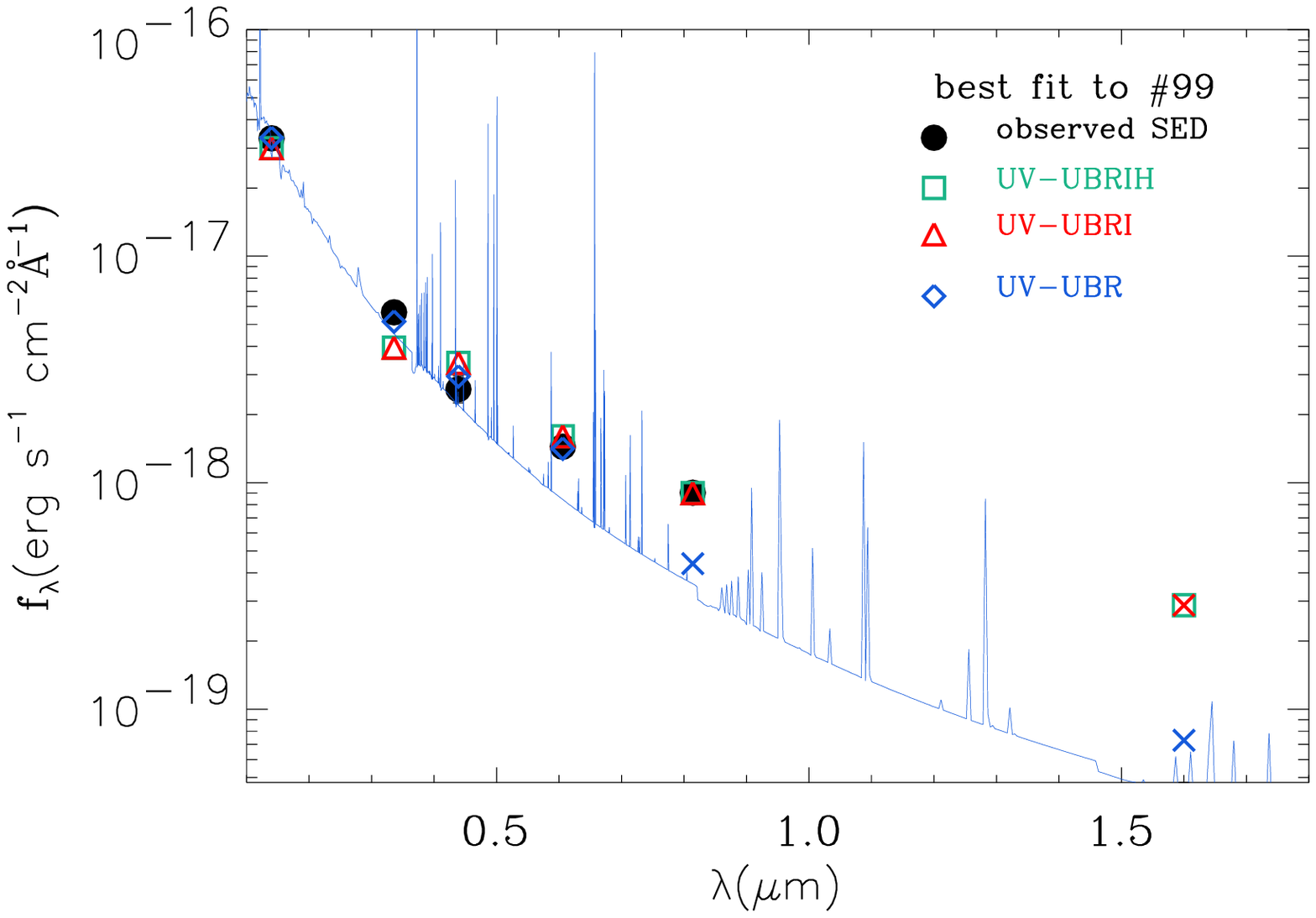}}} \\

\resizebox{0.48\hsize}{!}{\rotatebox{0}{\includegraphics{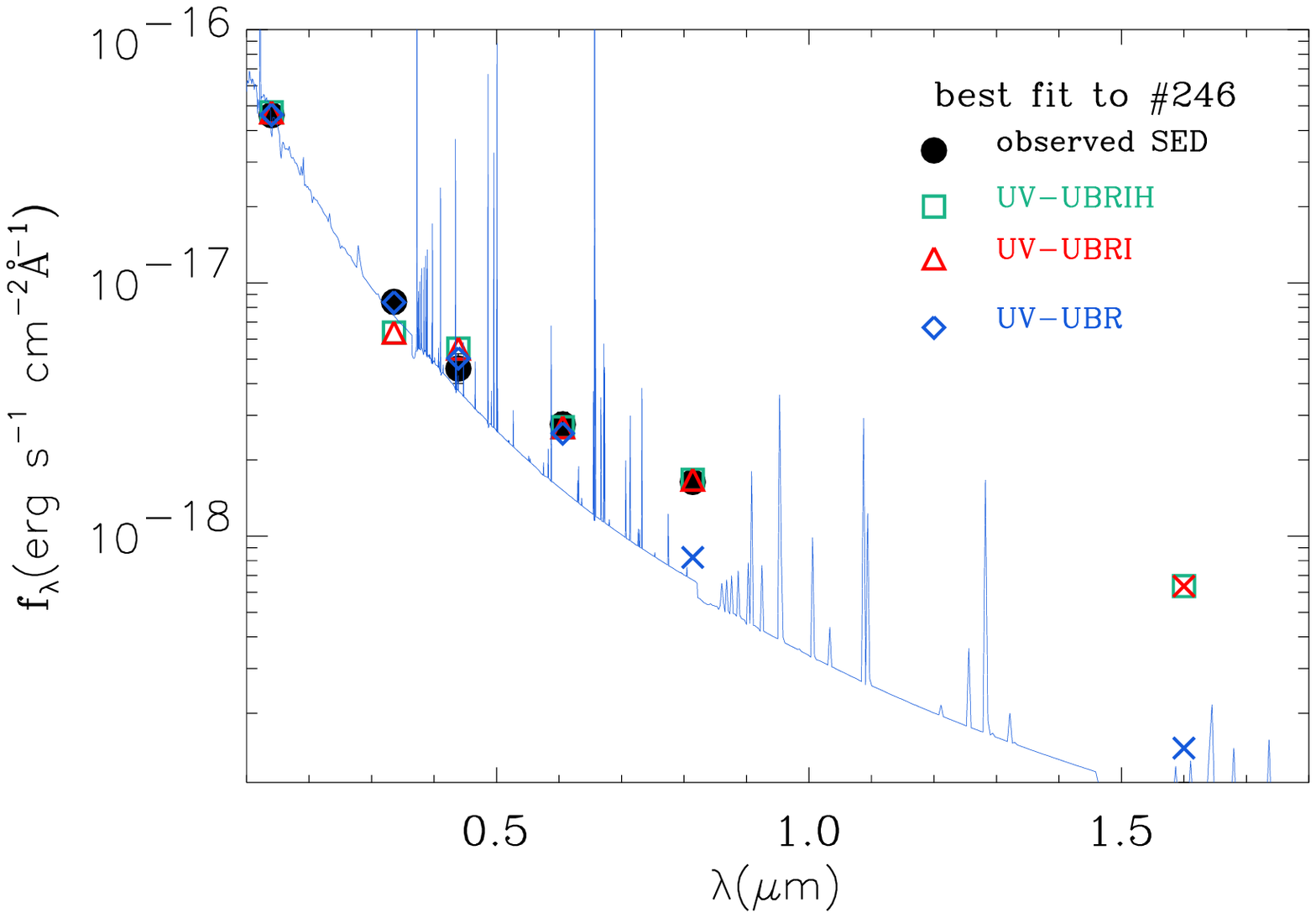}}}
\resizebox{0.48\hsize}{!}{\rotatebox{0}{\includegraphics{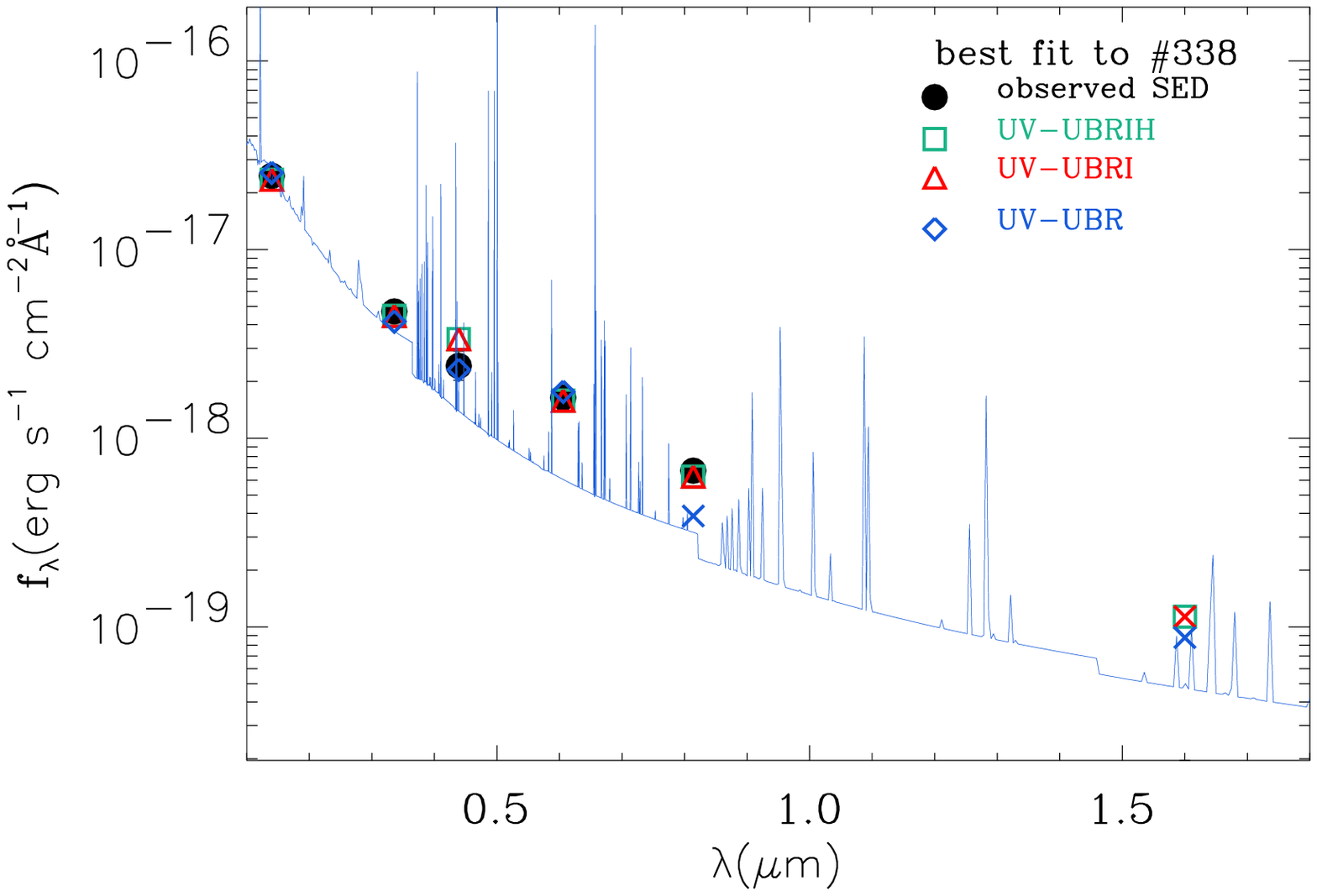}}} \\

\resizebox{0.48\hsize}{!}{\rotatebox{0}{\includegraphics{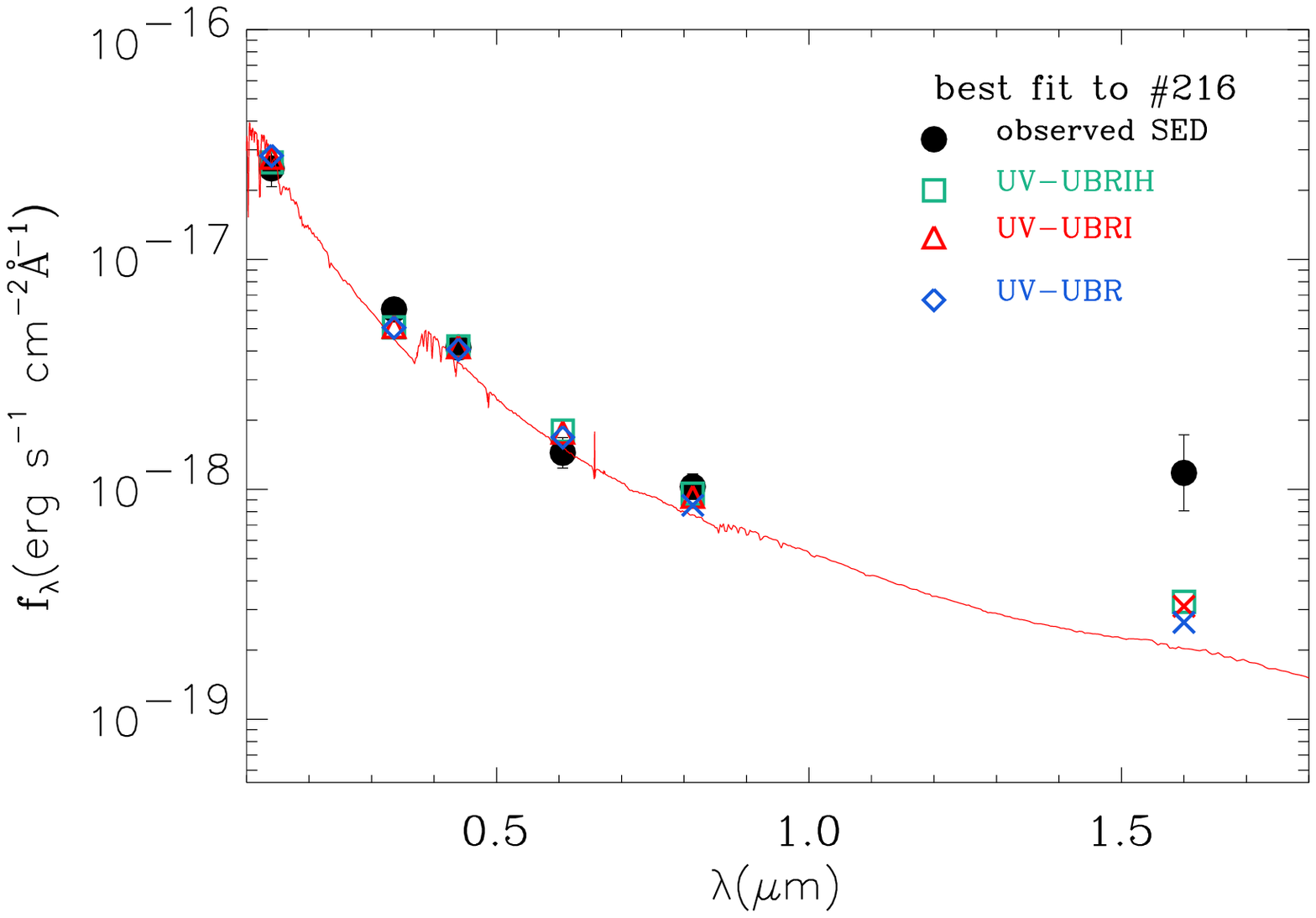}}}
\resizebox{0.48\hsize}{!}{\rotatebox{0}{\includegraphics{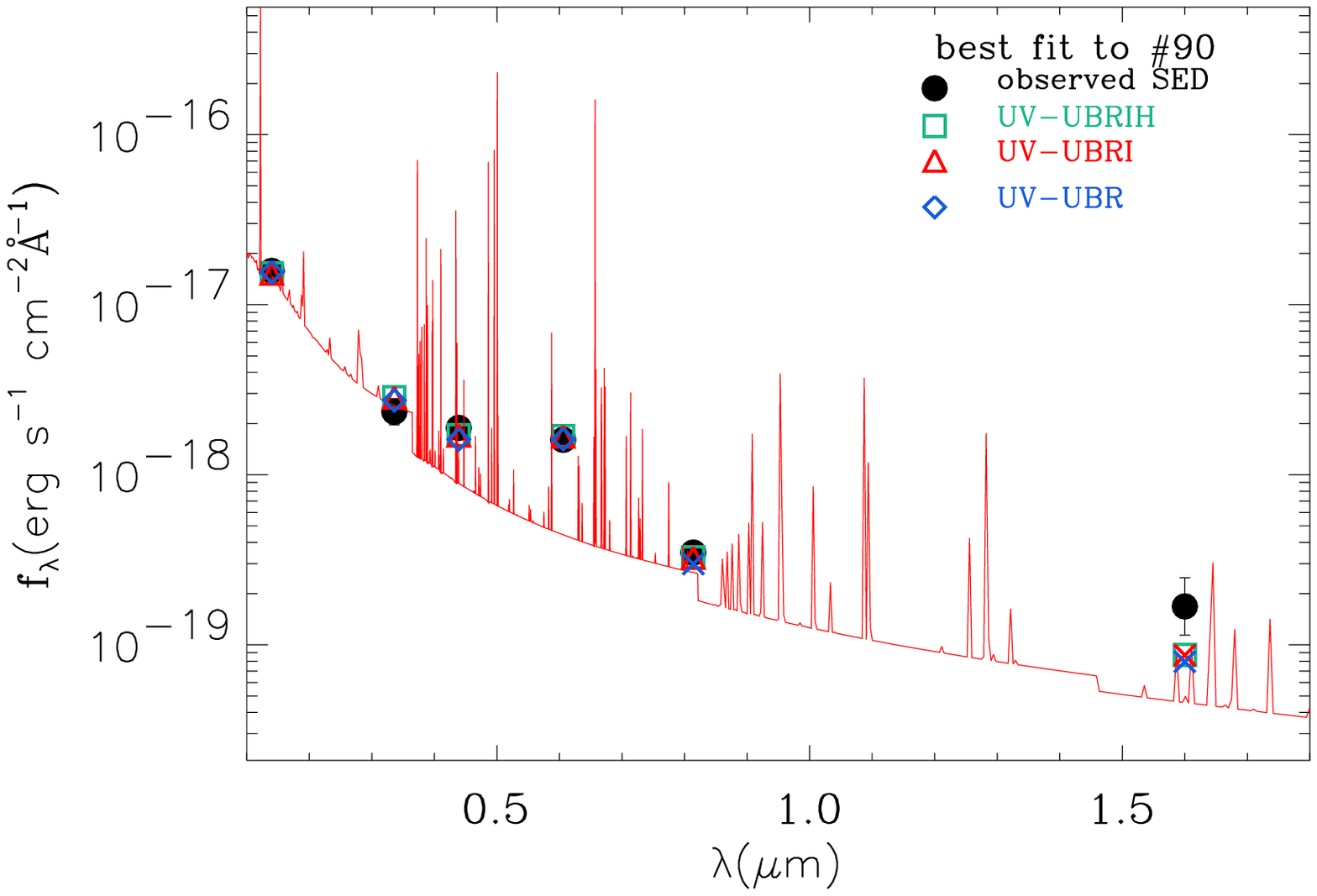}}} \\

\caption{Continue from Figure \ref{spec}}
\label{spec_app1}
\end{figure*}

\begin{figure*}
\resizebox{0.48\hsize}{!}{\rotatebox{0}{\includegraphics{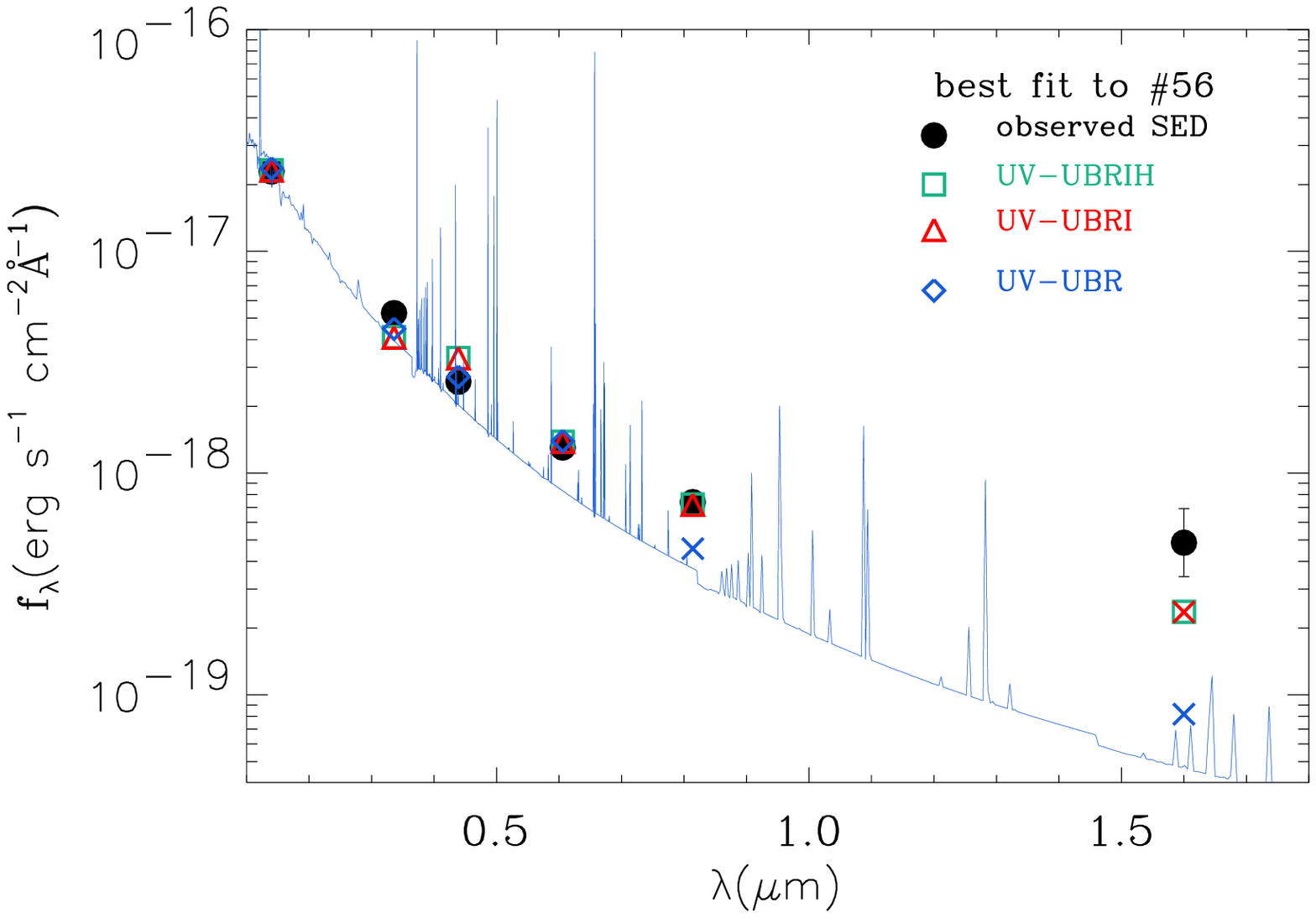}}}
\resizebox{0.48\hsize}{!}{\rotatebox{0}{\includegraphics{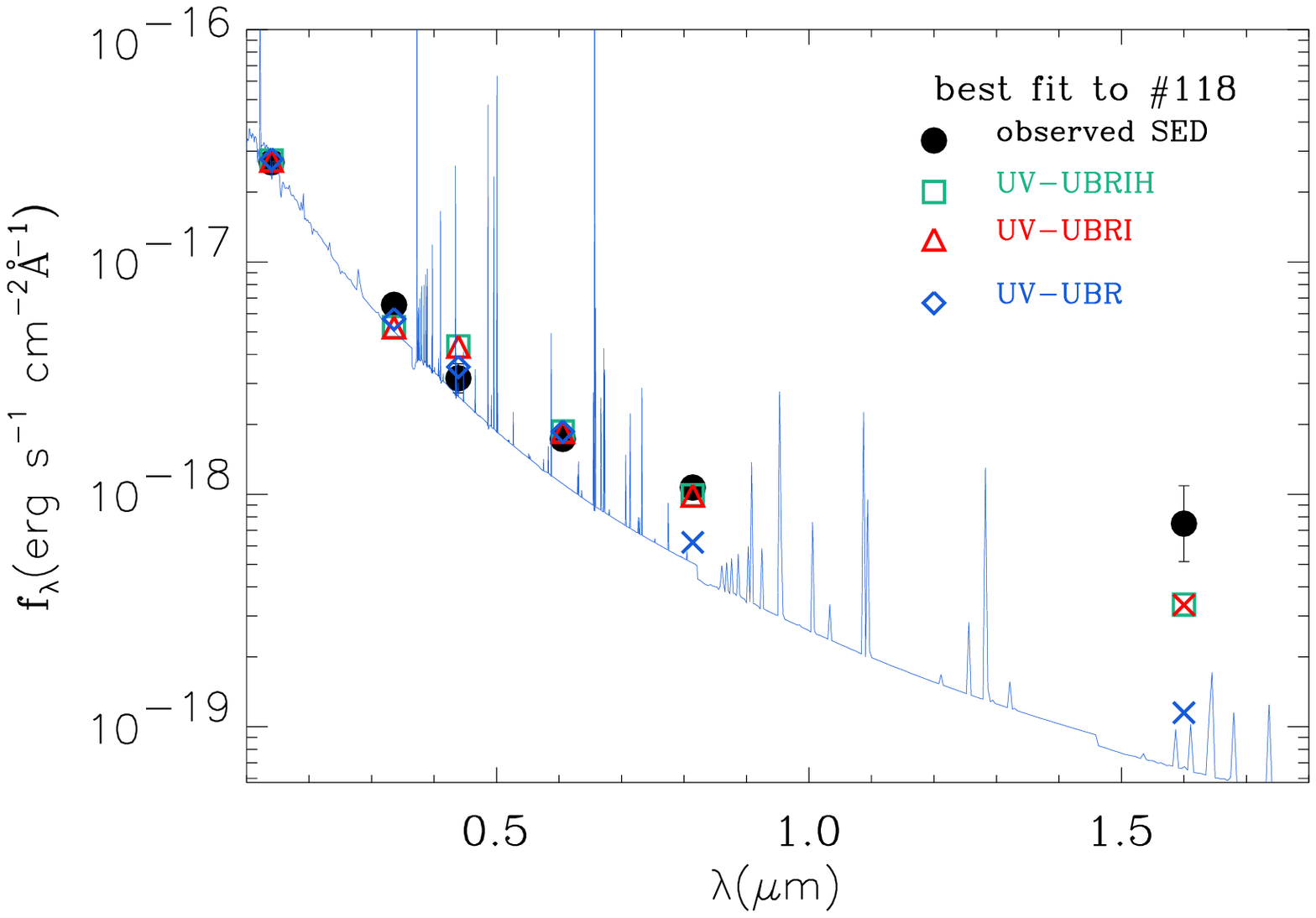}}} \\

\caption{Continue from Figure \ref{spec}}
\label{spec_app2}
\end{figure*}

\begin{table*}
  \caption{Continue from Table \ref{fit-output}}
\centering
  \begin{tabular}{|c|c|c|c|c|c|}
  \hline
  \hline
 IDs&$\chi_r^2$& $Q$&Myr&$10^5 \msun$&E(B-V)\\
 \hline
 &\multicolumn{4}{|c|}{{\it UV-UBRIH} fit}& \\  
 \hline  
      56   &   4.6   & 0.008 & 19.5  &    1.4   &  0.04\\
           90     &  1.4    &  0.392  & 1.5    &   0.2   &  0.08\\
       99      & 7.6   & 0.002   &8.5     &  0.4   &    0.0\\
   118    &   5.1   & 0.004   &19.5   &    2.0   &  0.06  \\
    141    & {\bf  0.6   }   &  {\bf 0.727   } & {\bf 9.5 }      &  {\bf 1.5  }  &  {\bf 0.05 }  \\
  146     & {\bf  2.5   }   &  {\bf 0.115  }  & {\bf 1.5    }  &  {\bf  0.5  }   &  {\bf 0.08 } \\
       147    & {\bf  0.6  }    &  {\bf 0.736  }   &{\bf 3.5   }    & {\bf  0.4 }     & {\bf 0.02 } \\
     216    &   2.9   &  0.066  & 19.5    &   2.0   &  0.06\\
  246    &   3.8    & 0.055  & 9.5     &  0.9  &  0.01\\
     338     &  2.9  &    0.1205 & 4.5   &    0.3   &  0.07 \\
 \hline
 &\multicolumn{4}{|c|}{{\it UV-UBRI} fit}&\\  
 \hline  
      56  &      5.8   &   0.009  &   19.5    &     1.4 &    0.04 \\
     90     &     {\bf 1.3   }  &    {\bf  0.460  } &  {\bf  1.5   }   &   {\bf   0.1 }    &    {\bf 0.08 } \\
       99    &    7.6   &   0.002  &   8.5     &    0.4   &      0.0 \\
   118    &      6.6   &   0.004  &   19.5    &     2.0   &    0.06 \\
    141    &    0.6    &    0.727  &   9.5    &     1.5   &    0.05\\
  146      &    3.6   &    0.063  &   1.5     &    0.6    &   0.09   \\
       147    &    0.6    &    0.736   &  3.5    &     0.4   &    0.02 \\
     216    &      {\bf 2.6   }  &     {\bf 0.160  }  &   {\bf 19.5  }     &    {\bf 1.9  }  &    {\bf 0.05 }\\
  246    &  3.8   &    0.055  &   9.5    &    0.9   &    0.01\\
     338     &   2.9  &     0.1205  &   4.5    &    0.3   &    0.07 \\

 \hline
 &\multicolumn{4}{|c|}{{\it UV-UBRIH} fit}&  \\  
 \hline  
      56   &       {\bf 3.7  }  &   {\bf  0.159  }  &  {\bf  3.5 }     &    {\bf  0.2  }   &   {\bf  0.07 } \\
     90     &       1.8 &    0.340     &    1.0    &     0.2   &    0.06\\
       99      &    {\bf  1.8 }     &   {\bf  0.410 }   &  {\bf 3.5 }      &   {\bf  0.1  }   &    {\bf 0.02 } \\
   118    &   {\bf     3.7  }    &    {\bf 0.160  }  &   {\bf 3.5  }    &    {\bf  0.2  }   &   {\bf  0.09 } \\
    141    &     1.3    &    0.524   &  9.5    &     1.5   &    0.05\\
  146     &     3.8    &    0.152   &     1.0    &    0.5  &    0.06\\
       147    &    0.8   &     0.654  &  4.5   &      0.4    &     0.0\\
     216    &      4.0  &     0.132   &  18.5  &      1.6   &    0.04\\
  246   &     {\bf 1.0  }   &    {\bf 0.601  }   & {\bf  3.5  }    &     {\bf 0.3 }    &    {\bf 0.06 } \\
     338     &   {\bf   1.7  }    &  {\bf   0.422  }  & {\bf  2.5 }     &     {\bf 0.2   }  &    {\bf 0.06 } \\
 \hline
 \hline
\end{tabular}
\label{tab-chi2}
\end{table*}

\begin{table*}
\caption{Photometric properties of the clusters showed in Figure~\ref{spec}, \ref{spec_app1}, and \ref{spec_app2}. Magnitudes are in Vega system throughput. The kind of excess found in the cluster is indicated in column 2. The errors gave in the table includes photometric errors and uncertainties due to aperture correction.}
\centering
  \begin{tabular}{|c|c|c|c|c|c|c|c|}
   \hline
IDs& Obs Excess &F140LP& F336W&F439W& F606W &F814W&F160W\\  
\hline
  \hline 
      56& I, H   &   21.10    $\pm$ 0.09    &  22.00    $\pm$ 0.10   &   23.52     $\pm$ 0.14   &   23.29     $\pm$ 0.11   &   22.96    $\pm$ 0.06    &  21.40     $\pm$ 0.38\\
      90&H      &21.51    $\pm$ 0.09     & 22.88     $\pm$ 0.19   &   23.87     $\pm$ 0.16   &   23.07     $\pm$ 0.11   &   23.78    $\pm$ 0.13   &   22.55     $\pm$ 0.42\\
      99& I   & 20.70     $\pm$ 0.10 &  21.92     $\pm$ 0.11   &   23.52     $\pm$ 0.15    &  23.19     $\pm$ 0.12   &   22.75    $\pm$ 0.07    &  - \\
     118& I, H    &  20.93   $\pm$ 0.11     & 21.76    $\pm$ 0.09&   23.31     $\pm$ 0.16   &  22.99     $\pm$ 0.12   &   22.56    $\pm$ 0.09  &    20.93     $\pm$ 0.41\\
     141& None     & 20.18    $\pm$ 0.08    &  21.48    $\pm$ 0.10   &   22.41   $\pm$ 0.08 &   22.01  $\pm$ 0.10   &   21.63  $\pm$ 0.05 & -\\
     146& None    & 20.17    $\pm$ 0.06     & 21.49    $\pm$ 0.08   &  22.59   $\pm$ 0.08  &   21.73    $\pm$ 0.10   &   22.44    $\pm$ 0.05    &  22.04     $\pm$ 0.39\\
     147& None     & 19.75    $\pm$ 0.06    & 21.08    $\pm$ 0.08  &  22.32    $\pm$ 0.09  &  22.28     $\pm$ 0.11   &   22.55   $\pm$ 0.07   &  -\\
     212& I, H     & 19.66    $\pm$ 0.06  &  20.76   $\pm$ 0.07&    22.21    $\pm$ 0.08   &   21.93$\pm$ 0.10   &   21.61   $\pm$ 0.05&    20.37     $\pm$ 0.39\\
     215& H     & 21.38     $\pm$ 0.18 & 21.87     $\pm$ 0.11 &   23.20    $\pm$ 0.10  &  22.65     $\pm$ 0.10   &   22.80   $\pm$ 0.07    &  21.04     $\pm$ 0.38\\
     216&H     & 21.01     $\pm$ 0.20   & 21.84     $\pm$ 0.12  &  23.01   $\pm$ 0.11  &   23.18 $\pm$ 0.16   &   22.61     $\pm$ 0.14    &  20.44     $\pm$ 0.41\\
     246&I     & 20.35     $\pm$ 0.11   &   21.49     $\pm$ 0.11&   22.90   $\pm$ 0.15   &   22.48    $\pm$ 0.12    &  22.10   $\pm$ 0.10   &   -\\
     286&None     & 18.98    $\pm$ 0.05  &  20.62   $\pm$ 0.06  &    21.63   $\pm$ 0.06  &   21.26  $\pm$ 0.09   &   20.90  $\pm$ 0.04    & 19.92     $\pm$ 0.39\\
     338&I     & 21.02 $\pm$ 0.14  & 22.12   $\pm$ 0.12  &  23.60     $\pm$ 0.19  &   23.04   $\pm$ 0.11  &  23.07     $\pm$ 0.10    &  -\\
     344&I     & 20.41    $\pm$ 0.10 &    21.56   $\pm$ 0.11   &  23.05$\pm$ 0.16  &   22.51$\pm$ 0.14  &    22.35    $\pm$ 0.10    &  -\\

 \hline
  \end{tabular}
  \label{tab-phot}
      \end{table*}

\bsp

\label{lastpage}

\end{document}